\documentclass[12pt]{article}
\usepackage{epsfig,amssymb,amsmath,psfrag,subfigure,rotate,slashed,float,caption}

\allowdisplaybreaks

\def \be  {\begin{equation}}
\def \ee  {\end{equation}}
\def \ba  {\begin{eqnarray}}
\def \ea  {\end{eqnarray}}
\def \baa {\begin{eqnarray*}}
\def \eaa {\end{eqnarray*}}
\def \lab #1 {\label{#1}}

\newcommand\re[1]{(\ref{#1})}
\def\d{\hbox{{d}\kern-.20em\hbox{l}}}

\def \matrix #1 {\left(\begin{array}{cc} #1 \end{array}\right)}

\def\II{\hbox{{1}\kern-.25em\hbox{l}}}

\newcommand \vev [1] {\langle{#1}\rangle}

\newcommand \ket [1] {|{#1}\rangle}
\newcommand \bra [1] {\langle {#1}|}
\newcommand{\bit}[1]{\mbox{\boldmath$#1$}}

\newcommand{\ft}[2]{{\textstyle\frac{#1}{#2}}}

\begin{document}

\begin{titlepage}

\thispagestyle{empty}

\vspace*{3cm}

\centerline{\large \bf Renormalization of twist-four operators in light-cone gauge}
\vspace*{1cm}

\centerline{\sc Yao Ji, A.V.~Belitsky}

\vspace{10mm}

\centerline{\it Department of Physics, Arizona State University}
\centerline{\it Tempe, AZ 85287-1504, USA}

\vspace{1cm}

\centerline{\bf Abstract}

\vspace{5mm}

We compute one-loop renormalization group  equations for non-singlet twist-four operators in QCD. The calculation heavily relies on the light-cone gauge formalism 
in the momentum fraction space that essentially rephrases the analysis of all two-to-two and two-to-three transition kernels to purely algebraic manipulations both for 
non- and quasipartonic operators. This is the first brute force calculation of this sector available in the literature. Fourier transforming our findings to the coordinate space, 
we checked them against available results obtained within a conformal symmetry-based formalism that bypasses explicit diagrammatic calculations and confirmed 
agreement with the latter.
\vspace{5cm}

\centerline{\it Dedicated to the memory of Eduard A. Kuraev}

\end{titlepage}

\setcounter{footnote} 0

\newpage

\pagestyle{plain}
\setcounter{page} 1

{\footnotesize \tableofcontents}

\newpage

\section{Introduction}
\label{Intro}

The leading power approximation to QCD processes with large momentum transfer, such as the deep-inelastic and deeply virtual Compton scattering, 
admits an intuitive probabilistic description in the framework of the Feynman parton model \cite{Fey71}. According to the latter, physical cross sections are 
expressible in terms of (generalized) parton distribution functions. The QCD improved picture arises via systematic inclusions of quantum corrections to 
probe-parton scattering amplitudes as well as renormalization effects of leading twist Wilson operators that parametrize Feynman parton densities. More subtle 
effects arise from power-suppressed contributions to hadronic cross sections since they encode information on interference of hadronic wave functions with 
different number of partons. On the one hand, these are of interest in their own right since they provide access to intricate QCD dynamics \cite{ShuVai82}. 
On the other hand, they can be regarded as a QCD contaminating background to high precision measurement of New Physics, see, e.g., \cite{EWDIS}. In 
either case, understanding these contributions quantitatively is indispensable at the precision frontier. Since data is typically taken at different values of the 
momentum transfer, at some point one has to incorporate effects of logarithmic scaling violation stemming from renormalization of higher twist operators. 
The task of their unravelling at twist-four level will be undertaken in the present study.

Until very recently, only partial results for certain subsets of operators were available in the literature \cite{Twist4}. A special class of operators out of all higher 
twists is known as quasipartonic. They can be characterized either as composite fields built from on-shell fields of the Feynman parton model or 
understood as operators with their twist equal to their length, i.e., the number of fields that form them. For this class of operators  a systematic approach to 
constructing high-twist evolution equations was developed about  three decades ago by Bukhvostov, Frolov, Lipatov and Kuraev in Ref.\ \cite{BFLK}. At leading 
order in QCD coupling, the evolution kernel for these was found as a sum of pairwise interaction kernels between elementary fields comprising the operators in 
question. The particle number-preserving nature allows one to map it to a Hamiltonian quantum mechanical problem. This advantage was explored in
a number of works at twist three level \cite{Twist3}\footnote{For a more recent discussion of operator renormalization arising in certain single-spin 
asymmetries, see \cite{SingleSpin}.} starting from \cite{Bukhvostov:1983te}. Eventually, the problem was mapped into an exactly solvable lattice model 
\cite{BraExact98,BelExact98}. However, while the quasipartonic operators form a subset closed under the renormalization group evolution \cite{BFLK,BukFro87}, 
they do not exhaust the set of all operators contributing at a given twist. The remaining ones are dubbed non-quasipartonic and they contain at least one bad field 
component in the formalism of light-cone quantization. These operators are characterized by the property that their twist is greater than their length. Their 
evolution does not preserve the number of fields in quantum transitions and thus their study is more elaborate. In the twist three case alluded to above, 
this was not a pressing issue since the use of QCD equations of motion allows one to remove all non-quasipartonic operators from the basis. For even 
higher twists, this is not sufficient and particle number changing transitions involved in the analysis of non-quasipartonic operators have to be addressed 
explicitly. 

The analysis of the renormalization problem for twist-four operators was completed recently in the coordinate space \cite{Braun:2009vc}, i.e., in terms of 
light-ray composite operators. The formalism is based on the use of conformal symmetry preserved by leading order QCD evolution equations, Poincar\'e 
transformations in the transverse plane and a minimal input from Feynman graphs. Presently we will perform a brute-force computation of Feynman 
diagrams in the light-cone gauge and rely on the momentum-space technique which makes the underlying calculation rather straightforward. For an 
exception of a few subtleties with the use of QCD equation of motion to recover the particle-number increasing transitions, it reduces to a few algebraic,
though rather tedious, steps.

The choice of the operator basis at higher twists is not unique due to multiple relations among a redundant set of operators via QCD equations of motion. 
Thus it is  driven by requirements of simpler transformation properties under residual (conformal) symmetry as well as simplicity of underlying calculations. 
In the current work we will adopt the basis of twist-four operators suggested in Ref.\ \cite{Braun:2009vc}. This will allow us to verify our results obtained 
by an independent calculation based on a different technique. Since we will focus on the twist-four sector, we have three types of building blocks at our 
disposal as two-particle elements of operators in question: good-good, good-bad and bad-bad light-cone field components. According to traditional 
classification, they possess twists two, (at least) three and (at least) four, respectively. We will address only the first two types, since the last one can be 
eliminated in hadronic matrix elements in favor of the other ones containing more fields via QCD equations of motion, as discussed below. Our consideration 
will be limited to QCD nonsinglet sector, though partial results for two-to-two transitions will be reported for the singlet sector as well.

Our subsequent presentation is organized as follows. In the next section, we spell out the operator basis used in the current calculation and provide a
dictionary between the twistor notations adopted in Ref.\ \cite{Braun:2009vc} and the light-cone conventions used in the present analysis. Then, we 
discuss the general structure of twist-four evolution equations and provide a Fourier transform bridge between the light-ray and momentum fraction 
space representations. In Sects. \ref{2to2quasi}, \ref{2to2nonquasi} and \ref{2to3transition}, we present evolution kernels for two-to-two quasipartonic, 
non-quasipartonic and two-to-three transitions, respectively. As a result 
of this analysis we find a simplified form of light-ray evolution kernels for certain evolution kernels which are reported in the Appendices. The latter also 
contain technical details on the calculation of Feynman diagrams defining operator mixing as well as singlet two-to-two transitions.

\section{Operator basis}
\label{Basis}

The light-cone dominated processes are parametrized by matrix elements of composite operators built up by fields localized on a light-cone ray defined by the 
vector $n_\mu = (1, 0, 0, 1)/\sqrt{2}$ that is reciprocal to the large light-cone component of the momentum transfer. Thus they have the following generic form
\begin{align}
\label{GenericO}
\mathbb{O} (z_1, \dots, z_N) 
=  
C_{I_1I_2 \dots I_N} [z_0^-, z_1^-]_{I_1 J_1} X_1^{J_1} (z_1^-)  [z_0^-, z_2^-]_{I_2 J_2} X^{J_2} (z_2^-) \dots  [z_0^-, z_N^-]_{I_N J_N} X_N^{J_N} (z_N^-)
\, ,
\end{align}
where the $X$-field cumulatively stands for certain components of quark and gluon fields as explained below. The positions $z_k^- = \bar{n} \cdot z_k$ of the 
fields on the light-cone are defined with the help of a tangent null vector $\bar{n}^\mu = (1, 0, 0, -1)/\sqrt{2}$ to the light-cone normalized such that $n \cdot \bar{n} = 1$.
The gauge invariance of $\mathbb{O}$ is achieved by means of an appropriate contraction of the color indices $I_k$ (either in the (anti-)fundamental 
$I_k = i_k$ or adjoint representation $I_k = a_k$ of the color group) into an $SU (N)$ singlet with a tensor $C_{I_1I_2 \dots I_N}$ and field coordinates 
parallel transported to an arbitrary position $z_0^-$ with the help of the Wilson lines
\begin{align}
[z_0^-, z_k^-]
=
P \exp 
\left( i g \int_{z_k^-}^{z_0^-} dz^- \, A^+ (z^-)
\right)
\, .
\end{align}
Here 
\begin{align}
A^+ = n \cdot A = \ft{1}{\sqrt{2}}(A^0 + A^3) 
\end{align}
is the light-cone projection of the gauge field. 

\subsection{Good and bad light-cone fields}

It is well-known that the light-cone gauge
\begin{align}
\label{LCgaugeCondition}
A^+ = 0
\, ,
\end{align}
has a number of advantages. First, we observe that the gauge links are gone in Eq.\ \re{GenericO} and, as a consequence, this results in reducing of 
the number of diagrams contributing to loop amplitudes. Second, the Feynman parton model arises naturally from the light-cone gauge QCD. Namely,
one decomposes the quark $\Psi$ and gluon fields $A^\mu$, 
\begin{align}
\Psi 
= \ft12 \gamma^- \gamma^+ \Psi + \ft12 \gamma^+ \gamma^- \Psi \equiv \Psi_+ + \Psi_-
\, , \qquad
A^\mu 
= n^\mu A^- -  \bar{e}_\perp^\mu A_\perp - e_\perp^\mu \bar{A}_\perp 
\, ,
\end{align}
in terms of the good $X_+ = \{ \Psi_+, A_\perp, \bar{A}_\perp \}$ and the bad $X_- = \{ \Psi_- , A^- \}$ components, respectively. 
Note that for the vector $A^\mu$ we defined its minus as follows
projection 
\begin{align}
A^- = \bar{n} \cdot A = \ft{1}{\sqrt{2}}(A^0 - A^3) 
\, ,
\end{align} 
and, in addition, we decomposed the transverse gauge field in terms of its anti- and holomorphic components 
\begin{align}
\bar{A}_\perp = \bar{e}_\perp \cdot A = \ft{1}{\sqrt{2}} (A^1 - i A^2)
\, , \qquad
A_\perp = e_\perp \cdot A=  \ft{1}{\sqrt{2}} (A^1 + i A^2)
\end{align}
with the help of the vector $e_\perp^\mu = (0,- 1, - i,0)/\sqrt{2}$ (and its complex conjugate $\bar{e} = e^\ast$). These possess helicity $h = \pm 1$, 
respectively, being eigenvalues of the helicity operator \cite{BelDerKorMan04}
\begin{align}
\label{Helicity}
H \equiv \bar{e}_\perp^\mu e_\perp^\nu \Sigma_{\mu\nu}
\, ,
\end{align}
that is built from the spin tensor $\Sigma_{\mu\nu}$ entering the Lorentz generators $i M_{\mu\nu}$. The bad components 
being non-dynamical in the light-cone time $z^+$ can be integrated out in the path integral and, thus, only the on-shell propagating modes $\Psi_+$, $A_\perp$
and $\bar{A}_\perp$ are left. We will not perform this step however and keep all non-propagating degrees of freedom in the QCD Lagrangian since the 
classification of operators will be easier in this case and moreover one does not loose Lorentz covariance. Finally, it is straightforward to construct an operator 
basis making use of the above building blocks, namely, the field $X$ in Eq.\ \re{GenericO} will have the following components (as well as their Hermitian conjugates 
$X^\dagger$)
\begin{align}
X = \{ X_+, X_-,  D_\perp X_+\}
\, ,
\end{align}
with $D_\perp = e_\perp \cdot D$ being the holomorphic covariant derivative $D_\mu = \partial_\mu - i g A_\mu$.

\subsection{Twistor representation}

To make a connection to the basis of Ref.\ \cite{Braun:2008ia}, let us recall the twistor formalism used there. We pass to the spinor representation for Lorentz vectors
by contracting them with the two-dimensional block $\sigma^\mu$ of four-dimensional Dirac matrices in the chiral representation $\gamma^\mu
= {\rm antidiag} (\bar\sigma^\mu, \sigma^\mu)$, e.g.,
\begin{align}
x_{\alpha\dot\alpha} = x_\mu \sigma^\mu_{\alpha\dot\alpha}
\, ,
\end{align}
where $\sigma^\mu = (1, \bit{\sigma})$ and $\bit{\sigma}$ is the three-vector of Pauli matrices, while $\bar\sigma^\mu =  (1, - \bit{\sigma})$. The light-cone vectors 
$n$ and $\bar{n}$ can be factorized into two twistors $\lambda_\alpha$ and $\mu_\alpha$
\begin{align}
n_{\alpha\dot\alpha} = \lambda_\alpha \bar\lambda_{\dot\alpha}
\, , \qquad
\bar{n}_{\alpha\dot\alpha} = \mu_\alpha \bar\mu_{\dot\alpha}
\, ,
\end{align}
where $\lambda_\alpha^\ast = \lambda_{\dot\alpha}$ and $\mu_\alpha^\ast = \mu_{\dot\alpha}$. For the light-cone vectors introduced in the
previous section, we can choose the two-dimensional spinors as $\lambda^\alpha = (0, \sqrt[4]{2})$ and $\mu^\alpha = (\sqrt[4]{2}, 0)$. These
twistors will allow us to construct good and bad fields for specific helicities. Namely, using the decomposition of the Dirac quark field in chiral 
representation
\begin{align}
\Psi = 
\left( {\psi_\alpha \atop \bar\chi^{\dot\alpha}} \right)
\, , 
\end{align}
we can introduce their good and bad components as follows
\begin{align}
\psi_+ = \vev{\lambda \psi}
\, , \qquad
\bar\psi_+ = [\bar\psi \bar\lambda]
\, , \qquad
\psi_- = \vev{\mu \psi}
\, , \qquad
\bar\psi_- =  [\bar\psi \bar\mu]
\, .
\end{align}
Identical relations hold for $\chi$ upon the obvious replacement $\psi \to \chi$. Here we introduced the bra and ket notations for undotted and
dotted $SL(2)$ indices, $\ket{\lambda} = \lambda_\alpha$, $\bra{\lambda} = \lambda^\alpha$ and $|\bar\lambda] = \lambda^{\dot\alpha}$,
$[ \bar{\lambda}| = \bar\lambda_{\dot\alpha}$ that allow us to uniformly contract undotted indices from upper-left to lower-right and dotted ones 
from lower-left to upper right, i.e.,  $\vev{\lambda \psi} = \lambda^\alpha \psi_\alpha$ and $[\bar\mu\bar\psi] = \bar\mu_{\dot\alpha} \bar\psi^{\dot\alpha}$.

In a similar fashion, the gluon field strength can be decomposed as
\begin{align}
F_{\mu\nu} \sigma^\mu_{\alpha\dot\alpha} \sigma^\nu_{\beta\dot\beta}
=
2 \varepsilon_{\dot\alpha\dot\beta} f_{\alpha\beta} - 2 \varepsilon_{\alpha\beta} \bar{f}_{\dot\alpha\dot\beta}
\, ,
\end{align}
in terms of its chiral $f_{\alpha\beta} = \ft{i}{4} \sigma^{\mu\nu}{}_{\alpha\beta} F_{\mu\nu}$ and anti-chiral $\bar{f}_{\dot\alpha\dot\beta} = \ft{i}{4} 
\bar\sigma^{\mu\nu}{}_{\dot\alpha\dot\beta} F_{\mu\nu}$ components with the help of the self-dual $\sigma_{\mu\nu} = \ft{i}{2} [\sigma_\mu \bar\sigma_\nu 
- \bar\sigma_\nu \bar\sigma_\mu]$ and anti-self-dual tensors $\bar\sigma_{\mu\nu} = \ft{i}{2} [\bar\sigma_\mu \sigma_\nu - \bar\sigma_\nu \sigma_\mu]$.
The plus and minus fields are found by projections
\begin{align}
f_{++} = - \bra{\lambda} f \ket{\lambda}
\, , \qquad
f_{+-} = - \bra{\lambda} f \ket{\mu}
\, , \qquad
\bar{f}_{++} = - [\bar\lambda| \bar{f} | \bar\lambda]
\, , \qquad
\bar{f}_{+-} = - [ \bar\lambda| \bar{f} | \bar\mu]
\, ,
\end{align}
etc.

Finally, as any four-vector, the covariant derivatives are decomposed in twistor components as follows
\begin{align}
D_{++} = \bra{\lambda} D | \bar\lambda]
\, , \quad
D_{+-} = \bra{\lambda} D | \bar\mu]
\, , \qquad
D_{--} = \bra{\mu} D | \bar\mu]
\, .
\end{align}

\subsection{Bridging light-cone and twistor projections}

The notations introduced in this and preceding sections allow us to establish a dictionary between the light-cone and twistor components.
They are summarized by the following set of relations:
\begin{align}
&
\psi_+ = \frac{\sqrt[4]{2}}{4} (1 + \gamma_5) \gamma^- \gamma^+ \Psi
\, , \qquad
\psi_- = \frac{\sqrt[4]{2}}{4} (1 + \gamma_5) \gamma^+ \gamma^- \Psi
\, , \qquad
\\
&
\chi_+ = \frac{\sqrt[4]{2}}{4} \bar{\Psi}(1 + \gamma_5) \gamma^+ \gamma^-
\, , \qquad
\chi_- = -\frac{\sqrt[4]{2}}{4} \bar{\Psi}(1 + \gamma_5) \gamma^- \gamma^+ 
\, , \qquad
\end{align}
for fermions, 
where $\gamma_5 = {\rm diag} (1,-1)$, and
\begin{align}
\label{ffields}
f_{++} = \sqrt{2}F^{+ \perp}
\, , \qquad
f_{+-}^a = - \frac{1}{2\sqrt{2}}
\left((\partial^+ A^{-})^a + (\bar{D}_\perp A_\perp)^a - (D_\perp \bar{A}_\perp)^a-gf^{abc}\bar{A}^b_{\perp}A^c_{\perp} \right)
\, ,
\end{align}
for gluons and, finally, covariant derivatives,
\begin{align}
&
D_{-+}=\bar{D}_{+-}= 2 \bar{D}_\perp \, , \qquad
D_{+-}=\bar{D}_{-+}=2 D_\perp \, , 
\\
&
D_{++}=\bar{D}_{++}=2 D^+ \, , \qquad
D_{--}=\bar{D}_{--}=2 D^-\, .
\end{align}
Making use if these conversion formulas, we can adopt the basis introduce in Ref.\ \cite{Braun:2008ia}, on the one hand, and use the momentum-space technique of Ref.\
\cite{Bukhvostov:1983te} that makes the calculation more concise while using conventional four-component notations for Lorentz vectors and Dirac matrices.

\subsection{$SL(2)$ invariance and basis primary fields}

Though massless QCD is not a conformal theory at quantum mechanical level since it induces a scale due to dimensional transmutation, the classical Lagrangian of 
the theory does enjoy $SO(4,2)$ invariance. The one-loop evolution equations that we are set to analyze in this work inherit the latter since the symmetry breaking 
graphs do not make their appearance till two-loop order. Since the light-cone operators \re{GenericO} involve fields localized on a light ray, the full conformal algebra
reduces to its  collinear conformal $SL(2)$ subalgebra that acts only on the minus projections $z^-_k \equiv z_k$ of the Minkowski space-time coordinates $z^\mu_k$. 
The differential representation of generators acting on the space spanned by the composite operators \re{GenericO} reads
\begin{align}
\label{SL2algebra}
S^+ = \sum_{n=1}^N (z_n^2 \partial_{z_n} + 2 j_n z_n)
\, , \qquad
S^- = - \sum_{n=1}^N \partial_{z_n}
\, , \qquad
S^0 = \sum_{n=1}^N (z_n \partial_{z_n} + j_n)
\, .
\end{align}
The irreducible representations are characterized by the value of the conformal spin $j_n = (d_n + s_n)/2$ determined by the canonical dimension $d_n$ and light-cone 
spin $s_n$ of the constituent fields $X_n$. These generators commute with the generator of helicity introduced in Eq.\ \re{Helicity} as well as twist $E = \sum_{n=1}^N 
(d_n - s_n)/2$, see, e.g., \cite{BraKorMul03,BelBraGorKor04}.

The field projections introduced in the previous section transform covariantly under $SL(2)$ transformations and can be organized into ``multiplets'' of the same twist.
Namely, the good $\Phi_+$ and bad $\Phi_-$ chiral fields
\begin{align}
\Phi_+ = \{ \psi_+, \chi_+, f_{++} \}
\, , \qquad
\Phi_- = \{ \psi_-, \chi_-, f_{+-} \}
\, ,
\end{align}
as well as their conjugate anti-chiral analogues $\bar\Phi_\pm = \Phi_\pm^\ast$, possess twist $E = 1$ and $E = 2$, respectively.

Since covariant derivatives $D_{++}$, $D_{\pm\mp}$ and $D_{--}$ carry twist zero, one and two, respectively, they can be used to generate additional high-twist 
``single-particle'' fields by acting on $\Phi_{\pm}$. Obviously, we can ignore $D_{++}$ since they just induce an infinitesimal shift along the light cone. The $D_{--}$ 
derivatives acting on $\Phi_+$ will produce a twist-three constituent, which when accompanied by another good field component, will form a twist-four operator. However, 
this operator can be safely dropped from the basis thanks to QCD equations of motion. Next, the transverse derivatives $D_{\pm\mp}$ can act either on good or 
bad fields. However, we can consider only their action on the former since we can always move derivatives from bad to good fields in a twist-four operator of the type 
$\Phi_+ D_{\pm\mp} \Phi_-$. Moreover, since it is desirable to deal with conformal primary fields as individual building blocks, as they obey simple $SL(2)$ 
transformations \re{SL2algebra} and thus yield  evolution equations with manifest conformal symmetry, one can reduce in half, as advocated in Ref.\ \cite{Braun:2008ia}, 
the basis of good fields with transverse covariant derivatives acting on them. This is achieved by eliminating the ones with non-covariant transformation properties.
The net result is that one introduces instead conformal primaries $D_{--} \Phi_+$ which can be safely neglected as alluded to above. This procedure allows us 
to trades $D_{-+}\Phi_+$ posing bad $SL(2)$ transformation properties in favor of the primary $D_{+-} \Phi_+$. Finally, two transverse derivatives acting on $\Phi_+$ 
can be again be reduced to the irrelevant primary $D_{--} \Phi_+$. This concludes the recapitulation of the reasoning behind the choice of the twist-one $X_+$ and 
twist-two $X_-$ primaries
\begin{align}
X_+ = \{ \Phi_+ , \bar\Phi_+ \}
\, , \qquad 
X_-
= \{ \Phi_-,  \bar\Phi_-, D_{+-} \Phi_+, D_{-+} \bar\Phi_+ \}
\, ,
\end{align} 
which build up the operator basis at twist four. The latter is thus spanned by quasipartonic and nonquasipartonic operators (that read schematically)
\begin{align}
\label{NonQuasiBasis}
\mathbb{O}_4
=
X_+ X_+ X_+ X_+ 
\, , \qquad
\mathbb{O}_3 
= 
X_- X_+ X_+
\, ,
\end{align}
respectively.

\section{Evolution equations}

The twist-four light-ray operators \re{NonQuasiBasis} mix under renormalization.  Their mixing matrix admits perturbative expansion in strong coupling
$\alpha_s = g^2/(4 \pi)$. The goal of our analysis is to calculate the leading term of the series, namely,
\begin{align}
\frac{d}{d \ln \mu}
\left(
\begin{array}{c}
\mathbb{O}_3 \\ 
\mathbb{O}_4
\end{array}
\right)
= 
- \frac{\alpha_s}{2 \pi}
\left(
\begin{array}{cc}
\mathbb{H}^{(3 \to 3)} & \mathbb{H}^{(3 \to 4)} \\ 
0                  & \mathbb{H}^{(4 \to 4)}
\end{array}
\right)
\left(
\begin{array}{c}
\mathbb{O}_3 \\ 
\mathbb{O}_4
\end{array}
\right)
+
O(\alpha_s^2)
\, .
\end{align}
Notice that the mixing matrix takes a triangular form (to all orders in coupling) since the quasipartonic operators form an autonomous set under renormalization 
group evolution. Here the transition kernels are some integral operators that shift  fields on the light-cone towards each other. Their form is highly contained 
by conformal symmetry and was the subject of recent analysis \cite{Braun:2009vc}. The distinguished feature of nonquasipartonic operators is that they can change 
the number of fields upon evolution. Thus, while $\mathbb{H}^{(N \to N)}$ for $N=3,4$ is merely given by the sum of pairwise transition kernels,
\begin{align}
\mathbb{H}^{(N \to N)} = \sum_{j<k} \mathbb{H}_{jk}^{(2 \to 2)}
\end{align}
the  kernel $\mathbb{H}^{(3\to4)}$ involves both two-to-two and two-to-three transitions 
\begin{align}
\mathbb{H}^{(3 \to 4)} = \sum_{j<k} \left( \mathbb{H}_{jk}^{(2 \to 2)} + \mathbb{H}_{jk}^{(2 \to 3)} \right)
\, ,
\end{align}
the latter exists whenever there is a bad field involved in a two-particle block, i.e., either $j$ or $k$ label belongs to a bad field. 

Extracting the color structures from these transitions
\begin{align}
\mathbb{H}_{12}^{(2 \to 2)} [X^{I_1}_1 (z_1) X^{I_2}_2 (z_2)] 
&
= \sum_c \sum_{J_1J_2} [C_{c}]_{I_1 I_2}^{J_1 J_2} 
\mathbb{H}_c [X^{J_1}_1 X^{J_2}_2] (z_1, z_2)
\, , \\
\mathbb{H}_{12}^{(2 \to 3)} [X^{I_1}_1 (z_1) X^{I_2}_2 (z_2)] 
&
= \sum_c \sum_{J_1J_2 J_3} [C_{c}]_{I_1 I_2}^{J_1 J_2 J_3} 
\mathbb{H}_c [X^{J_1}_1 X^{J_2}_2 X^{J_3}_3] (z_1, z_2)
\, ,
\end{align}
the reduced integral operators $\mathbb{H}_c$ for  two- and three-particle transitions are defined by their $H$-kernels
are given by
\begin{align}
[\mathbb{H}_c \mathbb{O}] (z_1, z_2)
&
=
z_{12}^\sigma
\int_0^1 d \alpha_1 d \alpha_2 \, H_c (\alpha_1, \alpha_2) \mathbb{O} (\bar\alpha_1 z_1 + \alpha_1 z_2, \bar\alpha_2 z_2 + \alpha_2 z_1)
\, , \\
\label{evolutioncoord2}
[\mathbb{H}_c \mathbb{O}] (z_1, z_2)
&
=
z_{12}^\sigma
\int_0^1 d \alpha_1 d \alpha_2 d \alpha_3 \, H_c (\alpha_1, \alpha_2, \alpha_3) 
\mathbb{O} (\bar\alpha_1 z_1 + \alpha_1 z_2, \bar\alpha_2 z_2 + \alpha_2 z_1, \bar\alpha_3 z_2 + \alpha_3 z_1)
\, ,
\end{align}
where $\sigma$ is a positive/negative integer reflecting the mismatch in the field dimension in a given operator transition.
The manifest $SL(2)$ covariance on conformal primaries building up the composite operators and the fact that one loop
transitions do not receive contributions from counter terms that break conformal invariance implies the commutativity of the 
kernels with the generators of the algebra 
\begin{align}
[S^{\pm, 0}, \mathbb{H}] = 0
\, ,
\end{align}
and thus impose sever constraints on the form of the kernels.

\subsection{Renormalization in momentum space}

Though the conformal symmetry is more explicit in the coordinate space, the acutual calculations of one-loop graphs determining the mixing matrix 
are far more straightforward and elementary in the reciprocal momentum space. As we pointed out in the introduction, a technique to perform this
analysis is available for quasipartonic operators from Ref.\ \cite{BFLK}. Presently we will get it generalized to nonquasipartnic operators as well. The 
formalism relies heavily on the light-cone gauge, where the gluon propagator takes the form
\begin{align}
\label{GluonPropagator}
G_{\mu\nu}^{ab} (k) = \frac{(-i)d_{\mu\nu} (k)}{k^2 + i 0} 
\, , \qquad
d_{\mu\nu} (k) = g_{\mu\nu} - \frac{k^\mu n^\nu + k^\nu n^\mu}{k^+}  
\, .
\end{align}
As we can see, the integration over the loop momentum $k$ decomposed in Sudakov variables $k^\mu = k^+ \bar{n}^\mu + k^- n^\mu + k^\mu_\perp$,
\begin{align}
\label{LoopMeasure}
\int \frac{d^4 k}{(2 \pi)^4} 
=
\int_{- \infty}^\infty \frac{d k^+}{2 \pi} \int_{- \infty}^\infty \frac{d k^-}{2 \pi} \int_{- \mu}^{\mu} \frac{d^2 \bit{k}_\perp}{(2 \pi)^2}
\, ,
\end{align}
will potentially produce  divergences in the longitudinal variable $k^+$ due to an extra pole in the propagator, in addition to the conventional 
ultraviolet singularities regularized by a cut-off $\mu$ in the transverse momentum. The former arise due to incomplete gauge fixing by the condition 
\re{LCgaugeCondition} that allows one for light-cone time-independent residual transformations. To complete gauge fixing, one has to impose a condition 
on how to go around the $1/k^+$ singularity. This will not be a pressing issue for the current work since we will focus on kinematics away from the phase 
space boundaries where these have to be treated properly. Let us point out however, that the advanced/retarded and symmetric boundary conditions on 
the gauge potential on the light-cone infinity impose $\pm i 0$ and principal value prescription \cite{Belitsky:2002sm}. While the condition consistent with 
the equal-time quantization yields the Mandelstam-Leibbrandt prescription \cite{BassettoBook}. 

Thus, we convert the light-ray operators to the momentum fraction space 
\begin{align}
\mathcal{O} (x_1, \dots, x_N)
=
\int \prod_{n=1}^N \frac{d^4 k_n}{(2 \pi)^4} \delta (k_n^+ - x_n) \, \mathcal{O} (k_1, \dots, k_N)
\, .
\end{align}
by means of the Fourier transform 
\begin{align}
\mathcal{O} (k_1, \dots, k_N)
=
\int \prod_{n=1}^N d^4 z_n {\rm e}^{i k_n \cdot z_n} \, \mathbb{O} (z_1, \dots, z_N)
\, .
\end{align}
Then, the evolution kernels arise from the $N$ to $M$-particle transition amplitude,
\begin{align}
\mathcal{O} (x_1, \dots, x_N)
&=\int \prod^M_{m=1} dy_m
\int \prod_{m=1}^M \frac{d^4 p_m}{(2 \pi)^4} \delta (p_m^+ - y_m) \mathcal{O} (p_1, \dots, p_M)
\nonumber\\
\label{GenericGraph}
&\times\int \prod_{n=1}^N \frac{d^4 k_n}{(2 \pi)^4} \delta (k_n^+ - x_n) 
\mathcal{G} (k_1, \dots, k_n | p_1, \dots, p_M)
\, , 
\end{align}
where $\mathcal{G} (k_1, \dots, k_n | p_1, \dots, p_M)$ is a sum of corresponding Feynman graphs. Extraction of the leading logarithmic divergence
from the momentum integrals results in the sought momentum-space evolution equation
\begin{align}
\mathcal{O} (x_1, \dots, x_N)
=
- \frac{\alpha_s}{2 \pi} \ln \mu \, [ \mathcal{K}^{(N \to M)} \mathcal{O}] (x_1, \dots, x_N)
\end{align}
where integral kernel in the momentum representation is
\begin{align}
\label{genericevolution}
[ \mathcal{K}^{(N \to M)} \mathcal{O}] (x_1, \dots, x_N)
=
\int [\mathcal{D}^M y]_N K (x_1, \dots, x_N | y_1, \dots, y_M) \mathcal{O} (y_1, \dots, y_M)
\, ,
\end{align}
with a  notation introduced for the measure
\begin{align}
\label{MomentumMeasure}
[\mathcal{D}^M y]_N \equiv \prod_{m=1}^M d y_m \delta \left( \sum_{m=1}^M y_m - \sum_{n=1}^N y_n \right)
\, .
\end{align}
The $N-1$ momentum  integrals in \re{GenericGraph} are eliminated by means of four-momentum conserving delta functions
stemming from Feynman rules leaving us with a single four-dimensional loop-momentum integration measure \re{LoopMeasure}. The 
extraction of $1/\bit{k}_\perp^2$ contribution in the integrand can be easily achieved by rescaling the $k^-$ component of the loop
momentum by introducing a new variable
\begin{align}
\label{beta}
\beta = 2 k^-/\bit{k}_\perp^2
\, , 
\end{align}
where $\bit{k}_\perp$ is the transverse loop momentum. We remove the $k^+$ integral making use of the momentum fraction delta functions
in Eq.\ \re{GenericGraph}, while the rescaled $k^-$-integrals are evaluated in terms of the generalized step-functions \cite{BFLK,BelRad}
\begin{align}
\label{Generalizedstep}
\vartheta^k_{\alpha_1, \dots, \alpha_n} (x_1, \dots, x_n) 
= 
\int_{-\infty}^{\infty} \frac{d \beta}{2 \pi i} \beta^k \prod_{\ell = 1}^n (x_\ell \beta - 1 + i 0)^{- \alpha_\ell}
\, .
\end{align}
These can be reduced to the simplest one $\vartheta^0_{11} (x_1, x_2) = [\theta (x_1) - \theta (x_2)]/(x_1 - x_2)$ making use of a set of known relations 
\cite{BFLK,BelRad}. So the advantage of this formalism is that there are no actual integrals to perform and the procedure of computing the evolution
kernels is reduced to straightforwards but tedious algebraic  manipulations with Dirac and Lorentz structures.

So all we need for the calculation is Fourier transforms of the conformal primary fields defining composite operators. When cast in four-dimensional light-cone 
notations, they read
\begin{align}
&\Phi_+
\stackrel{\rm\scriptscriptstyle FT}{\to}
\Phi_+ 
=
\left\{ \ft14 (1 + \gamma_5) \gamma^- \gamma^+ \Psi , \ft14 (1 - \gamma_5) \gamma^- \gamma^+ \Psi , - \ft{i}{2} k^+ A_\perp \right\}
\, , \\
&\bar{D}_{-+} \Phi_+
\stackrel{\rm\scriptscriptstyle FT}{\to}
- i \sqrt{2}
\left(
k_\perp + g A_\perp
\right)
\Phi_+
\, , \\
&\Phi_-
\stackrel{\rm\scriptscriptstyle FT}{\to}
\Phi_-
=
\left\{ 
\ft14 (1 + \gamma_5) \gamma^+ \gamma^- \Psi 
, 
\ft14 (1 - \gamma_5) \gamma^+ \gamma^- \Psi 
, - \ft{i}{2} k^+ A^- + \ft{i}{2} (k_\perp \bar{A}_\perp - \bar{k}_\perp A_\perp) - \ft{i}{4} g [\bar{A}_\perp, A_\perp]
\right\} 
\, .
\end{align}
The results for the antichiral fields $\bar\Phi$ are obtained from above by complex conjugation.

\subsection{From coordinate to momentum space}
\label{Fourier}

It is straightforward to relate evolution kernels in the coordinate and momentum space by a Fourier transformation. For two-to-two transitions,
we immediately find
\begin{align}
K (x_1, x_2 | y_1, y_2) = (- i \partial_{x_1})^\sigma \int_0^1 d \alpha_1 d \alpha_2 \, H (\alpha_1, \alpha_2) \delta (x_1 - \bar\alpha_1 y_1 - \alpha_2 y_2)
\, ,
\end{align}
which are subject to the momentum-fraction conservation condition $x_1 + x_2 = y_1 + y_2$.  Analogously, for two-to-three transitions, we
find
\begin{align}
\label{fouriertransform}
K (x_1, x_2 | y_1, y_2, y_3) = (- i \partial_{x_1})^\sigma \int_0^1 d \alpha_1 d \alpha_2 d \alpha_3 \, 
H (\alpha_1, \alpha_2, \alpha_3) \delta (x_1 - \bar\alpha_1 y_1 - \alpha_2 y_2 - \alpha_3 y_3)
\, ,
\end{align}
where we assume that $x_1 + x_2 = y_1 + y_2 + y_3$. Having results in one representation, one can easily obtain the other making use of the
following two elementary operations
\begin{align}
\int^1_0d\alpha f(\alpha)\delta(x-\alpha y)
&
= f\bigg(\frac{x}{y}\bigg)\vartheta^0_{11} (x,x-y)
\, , \\
\int^1_0 d\alpha \, \bar{\alpha}^n\vartheta^0_{11}(x_1-y_1\bar{\alpha}, x_1 - \eta \bar\alpha)
&
=
\frac{1}{n}\bigg\{\bigg[1-\bigg(\frac{x_1}{\eta}\bigg)^n\bigg]\vartheta^0_{11}(x_1-y_1,-x_2)
\nonumber\\
 &-\frac{y_1}{y_2}\bigg[\bigg(\frac{x_1}{\eta}\bigg)^n-\bigg(\frac{x_1}{y_1}\bigg)^n\bigg]\vartheta_{11}^0(x_1-y_1,x_1)\bigg\}
\, ,
\end{align}
where $\eta = x_1+x_2 = y_1+y_2$ implies momentum conservation.
Finally, the coordinate kernels possess integrable end-point singularties, that have to be regularized in the course of the Fourier transform.
We found the cut-off regularization to be the simplest one to handle the emerging divergencies
\begin{align}
\int^1_{\varepsilon} \frac{d \alpha}{\alpha}\theta(x-y\alpha)
&=\bigg[\ln\bigg(\frac{x}{y}\bigg)-\ln\varepsilon\bigg]\theta(x)-\ln\bigg(\frac{x}{y}\bigg)\theta(x-y)
\, , \\
\int^1_\varepsilon \frac{d\alpha}{\alpha^2}\theta(x-y\alpha)
&=\bigg(\frac{1}{\varepsilon}-\frac{y}{x}\bigg)\theta(x)+\bigg(\frac{y}{x}-1\bigg)\theta(x-y)
\, .
\end{align}
At the end of the calculation, all singular terms cancel in the limit $\varepsilon\rightarrow 0$  rendering the total result regular. We provide an
example of explicit transformation in Appendix \ref{FourierAppendix}.

\subsection{Conformal symmetry in momentum space}
\label{ConSymmSect}

Since conformal invariance played a crucial role for the coordinate-space calculations \cite{Braun:2009vc}, let us analyze its consequences in the momentum 
space. To this end, following the same reasoning leading to the expression of Eqs\ \ref{fouriertransform} and taking care of the ordering of $z$ and $\partial_{z}$, 
one observes the following identifications between the light-ray coordinates and the momentum fractions,
 \begin{align}
 z_n \xrightarrow{FT} -i\partial_{x_n},\qquad\qquad \partial_{z_n}\xrightarrow{FT} ix_n,
 \, .
 \end{align}
 where $x_n$ is the reciprocal momentum to the coordinate $z_n$. Thus the conformal generators shown in Eq.\ \re{SL2algebra} can be rewritten in the 
 momentum space as
 \begin{align}
 \label{Splus}
 \widetilde{S}^+\mathcal{O}(x_1, \dots, x_N)
 &
 =
i \sum^N_{n=1} \left(\partial_{x_n}^2x_n - 2 j_n \partial_{x_i} \right) \mathcal{O}(x_1, \dots, x_N)
\, ,\\
\widetilde{S}^0 \mathcal{O}(x_1, \dots , x_N)
&
= - \sum^n_{i=1} \left(\partial_{x_n} x_n - j_n \right) \mathcal{O} (x_1, \dots , x_N)
\, , 
\\
\widetilde{S}^- \mathcal{O} (x_1, \dots , x_N)
&
= - i \sum^n_{n=1} x_n \mathcal{O}(x_1,\dots ,x_N)
\, .
\end{align}
Imposing the condition of commutativity 
\begin{align}
 [\mathcal{K}, \tilde{S}^{\pm, 0}] \mathcal{O} (x_1, \dots x_N) = 0
 \, , 
\end{align}
we find that the evolution kernels  obey the following differential equations (away from kinematical boundaries)
\begin{align}
\label{Splusonkernel}
&
\left(
\sum^M_{m=1} \left( y_m \partial_{y_m}^2 + 2j_m\partial_{y_m} \right)
-
\sum^{N-1}_{n=1}\left(\partial_{x_n}^2x_n-2j_n\partial_{x_n} \right)
\right)
K (x_1, \dots, \bigg(\sum^M_{m=1}y_m-\sum^{N-1}_{n=1}x_n\bigg) | y_1, \dots, y_M)
=
0
\, , \\
\label{S0onkernel}
&
\left(
\sum^M_{m=1}
\left(y_m\partial_{y_m}+j_m\right)
+
\sum^{N-1}_{n=1}
\left(\partial_{x_n} x_n-j_n\right)-j_N
\right)
K (x_1, \dots, \bigg(\sum^M_{m=1}y_m-\sum^{N-1}_{n=1}x_n\bigg) | y_1, \dots, y_M)
=
0
\, , 
\end{align}
for $S^+$ and $S^0$, respectively.  In arriving at the expressions of Eqs.\ \re{Splusonkernel} and \re{S0onkernel}, we have made use of the 
momentum-conserving Dirac delta function factorized from \re{genericevolution}. Since essentially there are only $M+N-1$ independent variables 
in the game, we are required to rewrite one of the variables as a linear combination of the rest before differentiation. Above we chose to eliminate 
$x_N$. Finally, the $S^-$ simply yields the momentum fraction conservation condition which is trivially obeyed due to an overall delta function 
\re{MomentumMeasure} that accompanies the transition kernel,
\begin{align}
\label{Sminusonkernel}
\sum^M_{m=1} y_m
-
\sum^N_{n=1} x_n
=
0
\, , 
\end{align}
Since we are interested in two-to-two and two-to-three transitions in this work, the expressions in Eqs.\ \ref{Splusonkernel} and \ref{S0onkernel} 
simplify to
 \begin{align}
 \Big[y_1\partial_{y_1}+y_2\partial_{y_2}+j_{1'}+j_{2'}+\partial_{x_1}x_1-(j_{1}+j_{2})\Big]K(x_1,y_1+y_2-x_1;y_1,y_2)=0
 \\
 \Big[y_1\partial^2_{y_1}+y_2\partial^2_{y_2}+2j_{1'}\partial_{y_1}+2j_{2'}\partial_{y_2}
 -\partial^2_{x_1}x_1+2j_{1}\partial_{x_1}\Big]K(x_1,y_1+y_2-x_1;y_1,y_2)=0
 \end{align}
where we use $j_n$ and $j_{n'}$ to refer to the conformal spins of the incoming and out going particles, respectively. Similarly, for three-particle 
transitions, we get
 \begin{align}
 \label{S0MomDiff}
 &\Big[\sum^3_{i=1}\bigg(y_i\partial_{y_i}+j_{i'}\bigg)+\partial_{x_1}x_1-(j_{1}+j_{2})\Big]K(x_1,\sum^3_{i=1}y_i-x_1;y_1,y_2,y_3)=0
 \, , 
 \\
\label{S+MomDiff}
&\Big[\sum^3_{i=1}\bigg(y_i\partial^2_{y_i}+2j_{i'}\partial_{y_i}\bigg)
 -\partial^2_{x_1}x_1+2j_{1}\partial_{x_1}\Big]K(x_1,\sum^3_{i=1}y_i-x_1;y_1,y_2,y_3)=0
 \, .
 \end{align}
 
\section{One-loop kernels}
\label{1LoopSection}

In this section we will report on our findings of all nonsinglet transition kernels. The latter will be quoted away from kinematical boundaries, i.e., when some of 
the momentum fractions (or their sums) could coincide. This will be sufficient to compare our results with the Fourier transform of the light-ray evolution kernel 
derived in Ref.\ \cite{Braun:2009vc} by dropping all contact, i.e., delta-function, terms emerging from the latter. Of course, we can keep track of the latter as well 
and reproduce them from the momentum-fraction formalism by properly incorporating QCD field renormalization (as well as certain contact terms stemming 
from vertex graphs) into the game. Since the light-cone gauge explicitly breaks Lorentz symmetry, the good and bad components receive different renormalization 
constants as can be immediately seen from the quark and gluon propagators \cite{Bukhvostov:1983te,Belitsky:1996hg}
\begin{align}
\mathcal{P} (k) =  Z^{(q)}_1 (k) \frac{\slashed{k}}{k^2 + i 0}  Z^{(q)}_2 (k) 
\, , \qquad
G_{\mu\nu} (k) = Z^{(g)}_{\mu\rho} (k) \frac{d^{\rho \sigma} (k)}{k^2 + i 0} Z^{(g)}_{\sigma\nu} (k)
\, ,
\end{align}
computed to one-loop accuracy. Here the $Z$-factors become momentum-fraction dependent (contrary to covariant gauges) due to assumed principal value 
prescription for the $1/k_+$-pole in the gluon density matrix,
\begin{align}
Z^{(q)}_1 (k) &= \sqrt{ 1 - \Sigma_1} \left( 1 - (\Sigma_2 (k) - \Sigma_1) \frac{\slashed{k} \gamma^+ }{k^+} \right) 
\, , \\
Z^{(q)}_2 (k) &= \sqrt{ 1 - \Sigma_1} \left( 1 - (\Sigma_2 (k) - \Sigma_1) \frac{\gamma^+ \slashed{k}}{k^+} \right)
\, , \\
Z^{(g)}_{\mu\rho} (k)
&=
\sqrt{1 + \Pi_1 (k)}
\left(
g_{\mu\rho} - \frac{1}{2} \Pi_2 (k) \frac{k_\mu n_\rho + k_\rho n_\mu}{k^+}
\right)
\, .
\end{align}
where
\begin{align}
\Sigma_1 = \frac{\alpha_s C_F}{2 \pi} \ln \mu
\, , \qquad
\Sigma_2 (k) = \frac{\alpha_s C_F}{2 \pi} \ln \mu
\int d q^+ \frac{k^+}{k^+ -q^+} \vartheta_{11}^0 (q^+, q^+ - k^+)
\, , 
\end{align}
and 
\begin{align}
&\Pi_1 (k)
=
\frac{\alpha_s}{\pi} \ln \mu
\left[
C_A \int d q^+ \frac{[ (q^+)^2 - q^+ k^+ + (k^+)^2 ]^2}{q^+ (q^+ - k^+) (k^+)^2} \vartheta_{11}^0 (q^+, q^+ - k^+)
-
\frac{n_f}{3}
\right]
\, , \\
&\Pi_2 (k)
=
\frac{\alpha_s}{2 \pi} C_A \ln \mu
\int d q^+ \frac{5 q^+ (q^+ - k^+) (k^+)^2 + 6 (q^+)^2 (q^+ - k^+)^2 + 2 (k^+)^4}{q^+ (q^+ - k^+) (k^+)^2} \vartheta_{11}^0 (q^+, q^+ - k^+)
\, ,
\end{align}
for quark and gluon, respectively. Their contribution to the renormalization of the operator blocks reads
\begin{align}
\Gamma_{\mu_1 \dots \mu_n} \to Z^{(q)}_2 \Gamma_{\mu_1 \dots \mu_n} Z^{(q)}_1 Z^{(g)}_{\mu_1 \nu_1} \dots Z^{(g)}_{\mu_n \nu_n}
\, , 
\end{align}
where $\Gamma_{\mu_1 \dots \mu_n}$ is the Dirac-Lorentz tensor defining the composite operator  in question. The collinearly divergent integrals entering 
$\Sigma$'s and $\Pi$'s regulate the end-point singularities in the momentum-fraction kernels promoting them to conventional plus-distributions that 
become integrable over the entire range of momentum fractions \cite{Altarelli:1977zs}.

\subsection{Two-to-two transitions: quasipartonic operators}
\label{2to2quasi}

%%%%%%%%%%%%%%%%%%%%%%%%%%%%%%%%%%%%%%%%%%%%%%%%%%%%%%%%%%%%%%%%%%%%%%%%%%
%                                                                                                                   FIGURE                                                                                                                                      %
%%%%%%%%%%%%%%%%%%%%%%%%%%%%%%%%%%%%%%%%%%%%%%%%%%%%%%%%%%%%%%%%%%%%%%%%%%
\begin{figure}[t]
    \centering
         \psfrag{i}[bc][bc]{\scriptsize$i_1$}
	\psfrag{i'}[tc][tc]{\scriptsize$i'_1$}
	\psfrag{j}[bc][bc]{\scriptsize$i_2$}
	\psfrag{j'}[tc][tc]{\scriptsize$i'_2$}
	\psfrag{a'}[tc][tc]{\scriptsize$a'$}
	\psfrag{k1}[cl][cl]{\scriptsize$k_1$}
	\psfrag{k2}[cl][cl]{\scriptsize$k_2$}
	\psfrag{p1}[cl][cl]{\scriptsize$p_1$}
	\psfrag{p2}[cl][cl]{\scriptsize$p_2$}
	\psfrag{g1}[cc][cl]{\scriptsize$C_1$}
  	 \makebox[\textwidth]{\includegraphics[scale=0.45]{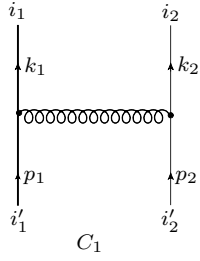}}
	  \caption{Feynman diagram determining the two-to-two transition of quark fields studied in Sect.\ \ref{quasiquarkquark} and Sect.\ \ref{412}.}
	  \label{fig:2to2quarkquark}
\end{figure}

To make our expressions more compact, we introduce a set color structures with open indices that show up in our expressions
\begin{align}
\label{2to2color}
[C_1]^{i_1i_2}_{i'_1i'_2}&=(t^a)_{i_1i'_1}(t^a)_{i_2i'_2},
& 
[C_2]^{ai}_{a'i'}&=f^{aa'c}(t^c)_{ii'},&
[C_3]^{ai}_{a'i'}&=(t^{a'}t^a)_{ii'}
\nonumber
\\
[C_4]^{i_1i_2}_{i'_1i'_2}&=(t^a)_{i_1i_2}(t^a)_{i'_1i'_2},
&[\tilde{C}_4]^{i_1i_2}_{i'_1i'_2}&=(t^a)_{i_1i_2}(\bar{t}^a)_{i'_1i'_2},
&[C_5]^{i_1i_2}_{ab}&=(t^at^b)_{i_1i_2},&
\nonumber
\\
 [C_6]^{i_1i_2}_{ab}&=(t^bt^a)_{i_1i_2},
 & 
 [C_7]^{ab}_{a'b'}&=f^{aa'c}f^{bb'c},
 & [C_8]^{ab}_{a'b'}&=f^{a'bc}f^{ab'c}.
\end{align}
where $f^{abc}$ is the $SU(N)$ structure constants while $t^a$ and $\bar{t}^a$ are the $SU(N)$ and $SU(\bar{N})$ generators in the fundamental
representation and its conjugate, respectively.

\subsubsection{
$\mathcal{O}^{i_1 i_2}(x_1, x_2)
=
\{
\psi_{+}^{i_1} \psi_{+}^{i_2 }
,\,
\psi_+^{i_1} \chi_+^{i_2}
,\,
\bar{\psi}_+^{i_1} \bar{\psi}_+^{i_2}
,\,
\bar{\psi}_+^{i_1}\bar{\chi}_+^{i_2}
,\,
\chi_+^{i_1}\chi_+^{i_2}
,\,
\bar{\chi}_+^{i_1}\bar{\chi}_+^{i_2}
\} (x_1, x_2)
$
}
\label{quasiquarkquark}
In this quark-quark sector the fields have open fundamental color indices $i_1$ and $i_2$. The operator renormalization kernel acts on them as follows
\begin{align}
[\mathcal{K} \, 
\mathcal{O}]^{i_1 i_2}(x_1, x_2)
= [C_1]^{i_1i_2}_{i'_1i'_2}\int [\mathcal{D}^2y]_2 K (x_1, x_2 | y_1, y_2) \mathcal{O}^{i'_1 i'_2}(y_1, y_2),
\end{align}
and its explicit expression arises from the graph shown in Fig. \ref{fig:2to2quarkquark}. It is given by
\begin{align}
K(x_1,x_2 | y_1,y_2) 
= - 2
\frac{x_1+x_2}{x_1-y_1} 
\vartheta^0_{111}(x_1,x_1-y_1,-x_2) - \frac{4 x_2}{x_1-y_1} \vartheta^0_{11}(x_1-y_1,-x_2).
\end{align}

\subsubsection{
$
\mathcal{O}^{i_1 i_2}(x_1, x_2)
=
\{
\psi_{+}^{i_1}\bar{\chi}_{+}^{i_2}
,\,
\bar{\psi}_+^{i_1}\chi_+^{i_2}
,\,
\psi_+^{i_1}\bar{\psi}_+^{i_2}
,\,
\bar{\chi}_+^{i_1}\chi_+^{i_2}
\} (x_1, x_2)
$
}
\label{412}

For the nonsinglet sector, the Feynman diagram responsible for the evolution is determined by the very same one-gluon exchange in Fig.\ 
\ref{fig:2to2quarkquark} so that
\begin{align}
\label{55}
[\mathcal{K}  \, \mathcal{O}]^{i_1 i_2}(x_1, x_2)
= 
[C_1]^{i_1i_2}_{i'_1i'_2}\int [\mathcal{D}^2 y]_2 K  (x_1, x_2 | y_1, y_2) \mathcal{O}^{i'_1 i'_2}(y_1, y_2)
\, ,
\end{align}
where 
\begin{align}
\label{56}
{K}(x_1,x_2 | y_1,y_2) 
&= - \frac{4x_2}{x_1-y_1} \vartheta^0_{11}(x_1-y_1,-x_2)
- 
\frac{x_2+y_1}{x_1-y_1}
\vartheta^0_{111}(x_1,x_1-y_1,-x_2)
\, .
\end{align}
For the quark-antiquark operators of the same flavor $\Phi^{i_1}(x_1)\Phi^{i_2}(x_2)=\{\psi_+^{i_1}(x_1)\bar{\psi}_+^{i_2}(x_2),\,\bar{\chi}^{i_1}_+(x_1)\chi^{i_2}_+(x_2)\}$, 
there are two extra transitions corresponding to  annihilation channels. Although we do not focus on the flavor singlet quark operators and the operators built up solely by 
gluon fields, we do  provide corresponding results for the $2 \rightarrow 2$ evolution kernels in Appendix \ref{appA}. 

\subsubsection{
$\mathcal{O}^{a i}(x_1, x_2)
=
\{f^a_{++} \psi_{+}^i
,\,
f^a_{++} \chi_{+}^i
,\,
\bar{f}^a_{++} \bar{\psi}_{+}^i
,\,
\bar{f}^a_{++} \bar{\chi}_{+}^i
\} (x_1, x_2)
$}
\label{quasigluonquark1}

\begin{figure}[t]
    \centering
         \psfrag{i}[bc][bc]{\scriptsize$i$}
	\psfrag{i'}[tc][tc]{\scriptsize$i'$}
	\psfrag{a}[bc][bc]{\scriptsize$a$}
	\psfrag{a'}[tc][tc]{\scriptsize$a'$}
	\psfrag{k1}[cl][cl]{\scriptsize$k_1$}
	\psfrag{k2}[cl][cl]{\scriptsize$k_2$}
	\psfrag{p1}[cl][cl]{\scriptsize$p_1$}
	\psfrag{p2}[cl][cl]{\scriptsize$p_2$}
	\psfrag{g1}[cc][cl]{\scriptsize$C_2$}
	\psfrag{g9}[cc][cl]{\scriptsize$C_3$}
  	 \makebox[\textwidth]{\includegraphics[scale=0.45]{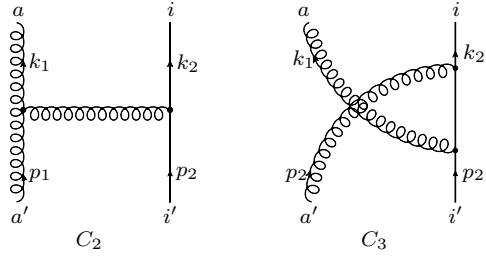}}
	  \caption{Feynman diagrams of two-to-two transition of quasipartonic gluon-quark fields in Sect.\ \re{quasigluonquark1} and Sect.\ \re{quasigluonquark2}.}
	  \label{fig:2to2quasigq}
\end{figure}

For the quark-gluon operator blocks, the renormalization opens up more than one color channel,
\begin{align}
[\mathcal{K} \, \mathcal{O}]^{a i}(x_1, x_2)
&
= - \int [\mathcal{D}^2y]_2 
\left\{ [C_2]^{ai}_{a'i'} K_1 + [C_3]^{ai}_{a'i'} K_2 \right\}
(x_1, x_2 | y_1, y_2) \mathcal{O}^{a' i'}(y_1, y_2)
\, ,
\end{align}
with corresponding transition kernels calculated from the graphs  shown in Fig.\ \ref{fig:2to2quasigq} being
\begin{align}
K_1(x_1,x_2;y_1,y_2)&= \frac{x_1}{y_1} \frac{2 x_1}{x_1-y_1}\vartheta^0_{11}(x_1,x_1-y_1)-\frac{2x_2}{x_1-y_1}\vartheta^0_{11}(x_1-y_1,-x_2)\, ,
\\
K_2(x_1,y_1;y_1,y_2)&=\frac{2x_2}{y_1}\vartheta^0_{11}(x_1-y_2,-x_2)\, .
\end{align}

%%%%%%%%%%%%%%%%%%%%%%%%%%%%%%%%%%%%%%%%%%%%%%%%%%%%%%%%%%%%%%%%%%%%%%%%%%
%                                                                                                                   FIGURE                                                                                                                                      %
%%%%%%%%%%%%%%%%%%%%%%%%%%%%%%%%%%%%%%%%%%%%%%%%%%%%%%%%%%%%%%%%%%%%%%%%%%

\subsubsection{
$\mathcal{O}^{a i}(x_1, x_2)
=
\{
f^a_{++} \bar{\psi}_{+}^i
,\,
f^a_{++} \bar{\chi}_{+}^i
,\,
\bar{f}^a_{++} \psi_{+}^i
,\,
\bar{f}^a_{++} \chi_{+}^i
\} (x_1, x_2)
$}
\label{quasigluonquark2}

Similar results are obtained by replacing the quark and an antiquark field,
\begin{align}
[ \mathcal{K} \, \mathcal{O} ]^{a i}(x_1, x_2)
&
= - \int [\mathcal{D}^2y]_2 
\left\{ [C_2]^{ai}_{a'i'} K_1 + [C_3]^{ai}_{a'i'} K_2 \right\}
(x_1, x_2 | y_1, y_2) \mathcal{O}^{a' i'}(y_1, y_2)
\, ,
\end{align}
with
\begin{align}
K_1(x_1,x_2|y_1,y_2)
&=\frac{2x_1x_2\vartheta^0_{11}(x_1,-x_2)}{y_1(x_1+x_2)}
-
\frac{2(x_1x_2+y_1^2)\vartheta^0_{111}(x_1,x_1-y_1,-x_2)}{(x_1-y_1)y_1}
\nonumber
\\
&
-\frac{2x_2(x_1+y_1)\vartheta^0_{11}(x_1-y_1,-x_2)}{(x_1-y_1)y_1}
\, , \\
K_2(x_1,x_2|y_1,y_2)
&=-2\frac{x_1-y_2}{y_1}\vartheta^0_{111}(x_1,x_1-y_2,-x_2)+\frac{2x_1x_2}{y_1(x_1+x_2)}\vartheta^0_{11}(x_1,-x_2) 
\, .
\end{align}
The Feynman graphs involved in the analysis differ from Fig. \ref{fig:2to2quasigq} only by the orientation of one of the quark lines.

All of these expressions agree with well-known particle transitions for quasipartonic operators \cite{BFLK,Bukhvostov:1983te,Twist3,BelExact98,Braun:2009vc}.

\subsection{Two-to-two transitions: non-quasipartonic operators}
\label{2to2nonquasi}

Now, we turn to the analysis of non-quasipartonic operators. According to the adopted basis \re{NonQuasiBasis}, the new two-particle blocks that we have to 
address contain a bad field component accompanied by a good one, namely
\begin{align}
\Phi_+(z_1)\otimes\Phi_-(z_2)\, , & \qquad \Phi_-(z_1)\otimes\Phi_+(z_2)
\nonumber
\\
\Phi_+(z_1)\otimes D_{-+}\Phi_+(z_2)\, , & \qquad D_{-+}\Phi_+(z_1)\otimes \Phi_+(z_2)
\, .
\end{align}

\subsubsection{Quark-quark transitions}

To start with, we consider the quark-quark transitions first. To this end we introduce non-quasipartonic two-particle operator built up from primary fields and
arranged as doublets since they mix under renormalization group evolution,
\begin{align}
\label{524}
\bit{\mathcal{O}}_+^{ij}
&=\Bigg\{\
\begin{pmatrix}
\psi_-^i \psi_+^j
\\
\psi_+^i \psi_-^j
\end{pmatrix}
,
\begin{pmatrix}
\psi_-^i \chi_+^j
\\
\psi_+^i \chi_-^j
\end{pmatrix}
,
\begin{pmatrix}
\chi_-^i \psi_+^j
\\
\chi_+^i \psi_-^j
\end{pmatrix}
,
\begin{pmatrix}
\chi_-^i \chi_+^j
\\
\chi_+^i \chi_-^j
\end{pmatrix}
\Bigg\},
\\
\bit{\mathcal{O}}_-^{ij}&=\Bigg\{
\begin{pmatrix}
\psi_{+}^i  \bar{D}_{-+}\psi_+^j
\\
\bar{D}_{-+}\psi_+^i \psi_{+}^j
\end{pmatrix}
,
\begin{pmatrix}
\psi_{+}^i  \bar{D}_{-+}\chi_+^j
\\
\bar{D}_{-+}\psi_+^i \chi_{+}^j
\end{pmatrix}
,
\begin{pmatrix}
\chi_{+}^i  \bar{D}_{-+}\psi_+^j
\\
\bar{D}_{-+}\chi_+^i \psi_{+}^j
\end{pmatrix}
,
\begin{pmatrix}
\chi_{+}^i  \bar{D}_{-+}\chi_+^j
\\
\bar{D}_{-+}\chi_+^i \chi_{+}^j
\end{pmatrix}
\Bigg\}
\, .
\end{align}
The Feynman diagram responsible for the mixing addressed in this section is the same one as in Fig. \ref{fig:2to2quarkquark}. We elaborate on an example of 
a specific transition in great detail in Appendix \ref{samplecal} to demonstrate the inner workings of the formalism. As a result, we find
\begin{align}
[ \mathcal{K} \, \bit{\mathcal{O}}_{+}]^{ij} (x_1,x_2)
 =
 - [C_1]^{ij}_{i'j'}\int [\mathcal{D}^2 y]_2
\bit{K} (x_1, x_2 | y_1, y_2)
\bit{\mathcal{O}}^{i'j'}_{+} (y_1,y_2)
\, ,
\end{align}
where the two-by-two mixing matrix
\begin{align}
\bit{K}
=
 \begin{pmatrix}
K_{11} & K_{12} \\
K_{21} & K_{22}
\end{pmatrix}
\end{align}
possesses the elements
 \begin{align}
 \label{528}
 K_{11}(x_1,x_2;y_1,y_2)&=\frac{2y_1}{x_1-y_1}\vartheta^0_{11}(x_1,x_1-y_1)-\frac{2x_2}{x_1-y_1}\vartheta^0_{11}(x_1-y_1,-x_2)\, ,
 \\
 K_{12}(x_1,x_2,y_1,y_2)&=2\vartheta^0_{11}(x_1,x_1-y_1)\, ,
 \\
 K_{21}(x_1,x_2;y_1,y_2)&=2\vartheta^0_{11}(x_1-y_1,-x_2)\, ,
 \\
 \label{531}
 K_{22}(x_1,x_2;y_1,y_2)&=\frac{2x_1}{x_1-y_1}\vartheta^0_{11}(x_1,x_1-y_1) -\frac{2y_2}{x_1-y_1}\vartheta^0_{11}(x_1-y_1,-x_2)\, .
 \end{align}
For $\bit{\mathcal{O}}_-^{ij}$ operator sets, we similarly get
\begin{align}
[ \mathcal{K} \, \bit{\mathcal{O}}_- ]^{ij}(x_1,x_2)
=
- [C_1]^{i_1i_2}_{i'_1i'_2}\int [\mathcal{D}^2 y]_2
\bit{K} (x_1, x_2 | y_1, y_2)
\bit{\mathcal{O}}_-^{i'j'} (y_1,y_2)\, ,
\end{align}
where 
\begin{align}
K_{11} (x_1,x_2;y_1,y_2)
&= \frac{4 x_2 \vartheta^0_{11}(x_1-y_1,-x_2)}{x_2-y_2}+\frac{2( x_1+x_2) \vartheta^0_{111} (x_1,x_1-y_1,-x_2)}{x_2-y_2}
\nonumber\\
&+\frac{4 x_2 \vartheta^0_{12} (x_1-y_1,-x_2)}{x_2-y_2}+\frac{2(x_1+x_2) \vartheta^0_{112} (x_1,x_1-y_1,-x_2)}{x_2-y_2}
\, , \\
K_{12} (x_1,x_2;y_1,y_2)&=2\frac{(y_1+y_2) (\vartheta^0_{112} (x_1,x_1-y_1,-x_2)+ \vartheta^0_{121} (x_1,x_1-y_1,-x_2))}{x_2-y_2}
\nonumber\\
&+\frac{2(x_1+y_1+y_2) \vartheta^0_{111} (x_1,x_1-y_1,-x_2)}{x_2-y_2}+\frac{4 x_2 \vartheta^0_{11} (x_1-y_1,-x_2)}{x_2-y_2}
\nonumber\\
&+\frac{4 x_2 (\vartheta^0_{12} (x_1-y_1,-x_2)+ \vartheta^0_{21} (x_1-y_1,-x_2))}{x_2-y_2}
\, , \\
K_{21}(x_1,x_2;y_1,y_2)&=\frac{2(y_1+y_2) \vartheta^0_{112} (x_1,x_1-y_1,-x_2)}{x_1-y_1}+\frac{2x_1 \vartheta^0_{111} (x_1,x_1-y_1,-x_2)}{x_1-y_1}
\nonumber\\
&+\frac{4 x_2 \vartheta^0_{12} (x_1-y_1,-x_2)}{x_1-y_1}
\, , \\
K_{22} (x_1,x_2;y_1,y_2)&=\frac{2(y_1+y_2) (\vartheta^0_{112} (x_1,x_1-y_1,-x_2)+\vartheta^0_{121} (x_1,x_1-y_1,-x_2))}{y_2-x_2}
\nonumber\\
&+\frac{2x_1 \vartheta^0_{111} (x_1,x_1-y_1,-x_2)}{y_2-x_2}+\frac{4 x_2 (\vartheta^0_{12} (x_1-y_1,-x_2)+\vartheta^0_{21} (x_1-y_1,-x_2))}{y_2-x_2}
\, .
\end{align}
This concludes our discussion on non-singlet operators. In Appendix \ref{nonquasigg1}, we also provide  transitions into gluonic operators 
involved in this class when it is generalized to the singlet channel as well.

\subsubsection{Quark-antiquark transitions}
\label{qantiq}

Next, we introduce doublets of quark-antiquark fields
\begin{align}
\bit{\mathcal{O}}^{ij}
=
\Bigg\{\
\begin{pmatrix}
\psi_-^i \bar{\psi}_+^j
\\
\psi_+^i \tfrac{1}{2}D_{-+}\bar{\psi}_+^j
\end{pmatrix},\,
\begin{pmatrix}
\psi_-^i \bar{\chi}_+^j
\\
\psi_+^i \tfrac{1}{2}D_{-+}\bar{\chi}_+^j
\end{pmatrix},\,
\begin{pmatrix}
\chi_-^i \bar{\psi}_+^j
\\
\chi_+^i \tfrac{1}{2}D_{-+}\bar{\psi}_+^j
\end{pmatrix},\,
\begin{pmatrix}
\chi_-^i \bar{\chi}_+^j
\\
\chi_+^i \tfrac{1}{2}D_{-+}\bar{\chi}_+^j
\end{pmatrix}
\Bigg\}\, ,
\end{align}
where we assume that the two-particle blocks possess different flavor such that they do not undergo annihilation transitions into gluon fields. Then the 
evolution equation can be written as
\begin{align}
[ \mathcal{K} \, \bit{\mathcal{O}} ]^{ij} (x_1,x_2)
= 
-
[C_1]^{ij}_{i'j'}\int [\mathcal{D}^2 y]_2
\bit{K} (x_1, x_2 | y_1, y_2) \bit{\mathcal{O}}^{i'j'} (y_1,y_2)\, .
\end{align}
with the elements of the evolution matrix  given by
 \begin{align}
 K_{11}&=\frac{2x_1y_1 (y_1+x_2)\vartheta^0_{11} (x_1,x_1-y_1)}{(y_1+y_2)^2 (y_2-x_2)}+\frac{2y_2 \vartheta^0_{11} (x_1,-x_2)}{x_1-y_1}
 \nonumber
 \\
 &+\frac{2\left(y_2^2 (y_1+y_2)+y_1 x_2 (y_1+2 y_2)+y_2 x_2^2\right)
   \vartheta^0_{11} (x_1-y_1,-x_2)}{(y_1+y_2)^2 (x_2-y_2)}\, ,
   \\
K_{12}
&=\frac{2x_1 (y_1+x_2) \vartheta^0_{11} (x_1,x_1-y_1)}{(y_1+y_2)^2 (x_2-y_2)}+\frac{2\vartheta^0_{11} (x_1,-x_2)}{x_1-y_1}
   \nonumber
   \\
   &-\frac{2\left(y_2^2 (y_1+y_2)+x_2^2 (y_1+2 y_2)-y_2 x_2 (y_1+2 y_2)\right) \vartheta^0_{11}
   (x_1-y_1,-x_2)}{y_2 (y_1+y_2)^2 (y_2-x_2)}
\, , \\
K_{21}
&= \frac{2x_1^2 y_1 (y_1+x_2) \vartheta^0_{11} (x_1,x_1-y_1)}{(y_1+y_2)^2 (x_2-y_2)}-\frac{2x_1 (x_1-2 y_1) \vartheta^0_{11} (x_1,-x_2)}{x_1-y_1}
   \nonumber
   \\
   &+\frac{2y_2 \left(x_1^3-x_1^2 (2 y_1+y_2)+x_1 (y_1+y_2)^2-y_1 (y_1+y_2)^2\right) \vartheta^0_{11} (x_1-y_1,-x_2)}{(y_1+y_2)^2
   (y_2-x_2)}\, ,
 \\
 K_{22}
 &= \frac{2x_1^2 (y_1+x_2) \vartheta^0_{11} (x_1,x_1-y_1)}{(y_1+y_2)^2 (x_2-y_2)}-\frac{2x_1(2 y_1+x_2)  \vartheta^0_{11} (x_1,-x_2)}{y_1 (x_2-y_2)}
   \nonumber
   \\
   &+\frac{2}{y_1 y_2 (y_1+y_2)^2 (y_2-x_2)}\big\{x_1^2 \left(2 y_1^3+6 y_1^2 y_2+4 y_1 y_2^2+y_2^3\right)-x_1^3 y_1 (y_1+2 y_2)
   \nonumber
   \\
   &\quad-x_1 (y_1+y_2)^4-y_1 y_2^2
   (y_1+y_2)^2\big\} \vartheta^0_{11} (x_1-y_1,-x_2)
\, .
 \end{align} 

\subsubsection{Quark-gluon transitions}
\label{nonquasiqg}

%%%%%%%%%%%%%%%%%%%%%%%%%%%%%%%%%%%%%%%%%%%%%%%%%%%%%%%%%%%%%%%%%%%%%%%%%%
%                                                                                                                   FIGURE                                                                                                                                      %
%%%%%%%%%%%%%%%%%%%%%%%%%%%%%%%%%%%%%%%%%%%%%%%%%%%%%%%%%%%%%%%%%%%%%%%%%%
 \begin{figure}[t]
    \centering
         \psfrag{i}[bc][bc]{\scriptsize$i$}
	\psfrag{i'}[tc][tc]{\scriptsize$i'$}
	\psfrag{a}[bc][bc]{\scriptsize$a$}
	\psfrag{a'}[tc][tc]{\scriptsize$a'$}
	\psfrag{k1}[cl][cl]{\scriptsize$k_1$}
	\psfrag{k2}[cl][cl]{\scriptsize$k_2$}
	\psfrag{p1}[cl][cl]{\scriptsize$p_1$}
	\psfrag{p2}[cl][cl]{\scriptsize$p_2$}
	\psfrag{g1}[cc][cl]{\scriptsize$C_2$}
	\psfrag{g9}[cc][cl]{\scriptsize$C_3$}
	\psfrag{g12}[cc][cl]{\scriptsize$C_2+C_3$}
  	 \makebox[\textwidth]{\includegraphics[scale=0.54]{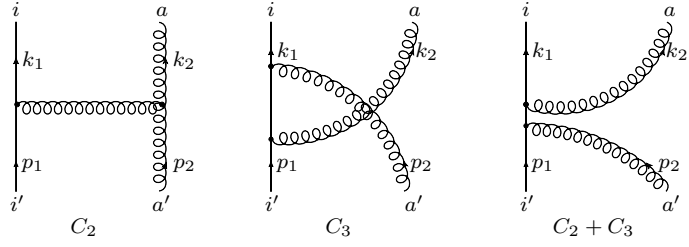}}
	  \caption{Feynman diagrams corresponding to two-to-two transition of non-quasipartonic quark-gluon blocks in Sect.\ \re{nonquasiqg}.}
	  \label{fig:2to2quarkgluonb}
\end{figure}

For the operators involving one quark and one gluon field, we introduce the following two-vectors
\begin{align}
\bit{\mathcal{O}}_+^{ia}
&=\Bigg\{
 \begin{pmatrix}
 \psi_-^i  f^a_{++}
 \\
 \psi_+^i  f_{+-}^a
 \end{pmatrix},\,
 \begin{pmatrix}
 \chi_-^i  f^a_{++}
 \\
 \chi_+^i  f^a_{+-}
 \end{pmatrix}\Bigg\}
\, , \\
\bit{\mathcal{O}}_-^{ia}
&=\Bigg\{
 \begin{pmatrix}
 \psi_+^i \big[\bar{D}_{-+}f_{++}\big]^a
 \\
 \big[D_{-+}\psi_+\big]^i  f^a_{++}
 \end{pmatrix},\,
 \begin{pmatrix}
 \chi_+^i \big[\bar{D}_{-+}f_{++}\big]^a
 \\
 \big[D_{-+}\chi_+\big]^i  f^a_{++}
 \end{pmatrix}
 \Bigg\}\, .
 \end{align}
Then, calculating Feynman diagrams responsible for their one-loop renormalization demonstrated in Fig. \ref{fig:2to2quarkgluonb},
we deduce that as in the quasipartonic case, there are two color-flow channels that induce the transitions
\begin{align}
[ \mathcal{K} \, \bit{\mathcal{O}}_+]^{ia} (x_1,x_2)
=
-
\int [\mathcal{D}^2y]_2
\left\{ [C_2]^{ai}_{a'i'} \bit{K} (x_1, x_2 | y_1, y_2) 
- [C_3]^{ai}_{a'i'} \widetilde{\bit{K}} (x_1, x_2 | y_1, y_2) 
\right\}
\bit{\mathcal{O}}_{+}^{i'a'}(y_1,y_2)\, ,
\end{align}
where the elements of the kernels $\bit{K}_{ij}$ and $\widetilde{\bit{K}}_{ij}$ admit the form 
 \begin{align}
 K_{11} &=\frac{y_1 (2 y_2^2 (y_1 + y_2)^2 - y_2 (y_1 + y_2) (3 y_1 + 2 y_2) x_2 + y_1 (2 y_1 + 3 y_2) x_2^2) \vartheta^0_{11}(x_1,x_1-y_1)}{y_2^2 (y_1 + y_2)^2 (y_2 - x_2)}
 \nonumber
 \\
 &+ \frac{
 x_2 ((y_1 + y_2) (3 y_1 + 2 y_2) + (y_1 + 2 y_2) x_2) \vartheta^0_{11}(x_1-y_1,-x_2)}{(y_1 + y_2)^2 (x_2 - y_2)}
 \nonumber
 \\
 &+ \frac{
 x_2 ( (2 y_1^2 + y_1 y_2 - 2 y_2^2) x_2-3 y_1 y_2 (y_1 + y_2) ) \vartheta^0_{11}(x_1,-x_2)}{
 y_2^2 (y_1 + y_2) (x_2 - y_2)} +\frac{2 x_2 \vartheta^0_{11}(x_1,-x_2)}{y_2}\, ,
 \\
 K_{12}
 &= \frac{2x_1 (y_2 (y_1+y_2)+y_1 x_2) \vartheta^0_{11} (x_1,x_1-y_1)}{(y_1+y_2)^2 (y_2-x_2)}+\frac{2x_2 (x_2-2 (y_1+y_2)) \vartheta^0_{11} (x_1,-x_2)}{(y_1+y_2) (y_2-x_2)}
 \nonumber
 \\
 &+\frac{2y_2 x_2 (x_2-2 (y_1+y_2)) \vartheta^0_{11}
   (x_1-y_1,-x_2)}{(y_1+y_2)^2 (x_2-y_2)}\, ,
   \\
   K_{21}
   &= -\frac{x_1^2 y_1 \vartheta^0_{11} (x_1,x_1-y_1)}{y_2 (x_1-y_1) (y_1+y_2)}+\frac{x_1^2 \vartheta^0_{11} (x_1,-x_2)}{y_2 (x_1-y_1)}
   \nonumber\\
&-\frac{\left(x_2^2 (2 y_1+3 y_2)-4 y_2 x_2 (y_1+y_2)+y_2
(y_1+y_2)^2\right) \vartheta^0_{11} (x_1-y_1,-x_2)}{y_2 (y_1+y_2) (y_2-x_2)}\, ,
\\
 K_{22}
 &= -\frac{\left(x_1^2 y_2+2 x_1 \left(y_2^2-y_1^2\right)+2 y_1 (y_1+y_2)^2\right) \vartheta^0_{11} (x_1-y_1,-x_2)}{y_1 (x_1-y_1) (y_1+y_2)}-\frac{x_1^2 \vartheta^0_{11} (x_1,x_1-y_1)}{(x_1-y_1)
   (y_1+y_2)}\
   \nonumber
   \\
   &+\frac{x_1 (x_1+2 (y_1+y_2)) \vartheta (x_1,-x_2)}{y_1 (x_1-y_1)}\, ,
   \\
   \widetilde{K}_{11}&=\frac{2x_2\vartheta^0_{11}(x_1,-x_2)}{y_2}+\frac{2y_1\vartheta^0_{111}(x_1,x_1-y_2,-x_2)}{y_2}\, ,
   \\
   \widetilde{K}_{12}&=2\vartheta^0_{12}(x_1,x_1-y_2)-\frac{2y_1\vartheta^0_{111}(x_1,x_1-y_2,-x_2)}{y_2}-\frac{2x_2\vartheta^0_{11}(x_1,-x_2)}{y_2}\, ,
   \\
   \widetilde{K}_{21}&=\frac{2x_1(y_1\vartheta^1_{111}(x_1,x_1-y_2,-x_2)-\vartheta^0_{11}(x_1,-x_2)}{y_2}\, ,
   \\
   \widetilde{K}_{22}&=0
\, .
\end{align}  
Similarly for the operators in the $\bit{\mathcal{O}}_{-} $ group, we get
\begin{align}
[ \mathcal{K} \, \bit{\mathcal{O}}_{-} ]^{ia}(x_1,x_2)
=
-
\int [\mathcal{D}^2y]_2
\Bigg\{ [C_2]^{ai}_{a'i'} \bit{K} (x_1, x_2 | y_1, y_2) 
+
[C_3]^{ai}_{a'i'} \widetilde{\bit{K}} (x_1, x_2 | y_1, y_2) 
\Bigg\}
\bit{\mathcal{O}}_{-}^{i'a'}(y_1,y_2)
\, ,
\end{align}
with
\begin{align}
 K_{11}&= -2\frac{x_1^2 \vartheta^0_{11} (x_1,x_1-y_1)}{(x_1-y_1) (y_1+y_2)}+2\frac{x_1 (x_1+y_1+y_2) \vartheta^0_{11} (x_1,-x_2)}{y_1 (x_1-y_1)}
 \nonumber
 \\
 &-2\frac{\left(2 y_2^3 (y_1+y_2)^2-3 y_2^3 x_2 (y_1+y_2)+y_1 x_2^3
   (y_1+y_2)+y_2^3 x_2^2\right) \vartheta^0_{11} (x_1-y_1,-x_2)}{y_1 y_2^2 (y_1+y_2) (y_2-x_2)}
\, , \\
   K_{12}&=\frac{2x_1(y_1(y_1+y_2)-x_1y_2)\vartheta^0_{11}(x_1,-x_2)}{(x_1-y_1)y_1^2}-\frac{2x_1^2\vartheta^0_{11}(x_1,x_1-y_1)}{(x_1-y_1)(y_1+y_2)}
   \nonumber\\
   &+\frac{2x_1y_2(x_1y_2-y_1(y_1+y_2))\vartheta^0_{11}(x_1-y_1,-x_2)}{(x_1-y_1)y_1^2(y_1+y_2)}\, ,
   \\
   K_{21}&= \frac{2x_1^2 \vartheta^0_{11} (x_1,x_1-y_1)}{(x_1-y_1) (y_1+y_2)}-\frac{2x_1^2 \vartheta^0_{11} (x_1,-x_2)}{y_1 (x_1-y_1)}
   \nonumber
   \\
   &+\frac{2(x_2^3y_1(y_1+y_2)+(x_1^2-x_2^2)y_2^3-x_2^2y_2(y_1^2+y_1y_2-y_2^2)) \vartheta^0_{11} (x_1-y_1,-x_2)}{y_1 y_2^2 (y_1+y_2) (y_2-x_2)}\, ,
   \\
   K_{22}&= \frac{2x_1^2 y_2 \vartheta^0_{11} (x_1,-x_2)}{y_1^2 (x_1-y_1)}+\frac{2x_1^2 \vartheta^0_{11} (x_1,x_1-y_1)}{(x_1-y_1) (y_1+y_2)}
   \nonumber
   \\
   &-\frac{2\left(x_2^2 \left(y_1^3+y_1^2 y_2+y_2^3\right)-2 y_2^3 x_2
   (y_1+y_2)+y_2^3 (y_1+y_2)^2\right) \vartheta (x_1-y_1,-x_2)}{y_1^2 y_2 (y_1+y_2) (y_2-x_2)}\, ,
   \\
   \widetilde{K}_{11}&=2\frac{x_1(\vartheta^0_{11}(x_1,x_1-y_2)+\vartheta^0_{12}(x_1,x_1-y_2))}{y_2}
   \, ,
   \\
   \widetilde{K}_{12}&=\frac{2x_1\vartheta^0_{11}(x_1,x_1-y_2)}{y_2}\, ,
   \\
   \widetilde{K}_{21}&=-\frac{2x_1\vartheta^0_{12}(x_1,x_1-y_2)}{y_2}\, ,
   \\
   \widetilde{K}_{22}&=0
\, .
 \end{align}
 
Having studied the operators generated by primary fields with the same chiralities, we now turn our attention to the cases where the operators are built up by 
fields of opposite chiralities, namely,
\begin{align}
\label{nonquasiqg2}
\bit{\mathcal{O}}^{ia}
=
\Bigg\{\
\begin{pmatrix}
\psi_-^i \bar{f}_{++}^a
\\
\psi_+^i \tfrac{1}{2}D_{-+}\bar{f}^a_{++}
\end{pmatrix},\,
\begin{pmatrix}
\chi_-^i \bar{f}_{++}^a
\\
\chi_+^i \tfrac{1}{2}D_{-+}\bar{f}^a_{++}
\end{pmatrix}
\Bigg\}
\, .
\end{align}
Their one-loop evolution equation is driven by
\begin{align}
[ \mathcal{K} \, \bit{\mathcal{O}}]^{ia}(x_1,x_2)
=
-
\int [\mathcal{D}^2 y]_2
\Bigg\{
[C_2]^{ai}_{a'i'} \bit{K} (x_1, x_2 | y_1, y_2) 
-
[C_3]^{ai}_{a'i'} \widetilde{\bit{K}} (x_1, x_2 | y_1, y_2) 
\Bigg\}
\bit{\mathcal{O}}^{i'a'} (y_1,y_2)
\, ,
\end{align}
and the transition kernels involved read
\begin{align}
K_{11}
&= \frac{2\left(y_1 x_2^3+y_2 (y_1+y_2)^3-y_1 x_2^2 (y_1+2 y_2)\right) \vartheta^0_{11} (x_1,-x_2)}{y_2 (y_1+y_2)^2 (y_2-x_2)}+\frac{2x_1x_2^2 y_1\vartheta^0_{11} (x_1,x_1-y_1)}{y_2 (x_1-y_1)
   (y_1+y_2)^2}
   \nonumber
   \\
   &-\frac{2\left(y_2^2 (y_1+y_2)^2+y_1 x_2^2 (y_1+y_2)+y_2 x_2^3\right) \vartheta^0_{11} (x_1-y_1,-x_2)}{y_2 (y_1+y_2)^2 (y_2-x_2)}\, ,
   \\
   K_{12}
   &=\frac{2x_2^2(x_1+y_2)\vartheta^0_{11}(x_1,-x_2)}{y_2(x_1-y_1)(y_1+y_2)^2}-\frac{2x_1x_2^2\vartheta^0_{11}(x_1,x_1-y_1)}{y_2(x_1-y_1)(y_1+y_2)^2}
   \nonumber
   \\
   &+\frac{2x_2^2(x_1(y_1+2y_2)-(y_1+y_2)^2)\vartheta^0_{11}(x_1-y_1,-x_2)}{y_2^2(x_1-y_1)(y_1+y_2)^2}\, ,
   \\
   K_{21}
   &=\frac{2x_1^2y_1 \left(y_1 y_2+x_2^2\right) \vartheta^0_{11}
   (x_1,x_1-y_1)}{y_2 (y_1+y_2)^2 (y_2-x_2)}
   \nonumber
   \\
   &-\frac{2x_1 \left(x_2^2
   \left(y_1^2+3 y_1 y_2+y_2^2\right)+y_2 (y_1-y_2) (y_1+y_2)^2-y_1 x_2^3\right) \vartheta^0_{11} (x_1,-x_2)}{y_2 (y_1+y_2)^2 (y_2-x_2)}
   \nonumber
   \\
   &+\frac{2\left(x_1^4-2 x_1^3 (y_1+y_2)+x_1^2 \left(y_1^2+3 y_1 y_2+y_2^2\right)+(y_1-x_1) y_2 (y_1+y_2)^2\right) \vartheta^0_{11} (x_1-y_1,-x_2)}{(y_1+y_2)^2 (y_2-x_2)}\, ,
 \\
   K_{22}
   &=  \frac{2x_1 \left(x_2^2 \left(y_1^2+3 y_1 y_2+y_2^2\right)+2 y_1 y_2 (y_1+y_2)^2-y_1 x_2^3\right) \vartheta^0_{11} (x_1,-x_2)}{y_1 y_2 (y_1+y_2)^2 (y_2-x_2)}
   \nonumber
   \\
   &+\frac{2x_1^2\left(y_1 y_2+x_2^2\right)
   \vartheta^0_{11} (x_1,x_1-y_1)}{y_2 (y_1+y_2)^2 (x_2-y_2)}-\frac{2}{y_1 y_2^2 (y_1+y_2)^2 (y_2-x_2)}\Big\{x_2^3 (y_1+y_2)\left(y_1^2+2 y_1 y_2-y_2^2\right)
   \nonumber
   \\
   &\quad+y_2^2 x_2^2 \left(y_1^2+3 y_1 y_2+y_2^2\right)-2 y_1 y_2^3 x_2 (y_1+y_2)+2 y_1
   y_2^3 (y_1+y_2)^2
   \nonumber
   \\
   &\quad-y_1 x_2^4 (y_1+2 y_2)\Big\} \vartheta^0_{11} (x_1-y_1,-x_2)\, ,
   \\
   \widetilde{K}_{11}
   &= \frac{2x_1^2 y_1^2 \vartheta^0_{11} (x_1-y_2,-x_2)}{y_2^2 (y_1+y_2)^2}-\frac{2 \left(x_1^2 y_1 (y_1+y_2)
+y_2^2 x_2^2\right)\vartheta^0_{11} (x_1,-x_2)}{y_2^2 (y_1+y_2)^2}+\frac{2x_1^2 y_1 \vartheta^0_{11} (x_1,x_1-y_2)}{y_2 (y_1+y_2)^2}
 \, ,  \\
   \widetilde{K}_{12}
 &=  -2\frac{x_1^2 \vartheta^0_{11} (x_1,x_1-y_2)}{y_2 (y_1+y_2)^2}
   \nonumber
   \\
   &+\frac{2\left(x_2^2 \left(y_1^2+y_1 y_2+y_2^2\right)-2 y_1^2 x_2 (y_1+y_2)+y_1 (y_1-y_2) (y_1+y_2)^2\right) \vartheta^0_{11} (x_1-y_2,-x_2)}{y_2^3 (y_1+y_2)^2}
   \nonumber
   \\
   &+\frac{2\left(2 y_1 x_2 (y_1+y_2)^2-x_2^2 \left(y_1^2+2 y_1 y_2+2 y_2^2\right)-(y_1-y_2) (y_1+y_2)^3\right) \vartheta^0_{11}
   (x_1,-x_2)}{y_2^3 (y_1+y_2)^2}\, ,
\\
   \widetilde{K}_{21}&=\frac{2(x_1-y_2)\big((2y_1-x_2)\vartheta^0_{111}(x_1,x_1-y_2,-x_2)+y_1\vartheta^0_{112}(x_1,x_1-y_2,-x_2)\big)}{y_2}
   \nonumber
   \\
   &+\frac{2x_1y_1x_2^2\vartheta^0_{11}(x_1,-x_2)}{y_2(y_1+y_2)^2}\, ,
   \\
  \widetilde{K}_{22}
  &= \frac{2(x_1-y_2) \left(x_1^2 \left(y_1^2+y_1 y_2+y_2^2\right)-y_2 (y_1+y_2)^2 (2 y_1+y_2-2 x_2)\right) \vartheta^0_{11}
   (x_1-y_2,-x_2)}{y_2^3 (y_1+y_2)^2}
  \nonumber
  \\
& -\frac{2x_1 \left(x_1^2 \left(y_1^2+2 y_1 y_2+2 y_2^2\right)-x_1 y_2 (y_1+y_2) (3 y_1+5 y_2)+3 y_2^2 (y_1+y_2)^2\right) \vartheta^0_{11} (x_1,-x_2)}{y_2^3 (y_1+y_2)^2}
   \nonumber
   \\
   &+\frac{2x_1^2 (y_2-x_1) \vartheta^0_{11} (x_1,x_1-y_2)}{y_2 (y_1+y_2)^2}
   \, .
 \end{align}
   
\subsubsection{Antiquark-gluon transitions}
\label{nonquasiaqg}

\begin{figure}[t]
    \centering
         \psfrag{i}[bc][bc]{\scriptsize$i$}
	\psfrag{i'}[tc][tc]{\scriptsize$i'$}
	\psfrag{a}[bc][bc]{\scriptsize$a$}
	\psfrag{d}[tc][tc]{\scriptsize$d$}
	\psfrag{a'}[tc][tc]{\scriptsize$a'$}
	\psfrag{k1}[cl][cl]{\scriptsize$k_1$}
	\psfrag{k2}[cl][cl]{\scriptsize$k_2$}
	\psfrag{p1}[cl][cl]{\scriptsize$p_1$}
	\psfrag{p2}[cl][cl]{\scriptsize$p_2$}
        \psfrag{g1}[cc][cl]{\scriptsize$C_2$}
	\psfrag{g9}[cc][cl]{\scriptsize$C_3$}
	\psfrag{g15}[cc][cl]{\scriptsize$C_2+C_3$}
	\makebox[\textwidth]{\includegraphics[scale=0.54]{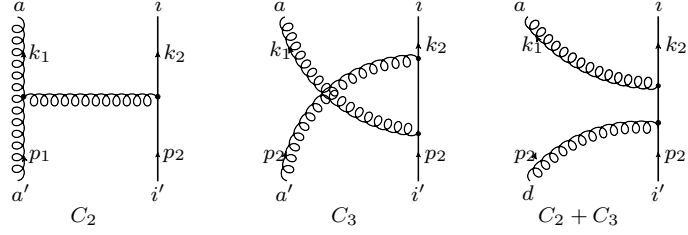}}
	  \caption{Feynman diagrams corresponding to two-to-two transitions of nonquasipartonic antiquark-gluon operator blocks in Sect.\ \ref{nonquasiaqg}.}
	  \label{fig:2to2gluonquarkb}
\end{figure}

For antiquark-gluon blocks, we introduce the doublets
\begin{align}
\bit{\mathcal{O}}^{ai}
=
\Bigg\{\
\begin{pmatrix}
f_{+-}^a \bar{\psi}_{+}^i
\\
f_{++}^a \tfrac{1}{2}D_{-+}\bar{\psi}^i_+
\end{pmatrix},\,
\begin{pmatrix}
f_{+-}^a \bar{\chi}_{+}^i
\\
f_{++}^a \tfrac{1}{2}D_{-+}\bar{\chi}^i_+
\end{pmatrix}
\Bigg\}
\, ,
\end{align}
whose transitions 
\begin{align}
[ \mathcal{K} \, \bit{\mathcal{O}}]^{ai}(x_1,x_2)
=
-
\int [\mathcal{D}^2 y]_2
\left\{
[C_2]^{ai}_{a'i'} \bit{K} (x_1, x_2 | y_1, y_2)  +
[C_3]^{ai}_{a'i'} \widetilde{\bit{K}} (x_1, x_2 | y_1, y_2) 
\right\}
\bit{\mathcal{O}}^{a'i'}(y_1,y_2)
\, ,
\end{align}
are determined by computing Feynman diagrams in Fig. \ref{fig:2to2gluonquarkb} and yield
 \begin{align}
 K_{11}&= \frac{2x_1 \left(x_1 (y_1+y_2)^2-x_1^2 y_1+y_1 \left(y_1^2+y_1 y_2+y_2^2\right)\right) \vartheta^0_{11} (x_1,x_1-y_1)}{y_1 (y_1+y_2)^2 (y_2-x_2)}
 \nonumber
 \\
 &+\frac{2x_1 \left(x_1^2 y_1 (y_1-y_2)+x_1 \left(3
   y_1^2 y_2-2 y_1^3+4 y_1 y_2^2+y_2^3\right)+y_1^2 (y_1-2 y_2) (y_1+y_2)\right) \vartheta^0_{11} (x_1,-x_2)}{y_1^2 (y_1+y_2)^2 (x_2-y_2)}
   \nonumber
   \\
   &+\frac{2\left(x_1^2 (y_1+y_2)^3-x_1^3 y_1 y_2-x_1 y_1^2 \left(y_1^2+2 y_1
   y_2+2 y_2^2\right)-y_1^2 y_2 (y_1+y_2)^2\right) \vartheta^0_{11} (x_1-y_1,-x_2)}{y_1^2 (y_1+y_2)^2 (y_2-x_2)}\, ,
   \\
   K_{12}&= \frac{2x_1 \left(x_1 (2 y_1+y_2) (3 y_1+y_2)-2 x_1^2 y_1-3 y_1^2 (y_1+y_2)\right) \vartheta^0_{11} (x_1,-x_2)}{y_1^2 (y_1+y_2)^2 (x_2-y_2)}
   \nonumber
   \\
   &+\frac{2x_1 \left(y_2^2 \left(y_1^2+3 y_1 y_2+y_2^2\right)-x_2 \left(y_2^3-y_1^2 y_2\right)-y_1 x_2^2 (y_1+2 y_2)\right) \vartheta^0_{11}
   (x_1-y_1,-x_2)}{y_1^2 y_2 (y_1+y_2)^2 (y_2-x_2)}
   \nonumber
   \\
   &+\frac{2x_1 \left(x_1^2+y_1 y_2\right) \vartheta^0_{11}
   (x_1,x_1-y_1)}{y_1 (x_1-y_1) (y_1+y_2)^2}\, ,
   \\
   K_{21}&=\frac{2x_1^2 \left(x_1 (y_1+y_2)-x_1^2+y_1^2\right) \vartheta^0_{11} (x_1,x_1-y_1)}{(y_1+y_2)^2 (y_2-x_2)}
   \nonumber
   \\
   &+\frac{2x_1^2 \left(y_2^2 (x_2-y_1)+y_2 x_2 (y_1-x_2)+y_1 (y_1+x_2)^2\right) \vartheta^0_{11}
   (x_1,-x_2)}{y_1 (y_1+y_2)^2 (x_2-y_2)}
   \nonumber
\\
   &-\frac{2y_2 \left(x_1^3 (y_1+y_2)-x_1^4+x_1^2 y_1^2+y_1 (y_1+y_2)^2 (x_2-y_2)\right) \vartheta^0_{11} (x_1-y_1,-x_2)}{y_1 (y_1+y_2)^2 (x_2-y_2)}\, ,
   \\
   K_{22}&=  \frac{2x_1^2 \left(y_1^2+x_1x_2 \right) \vartheta^0_{11} (x_1,x_1-y_1)}{y_1 (y_1+y_2)^2 (x_2-y_2)}
   \nonumber
   \\
   &-\frac{2}{y_1^2 y_2 (y_1+y_2)^2 (y_2-x_2)}\Big\{x_1^2
   \left(y_2^2 \left(6 y_1^2+6 y_1 y_2+y_2^2\right)-4 y_1 y_2^2 x_2+y_1 x_2^2 (y_1+2 y_2)\right)
   \nonumber
   \\
   &\quad-x_1^3 y_2^2 (5 y_1+y_2)+y_1^2 y_2^2 (y_1+y_2)^2\Big\}\vartheta^0_{11} (x_1-y_1,-x_2)
   \nonumber
   \\
   &+\frac{2(x_1^2 \left(x_2
   \left(3 y_1^2+2 y_1 y_2+y_2^2\right)+2 y_1^2 (y_1+y_2)+2 y_1 x_2^2\right) \vartheta^0_{11} (x_1,-x_2)}{y_1^2 (y_1+y_2)^2 (y_2-x_2)}\, ,
\\
   \widetilde{K}_{11}&=\frac{2x_1 (x_1 (y_1+3 y_2)-2 y_2 (y_1+y_2)) \vartheta^0_{11} (x_1,-x_2)}{y_2
   (y_1+y_2)^2} -\frac{2x_1^2 \vartheta^0_{11} (x_1,x_1-y_2)}{(y_1+y_2)^2}
   \nonumber
   \\
   &-\frac{2x_1^2 y_1 \vartheta^0_{11} (x_1-y_2,-x_2)}{y_2 (y_1+y_2)^2}\, ,
   \\
   \widetilde{K}_{12}&=\frac{2x_1^2 \vartheta^0_{11} (x_1,x_1-y_2)}{y_1 (y_1+y_2)^2}+\frac{2x_1  (x_1 (y_1^2-y_2^2)-2  x_2y_1 y_2)\vartheta^0_{11} (x_1,-x_2)}{y_1 y_2^2 (y_1+y_2)^2} 
 \\
  &-\frac{2\left(x_2^2 \left(y_1^2+y_1
   y_2+y_2^2\right)-2 y_1^2 x_2 (y_1+y_2)+y_1 (y_1-y_2) (y_1+y_2)^2\right) \vartheta^0_{11} (x_1-y_2,-x_2)}{y_1 y_2^2 (y_1+y_2)^2}\, ,
   \\
   \widetilde{K}_{21}&= 2\frac{(x_1-y_2) ((x_2-2 y_1) \vartheta^0_{111} (x_1,x_1-y_2,-x_2)-y_1 \vartheta^0_{112} (x_1,x_1-y_2,-x_2))}{y_1}
   \nonumber
   \\
   &+\frac{2x_1x_2^2 (y_1-y_2) \vartheta^0_{11} (x_1,-x_2)}{y_1 (y_1+y_2)^2}
\, , \\
   \widetilde{K}_{22}&= \frac{2(y_2-x_1) \left(x_1^2 \left(y_1^2+y_1 y_2+y_2^2\right)-y_2 (y_1+y_2)^2 (2 y_1+y_2-2 x_2)\right) \vartheta^0_{11} (x_1-y_2,-x_2)}{y_1 y_2^2 (y_1+y_2)^2}
   \nonumber
   \\
   &+\frac{2x_1^2 (x_1-y_2) \vartheta^0_{11}
   (x_1,x_1-y_2)}{y_1 (y_1+y_2)^2}+\frac{2x_1^2 \left(y_1^3-x_2 \left(y_1^2+2 y_1 y_2-y_2^2\right)-y_1 y_2^2\right) \vartheta^0_{11} (x_1,-x_2)}{y_1 y_2^2 (y_1+y_2)^2}
   \, .
 \end{align}
 
%%%%%%%%%%%%%%%%%%%%%%%%%%%%%%%%%%%%%%%%%%%%%%%%%%%%%%%%%%%%%%%%%%%%%%%%%%
%                                                                                                                   FIGURE                                                                                                                                      %
%%%%%%%%%%%%%%%%%%%%%%%%%%%%%%%%%%%%%%%%%%%%%%%%%%%%%%%%%%%%%%%%%%%%%%%%%%

This concludes our analysis of non-singlet two-to-two transitions of nonquasipartonic operators. They agree with corresponding findings in \cite{Bukhvostov:1983te},
the last paper of Ref.\ \cite{Twist3} and \cite{Braun:2009vc} after the Fourier transformations. For a partial result in the singlet channel, we refer the reader to
 Appendix \ref{appA}.

\subsection{Two-to-three transitions: non-quasipartonic operators}
 \label{2to3transition}

Finally, we come to the analysis of particle number-changing transitions. This is the most elaborate sector of twist-four operators. Apart from proliferation
of Feynman graphs, there are also subtle effects related to transitions induced by the use of QCD equations of motion. We provide for the latter a diagrammatic
representation that puts it on the same footing as the rest of the calculation and thus reduces the procedure to tedious algebraic manipulations. An example
exhibiting the formalism is worked out in Appendix \ref{EOMAppendix}.

%%%%%%%%%%%%%%%%%%%%%%%%%%%%%%%%%%%%%%%%%%%%%%%%%%%%%%%%%%%%%%%%%%%%%%%%%%
%                                                                                                                   FIGURE                                                                                                                                      %
%%%%%%%%%%%%%%%%%%%%%%%%%%%%%%%%%%%%%%%%%%%%%%%%%%%%%%%%%%%%%%%%%%%%%%%%%%
\begin{figure}[t]
    \centering
         \psfrag{i}[bc][bc]{\tiny$i$}
	\psfrag{i'}[tc][tc]{\tiny$i'$}
	\psfrag{j}[bc][bc]{\tiny$j$}
	\psfrag{j'}[tc][tc]{\tiny$j'$}
	\psfrag{d}[tc][tc]{\tiny$d$}
	\psfrag{k1}[cl][cl]{\tiny$k_1$}
	\psfrag{k2}[cl][cl]{\tiny$k_2$}
	\psfrag{p1}[cl][cl]{\tiny$p_1$}
	\psfrag{p2}[cl][cl]{\tiny$p_2$}
	\psfrag{p3}[c][]{\tiny$p_3$}
	\psfrag{k1mp1}[cl][cl]{\tiny$k_1-p_1$}
	\psfrag{k1mp3}[cl][cl]{\tiny$k_1-p_3$}
	\psfrag{g1}[cc][cl]{\tiny$C_1$}
	\psfrag{g2}[cc][cl]{\tiny$C_1+C_2$}
	\psfrag{g3}[cc][cl]{\tiny$C_1+C_3$}
	\psfrag{g5}[cc][cl]{\tiny$C_2$}
	\psfrag{g6}[cc][cl]{\tiny$C_3$}
	\makebox[\textwidth]{\includegraphics[width=0.85\textwidth]{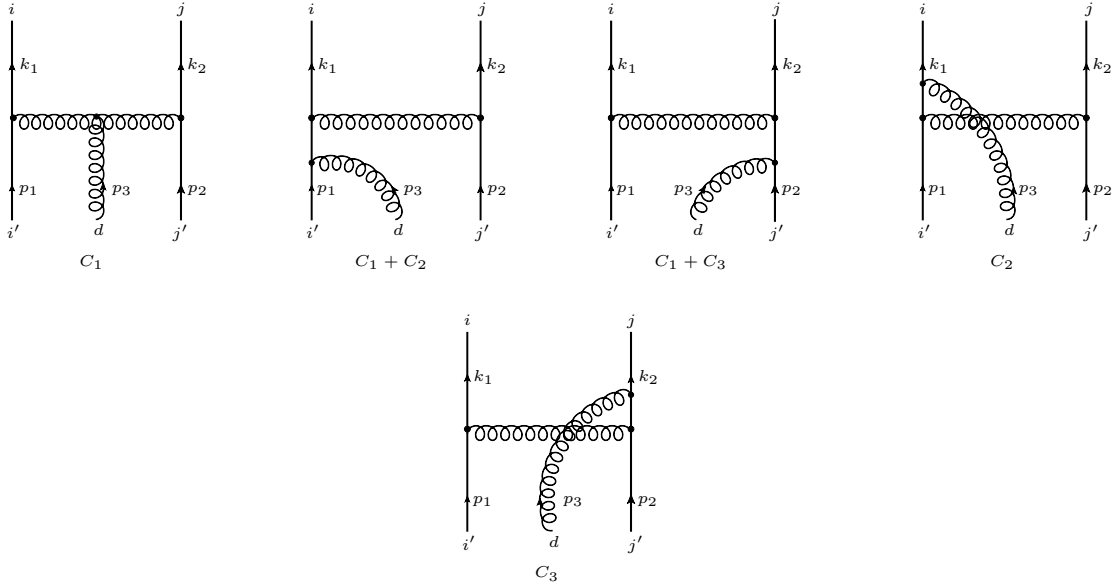}}
\caption{Feynman diagrams responsible for the evolution kernels of $\psi_{-}^i(x_1)\psi_{+}^j(x_2)\rightarrow \psi_{+}^{i'}(y_1)\psi_{+}^{j'}(y_2)\bar{f}^a_{++}(y_3)$ 
in Sect.\ \ref{2to3qq} and $\bar{\psi}_+(x_1)\psi_-(x_1)\rightarrow \bar{\psi}^{i'}(y_1)\psi_+^{j'}\bar{f}^d_{++}(y_3)$ in Sect.\ \ref{sec436}. $C_c$'s are the color 
structures defined in Eq.\ \re{63}.} 
	 \label{fig:2to3qq}
\end{figure}
 
 \subsubsection{$\psi_{-}\psi_{+}\,$ and $\,\frac{1}{2}D_{-+}\bar{\psi}_{+}\bar{\psi}_{+}$}
 \label{2to3qq}
 
Let us start our study with the quark-quark bad-good and good-good operators with a transverse derivative, respectively,
\begin{align}
\mathcal{O}^{ij}(x_1,x_2)=\psi_{-}^i(x_1)\psi_{+}^j(x_2) 
\, , \qquad
\mathcal{O}^{ij}(x_1,x_2)=\tfrac{1}{2}D_{-+}\psi^i_{+}(x_1) \psi^j_{+}(x_2)
\, ,
\end{align}
mixing with the following three-particle operator constructed from good field components
\begin{align}
\mathcal{O}^{ija} (y_1, y_2, y_3) = g\sqrt{2} \psi_{+}^i(y_1)\psi_{+}^j(y_2)\bar{f}^a_{++}(y_3)
\, . 
\end{align}
In both cases, there are three nontrivial color-flow channels
\begin{align}
\label{63}
[C_1]^{ij}_{i'j'd}=f^{dbc}t_{ii'}^bt^c_{jj'}
, \qquad 
[C_2]^{ij}_{i'j'd}=i(t^dt^b)_{ii'}t^b_{jj'}
, \qquad 
[C_3]^{ij}_{i'j'd}=it^b_{ii'}(t^dt^b)_{jj'}
\, ,
\end{align}
such that the evolution equation takes the form
\begin{align}
[ \mathcal{K} \, \mathcal{O}]^{ij} (x_1,x_2)
=
\int [\mathcal{D}^3y ]_2 \sum_{c=1}^{3} [C_c]^{ij}_{i'j'a} K_c (x_1, x_2 | y_1, y_2, y_3) \mathcal{O}^{i'j'a} (y_1,y_2,y_3)
\, .
\end{align}
First, for the $\mathcal{O}^{ij}(z_1,z_2)=\psi_{-}^i(z_1)\psi_{+}^j(z_2)$ case, the evolution kernels computed from the diagrams in Fig.\ \ref{fig:2to3qq} read
\begin{align}
K_1&=\frac{\theta (x_1) }{y_1 (y_1+y_3)^2}-\frac{\theta(x_1- y_1) }{ y_1 y_3^2}
+
\frac{( y_1+2 y_3) \theta (x_1- y_1- y_3) }{ y_3^2 ( y_1+ y_3)^2}\, ,
\\
K_2 &=\frac{y_3 (y_1+y_3)-x_1 (y_1+2 y_3) \theta (x_1)}{y_3^2 (y_1+y_3)^2
(y_1+y_3-x_1)}+\frac{(x_1-y_3) \theta(x_1-y_3)}{y_1 y_3^2 (y_1+y_3-x_1)}+\frac{\theta(x_1-y_1-y_3)}{y_1(y_1+y_3)^2}\, ,
\\
K_3&=0
\, ,
\end{align}
while for $\mathcal{O}^{ij}(x_1,x_2)=\tfrac{1}{2}D_{-+}\psi_{+}^i(x_1)\psi_{+}^j(x_2)$ they are
\begin{align}
K_1&=\frac {x_ 1^2\theta (x_ 1)} {y_ 1 (y_ 1 + y_ 3)^2 ( x_2-y_2)}
+
\left( \frac{ y_ 2-x_2}{y_ 2 y_3^2}+
\frac{y_ 1 (y_ 1 + y_ 3) -  x_ 1 (y_ 1 + 2 y_ 3)}{(y_ 1 + y_ 3)^2 y_3^2}- \frac{1}{y_3^2}\right)\theta (y_2-x_2)
\nonumber
\\
&+
\left( \frac{1}{y_3^2} -\frac{(x_ 1 - y_ 1)^2}{y_ 1 (x_ 2-y_2) y_3^2}+\frac{y_ 3 (y_ 2 + y_ 3) - x_ 1 (y_ 2 + 2 y_ 3) +  y_ 1 (y_ 2 + 2 y_ 3)} {(y_ 2 + y_ 3)^2 y_3^2}\right)
\theta (x_ 1 -  y_ 1)
\nonumber\\
&+\frac {x_2 \theta (x_1-y_1-y_2-y_3)} {y_ 2 (y_ 2 + y_ 3)^2}
\, ,
\\
K_2&=\frac{ x_1^2 ( y_1 + 2 y_3)\theta (x_1) }{y_3^2 ( y_1 + y_3)^2 ( x_2-y_2) } +\frac{(y_3^2- x_1^2)\theta ( x_1 - y_3) }{y_1 y_3^2 (y_1 + y_3- x_1 ) } 
-
\frac{( x_1 + y_1 + y_3)\theta (y_2-x_2) }{ y_1 ( y_1 + y_3)^2}\, ,
\\
K_3&=-\frac{x_2\theta(x_1-y_1)}{y_2(y_2+y_3)^2}+\frac{x_2\theta(x_1-y_1-y_2)}{y_2y_3^2}-\frac{x_2(y_2+2y_3)\theta(-x_2)}{y_3^2(y_2+y_3)^2}
\, .
\end{align}
Making use of the differential operators introduced in section \ref{ConSymmSect}, it is straightforward to verify that these kernels are all conformally invariant.

%%%%%%%%%%%%%%%%%%%%%%%%%%%%%%%%%%%%%%%%%%%%%%%%%%%%%%%%%%%%%%%%%%%%%%%%%%
%                                                                                                                   FIGURE                                                                                                                                      %
%%%%%%%%%%%%%%%%%%%%%%%%%%%%%%%%%%%%%%%%%%%%%%%%%%%%%%%%%%%%%%%%%%%%%%%%%% 
\begin{figure}[t]
    \centering
        \psfrag{i}[bc][bc]{\tiny$i$}
	\psfrag{i'}[tc][tc]{\tiny$i'$}
	\psfrag{j}[bc][bc]{\tiny$j$}
	\psfrag{j'}[tc][tc]{\tiny$j'$}
	\psfrag{a'}[tc][tc]{\tiny$j'$}
	\psfrag{d}[tc][tc]{\tiny$d$}
	\psfrag{k1}[cl][cl]{\tiny$k_1$}
	\psfrag{k2}[cl][cl]{\tiny$k_2$}
	\psfrag{p1}[cl][cl]{\tiny$p_1$}
	\psfrag{p2}[cl][cl]{\tiny$p_2$}
	\psfrag{p3}[c][]{\tiny$p_3$}
	\psfrag{k1mp1}[cl][cl]{\tiny$k_1-p_1$}
	\psfrag{k1mp3}[cl][cl]{\tiny$k_1-p_3$}
	\psfrag{k1-p1}[cl][cl]{\tiny$k_1-p_1$}
	\psfrag{k1-p3}[cl][cl]{\tiny$k_1-p_3$}
	\psfrag{g1}[cc][cl]{\tiny$C_1$}
	\psfrag{g2}[cc][cl]{\tiny$C_1+C_2$}
	\psfrag{g3}[cc][cl]{\tiny$C_1+C_3$}
	\psfrag{g5}[cc][cl]{\tiny$C_2$}
	\psfrag{g6}[cc][cl]{\tiny$C_3$}
	\makebox[\textwidth]{\includegraphics[width=0.85\textwidth]{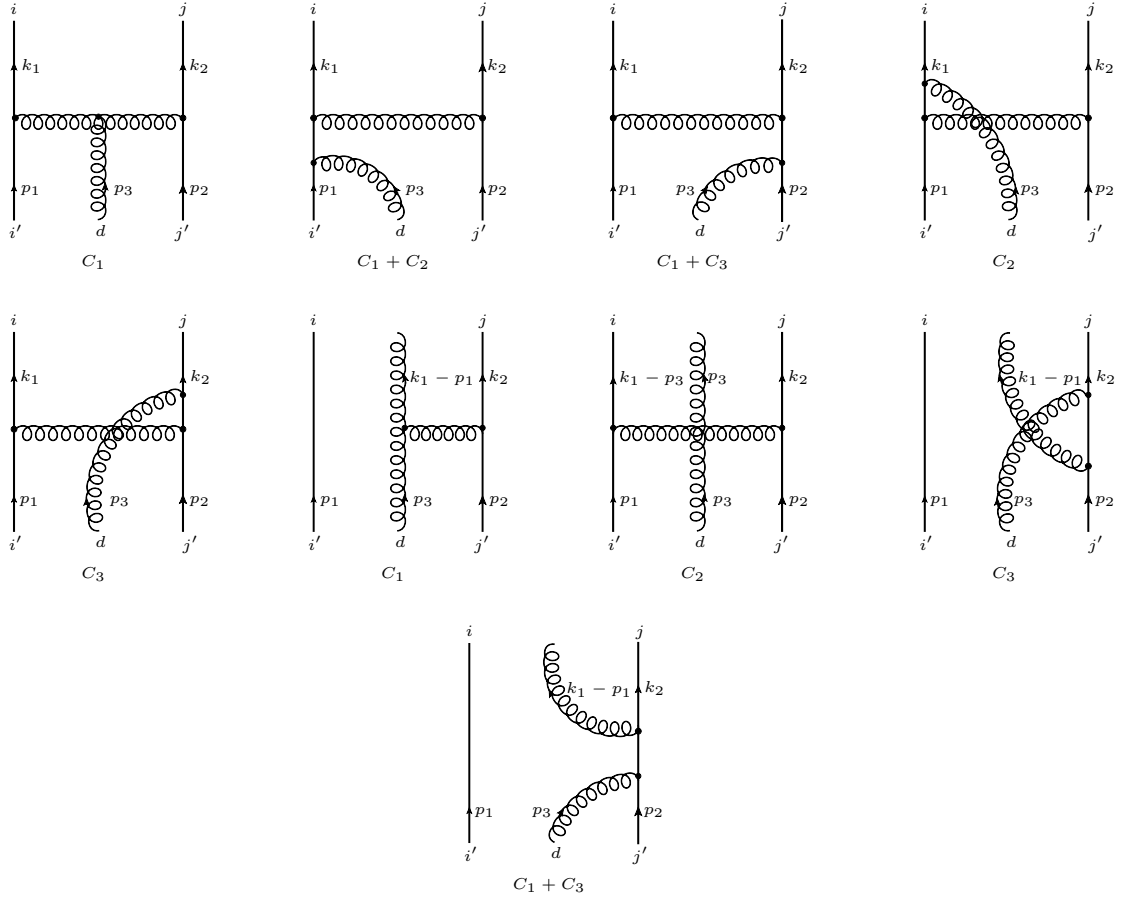}}
	\caption{Feynman diagrams defining the evolution kernels for transitions $D_{-+}\bar{\psi}_{+}^i(x_1)\bar{\psi}_{+}^j(x_2)\rightarrow 
	\psi_{+}^{i'}(y_1)\psi_{+}^{j'}(y_2)\bar{f}^a_{++}(y_3)$ in Sect.\ \ref{2to3qq} and $D_{-+}\bar{\psi}_+^i(x_1)\psi_+^j(x_1)\rightarrow 
	\bar{\psi}^{i'}(y_1)\psi_+^{j'}\bar{f}^d_{++}(y_3)$ in Sect.\ \ref{sec436} where $C_c$ are the color structures defined in Eq.\ \re{63}. 
	The last four diagrams correspond to the contribution of the gauge fieled in the covariant derivative $D_{-+}$.} 
	 \label{fig:2to3dqq}
\end{figure}
 
\subsubsection{$\tfrac{1}{2}D_{-+}\bar{\psi}_{+}\bar{f}_{++}$ and $\bar{\psi}_{+}\tfrac{1}{2}D_{-+}\bar{f}_{++}$}
 \label{sec62}

%%%%%%%%%%%%%%%%%%%%%%%%%%%%%%%%%%%%%%%%%%%%%%%%%%%%%%%%%%%%%%%%%%%%%%%%%%
%                                                                                                                   FIGURE 7                                                                                                                                    %
%%%%%%%%%%%%%%%%%%%%%%%%%%%%%%%%%%%%%%%%%%%%%%%%%%%%%%%%%%%%%%%%%%%%%%%%%%
\begin{figure}[p!]
    \centering
         \psfrag{i}[bc][bc]{\tiny$i$}
	\psfrag{i'}[tc][tc]{\tiny$i'$}
	\psfrag{a}[bc][bc]{\tiny$a$}
	\psfrag{d}[tc][tc]{\tiny$d$}
	\psfrag{a'}[tc][tc]{\tiny$a'$}
	\psfrag{k1}[cl][cl]{\tiny$k_1$}
	\psfrag{k2}[cl][cl]{\tiny$k_2$}
	\psfrag{p1}[cl][cl]{\tiny$p_1$}
	\psfrag{p2}[cl][cl]{\tiny$p_2$}
	\psfrag{p3}[c][]{\tiny$p_3$}
	\psfrag{k1mp1}[cl][cl]{\tiny$k_1-p_1$}
	\psfrag{k1mp3}[cl][cl]{\tiny$k_1-p_3$}
	\psfrag{g1}[cc][cl]{\tiny$C_1$}
	\psfrag{g2}[cc][cl]{\tiny$C_1+C_2$}
	\psfrag{g3}[cc][cl]{\tiny$-(C_1+C_3)$}
	\psfrag{g4}[cc][cl]{\tiny$(C_1+C_3);C_1;(-C_3)$}
	\psfrag{g5}[cc][cl]{\tiny$C_2$}
	\psfrag{g6}[cc][cl]{\tiny$C_3$}
	\psfrag{g7}[cc][cl]{\tiny$C_3+C_6-C_5$}
	\psfrag{g8}[cc][cl]{\tiny$C_4$}
	\psfrag{g9}[cc][cl]{\tiny$C_4-C_5$}
	\psfrag{g10}[cc][cl]{\tiny$C_5$}
	\psfrag{g11}[cc][cl]{\tiny$C_6-C_5$}
	\psfrag{g12}[cc][cl]{\tiny$C_2+C_4$}
	\psfrag{g13}[cc][cl]{\scriptsize$C_6$}
	\psfrag{g14}[cc][cl]{\scriptsize$C_6+C_3-C_1$}
	\psfrag{g15}[cc][cl]{\scriptsize$C_3+C_4+C_6-C_2-C_5$}
	\psfrag{g16}[cc][cl]{\scriptsize$C_2+C_5-C_1-C_4$}
	\makebox[\textwidth]{\includegraphics[width=0.85\textwidth]{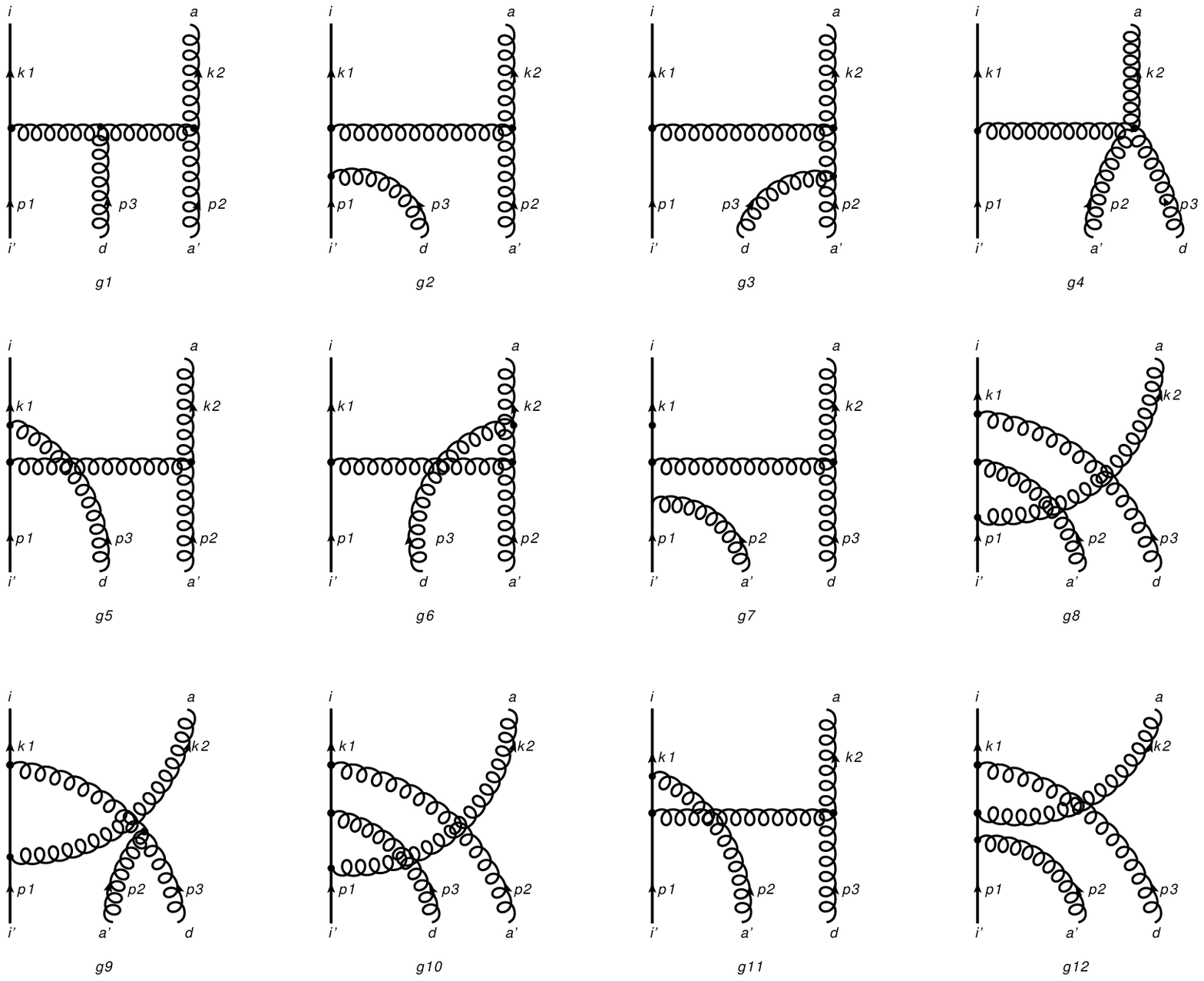}}
	\makebox[\textwidth]{\includegraphics[width=0.85\textwidth]{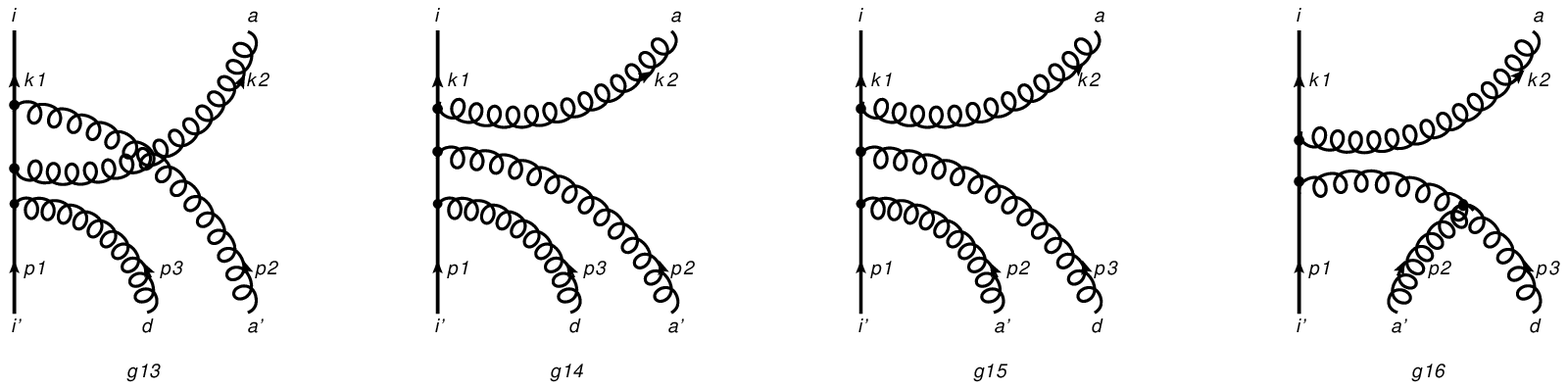}}
        \caption{Feynman diagrams that induce the transitions in Sects.\ \ref{sec62} and \ref{sec435}. For the operators in Sect.\ \ref{sec62} and 
        operator $\tfrac{1}{2}D_{-+}\bar{\psi}f_{++}$ in Sect.\ \ref{sec435}, these diagrams correspond to the flat derivative $\bar{\partial}_{\perp}$ residing  
        in the covariant derivative $D_{-+}$, while for the operator $\bar{\psi}_+f_{+-}$ in Sect.\ \ref{sec435}, they account for the contribution of 
        $\partial^+A^-\, , \bar{\partial}_{\perp}A_{\perp}$ and $\bar{\partial}_{\perp}\bar{A}_{\perp}$ originating from Eq.\ \re{ffields}.}
        \label{fig:qggdiff}
\end{figure}

\begin{figure}[t]
    \centering
         \psfrag{i}[bc][bc]{\tiny$i$}
	\psfrag{i'}[tc][tc]{\tiny$i'$}
	\psfrag{a}[bc][bc]{\tiny$a$}
	\psfrag{d}[tc][tc]{\tiny$d$}
	\psfrag{a'}[tc][tc]{\tiny$a'$}
	\psfrag{k1}[cl][cl]{\tiny$k_1$}
	\psfrag{k2}[cl][cl]{\tiny$k_2$}
	\psfrag{p1}[cl][cl]{\tiny$p_1$}
	\psfrag{p2}[cl][cl]{\tiny$p_2$}
	\psfrag{p3}[c][]{\tiny$p_3$}
	\psfrag{k1mp1}[cl][cl]{\tiny$k_1-p_1$}
	\psfrag{k1mp3}[cl][cl]{\tiny$k_1-p_3$}
	\psfrag{k1mp2}[cl][cl]{\tiny$k_1-p_2$}	
	\psfrag{g1}[cc][cl]{\tiny$C_1$}
	\psfrag{g2}[cc][cl]{\tiny$C_1+C_2$}
	\psfrag{g3}[cc][cl]{\tiny$-(C_1+C_3)$}
	\psfrag{g4}[cc][cl]{\tiny$(C_1+C_3);C_1;(-C_3)$}
	\psfrag{g5}[cc][cl]{\tiny$C_2$}
	\psfrag{g6}[cc][cl]{\tiny$C_3$}
	\psfrag{g7}[cc][cl]{\tiny$C_3+C_6-C_5$}
	\psfrag{g8}[cc][cl]{\tiny$C_4$}
	\psfrag{g9}[cc][cl]{\tiny$C_4-C_5$}
	\psfrag{g10}[cc][cl]{\tiny$C_5$}
	\psfrag{g11}[cc][cl]{\tiny$C_6-C_5$}
	\psfrag{g12}[cc][cl]{\tiny$C_6$}
        \makebox[\textwidth]{\includegraphics[scale=0.85]{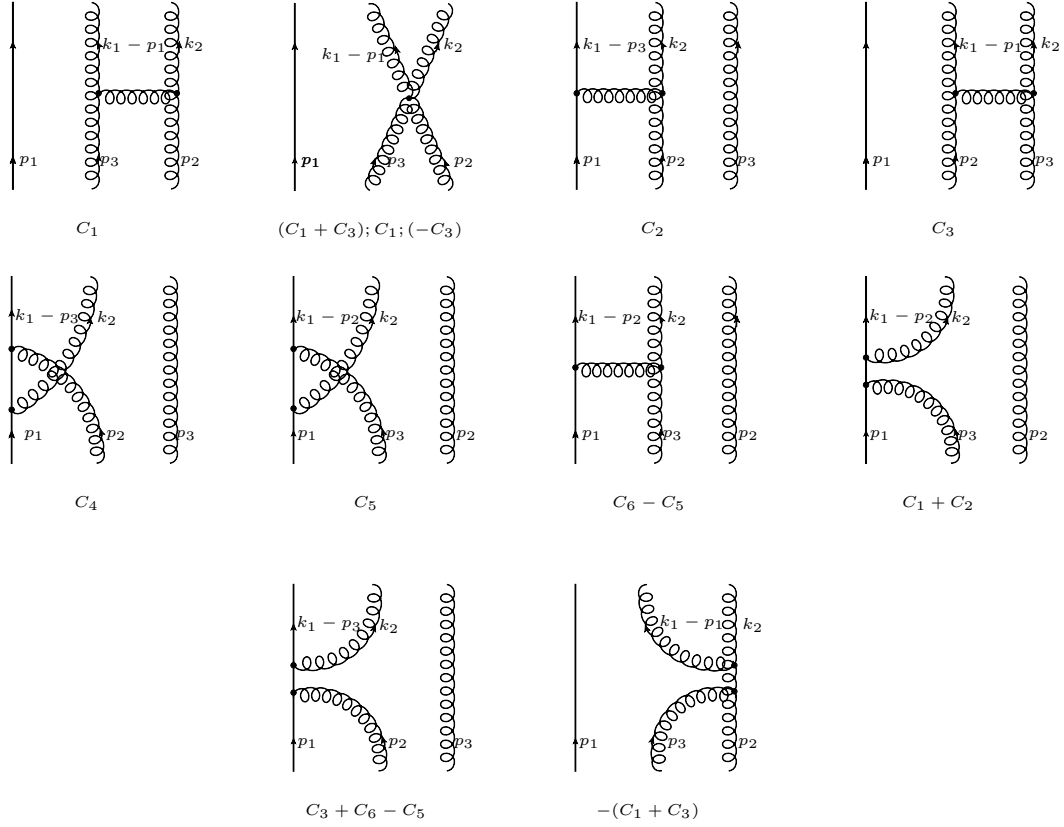}}
	  \caption{Feynman diagrams corresponding to the contributions of the gauge fields nested inside the covariant derivatives for the transitions 
	  of $\tfrac{1}{2}D_{-+}\bar{\psi}^i_{+}\bar{f}_{++}^a\rightarrow\bar{\psi}_{+}^i\bar{f}_{++}^a\bar{f}^d_{++}$ and $\tfrac{1}{2}D_{-+}\bar{\psi}^i_{+}{f}_{++}^a
	  \rightarrow\bar{\psi}_{+}^i{f}_{++}^a\bar{f}^d_{++}$ in Sects. \ref{sec62} and \ref{sec435} respectively.}
	  \label{fig:qggA}
\end{figure}

Next, we address the good-good quark-gluon operators with transverse derivatives 
\begin{align}
\mathcal{O}^{ia}(x_1,x_2)=\tfrac{1}{2}D_{-+}\bar{\psi}^i_{+}(x_1)\bar{f}_{++}^a(x_2)
\, , \qquad
\mathcal{O}^{ia}(x_1,x_2)= \tfrac{1}{2} \bar{\psi}^i_{+}(x_1) D_{-+}\bar{f}_{++}^a(x_2)
\end{align}
evolving into 
\begin{align}
\label{432incoming}
\mathcal{O}^{iad}=g \sqrt{2}\bar{\psi}_{+}^i(y_1)\bar{f}_{++}^a(y_2)\bar{f}^d_{++}(y_3)
\, .
\end{align}
Their one-loop evolution equation 
\begin{align}
\label{432transition}
[ \mathcal{K} \, \mathcal{O} ]^{ia}(x_1,x_2)
=
\int [\mathcal{D}^3 y]_2
\sum\limits_{c=1}^{6} [C_c]^{ia}_{i'a'd}
K_c (x_1, x_2 | y_1, y_2, y_3) \mathcal{O}^{a'i'd}(y_1,y_2,y_3)\, ,
\end{align}
develops six independent color structures
\begin{align}
\label{612}
[C_1]^{ia}_{i'a'd}&=-i(t^c)_{ii'}f^{cde}f^{aa'e}\, ,&[C_2]^{ia}_{i'a'd}&=-(t^dt^e)_{ii'}f^{aa'e}\, ,& [C_3]^{ia}_{ia'd'}&=-i(t^c)_{ii'}f^{ade}f^{ca'e}\, ,
\nonumber
\\
[C_4]^{ia}_{ia'd}&=i(t^dt^{a'}t^a)_{ii'}\, ,& [C_5]^{ia}_{i'a'd}\, ,&=i(t^{a'}t^dt^a)_{ii'}& [C_6]^{ia}_{i'a'd}&=i(t^{a'}t^at^d)_{ii'}\, .
\end{align}
Their computation of the diagrams in Figs.\ \ref{fig:qggdiff} and \ref{fig:qggA} yields for the $\tfrac{1}{2}D_{-+}\bar{\psi}^i_{+}(x_1)\bar{f}_{++}^a(x_2)$ operator
\begin{align}
K_1&=\frac{x_1^2\theta(x_1)}{y_1 (y_1+y_3)^2 (x_2-y_2)}+\frac{1}{y_1 y_3^2 (x_2-y_2) (y_2+y_3)^3}\Big\{x_2^3y_1 (y_2+3 y_3)-x_2^2 \big(y_1 y_2 (y_2+3 y_3)
\nonumber
\\
&\ \ \ \ +(y_2+y_3)^3\big)+2 x_2 (y_2+y_3)^4-(y_2+y_3)^5\Big\}\theta(x_1-y_1)+\frac{1}{y_2^2 y_3^2 (y_1+y_3)^2}\Big\{y_2^2 x_2 (y_1+2 y_3)
\nonumber
\\
&\ \ \ \ +y_2^2 (2 y_3 (y_2+y_3)-(y_1 (y_2+2 y_3)))-x_2^2 (y_1+y_3)^2\Big\}\theta(y_2-x_2)+\frac{x_2^2(3y_2+y_3)\theta(-x_2)}{y_2^2(y_2+y_3)^3}\, ,
\\
K_2&=\frac{x_1^2 (y_1+2 y_3)\theta(x_1) }{y_3^2 (y_1+y_3)^2 (y_2-x_2)}-\frac{(x_1^2-y_3^2)\theta(x_1-y_3)}{y_1 y_3^2 (x_2-y_2)}-\frac{(x_1+y_1+y_3) \theta(y_2-x_2)}{y_1 (y_1+y_3)^2}\, ,
\\
K_3&=\frac{x_1^2 \theta (x_1)}{y_1 (y_1+y_2)^2 (y_3-x_2)}+\frac{\left(y_3^2 (x_1 (y_1+2 y_2)-y_1 (y_1+y_2))+x_2^2 (y_1+y_2)^2\right)\theta(y_3-x_2) }{y_2^2 y_3^2
   (y_1+y_2)^2}
\nonumber
\\
& +\frac{1}{y_1 y_2^2
   (y_2+y_3)^3 (x_2-y_3)}\Big\{x_2^2 \left(y_3^2 (y_1+3 y_2)+3 y_2 y_3 (y_1+y_2)+y_2^3+y_3^3\right)
\nonumber
\\
& -y_1 x_2^3 (3 y_2+y_3)-2 x_2 (y_2+y_3)^4+(y_2+y_3)^5\Big\}\theta(x_1-y_1) -\frac{x_2^2 (y_2+3 y_3) \theta(-x_2)}{y_3^2 (y_2+y_3)^3}\, ,
\\
K_4&=\frac{x_1^2(y_2+3 y_3)\theta(x_1)}{y_3^2 (y_2+y_3)^3}+\frac{\left(x_1^2 (3 y_2+y_3)-(y_2+y_3)^3\right) \theta(y_1-x_2)}{y_2^2 (y_2+y_3)^3}+\frac{\left(y_3^2-x_1^2\right) \theta(x_1-y_3)}{y_2^2
   y_3^2}\, ,
\\
K_5&=\frac{x_1^2}{2 y_2^2}\Big\{\frac{y_1+2 y_2}{(y_1+y_2)^2 (x_2-y_3)}-\frac{2 y_2}{(y_2+y_3)^3}-\frac{1}{(y_2+y_3)^2}\Big\}\theta(x_1) -\frac{(x_1+y_1+y_2)\theta(y_3-x_2)}{2 y_1 (y_1+y_2)^2}
\nonumber
\\
&+\frac{\left((y_2+y_3)^3-x_1^2 (y_2+3
   y_3)\right) \theta(y_1-x_2)}{2 y_3^2 (y_2+y_3)^3}+\frac{(x_1^2-y_2^2)(y_1 x_2-y_3 (y_1+y_3))\theta(x_1-y_2)}{2 y_1 y_2^2 y_3^2 (x_2-y_3)}\, ,
\\
K_6&=\frac{x_1^2(y_1+2 y_2)\theta(x_1) }{2 y_2^2 (y_1+y_2)^2 (y_3-x_2)}+\frac{(x_1^2-y_2^2)\theta(x_1-y_2)}{2 y_1 y_2^2 (x_2-y_3)}+\frac{(x_1+y_1+y_2) \theta(y_3-x_2)}{2 y_1 (y_1+y_2)^2}
\, .
\end{align}

%%%%%%%%%%%%%%%%%%%%%%%%%%%%%%%%%%%%%%%%%%%%%%%%%%%%%%%%%%%%%%%%%%%%%%%%%%
%                                                                                                                   FIGURE   9                                                                                                                                      %
%%%%%%%%%%%%%%%%%%%%%%%%%%%%%%%%%%%%%%%%%%%%%%%%%%%%%%%%%%%%%%%%%%%%%%%%%%
\begin{figure}[t]
    \centering
         \psfrag{i}[bc][bc]{\tiny$i$}
	\psfrag{i'}[tc][tc]{\tiny$i'$}
	\psfrag{a}[bc][bc]{\tiny$a$}
	\psfrag{d}[tc][tc]{\tiny$d$}
	\psfrag{a'}[tc][tc]{\tiny$a'$}
	\psfrag{k1}[cl][cl]{\tiny$k_1$}
	\psfrag{k2}[cl][cl]{\tiny$k_2$}
	\psfrag{p1}[cl][cl]{\tiny$p_1$}
	\psfrag{p2}[cl][cl]{\tiny$p_2$}
	\psfrag{p3}[c][]{\tiny$p_3$}
	\psfrag{k2-p2}[cl][cl]{\tiny$k_2-p_2$}
	\psfrag{k2-p3}[cl][cl]{\tiny$k_2-p_3$}
	\psfrag{g1}[cc][cl]{\scriptsize$C_1$}
	\psfrag{g2}[cc][cl]{\scriptsize$C_1+C_2$}
	\psfrag{g3}[cc][cl]{\scriptsize$-(C_1+C_3)$}
	\psfrag{g4}[cc][cl]{\scriptsize$(C_1+C_3);C_1;(-C_3)$}
	\psfrag{g5}[cc][cl]{\scriptsize$C_2$}
	\psfrag{g6}[cc][cl]{\scriptsize$C_2+C_4$}
	\psfrag{g7}[cc][cl]{\scriptsize$C_3$}
	\psfrag{g8}[cc][cl]{\scriptsize$C_3+C_6-C_5$}
	\psfrag{g9}[cc][cl]{\scriptsize$C_4$}
	\psfrag{g10}[cc][cl]{\scriptsize$C_4-C_5$}
	\psfrag{g11}[cc][cl]{\scriptsize$C_5$}
	\psfrag{g12}[cc][cl]{\scriptsize$C_6-C_5$}
	\psfrag{g13}[cc][cl]{\scriptsize$C_6$}
	  \makebox[\textwidth]{\includegraphics[width=0.85\textwidth]{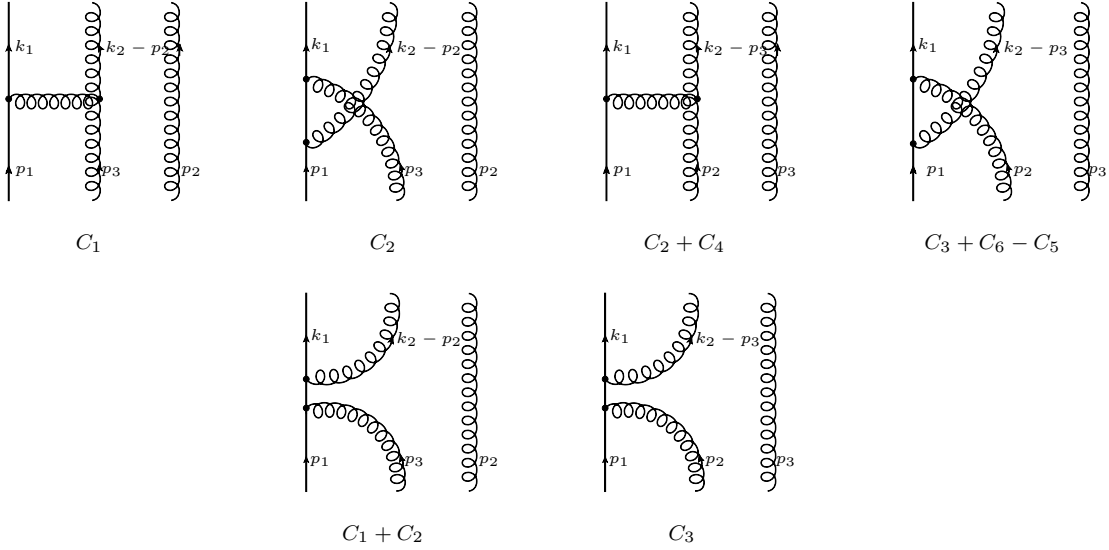}}
	  \caption{Additional Feynman diagrams contributing to the transition $\frac{1}{2}D_{-+}\bar{\psi}_{+}^i(x_1)\bar{f}^a_{++}(x_2)\rightarrow 
	  \bar{\psi}_{+}^{i'}(y_1)\bar{f}_{++}^{a'}(y_2)\bar{f}^d_{++}(y_3)$ in Sect.\ \ref{sec62}.}
	  \label{fig:qgAg}
\end{figure}

While for the $\bar{\psi}^i_{+}(z_1)\tfrac{1}{2}D_{-+}\bar{f}_{++}^a(z_2)$, the contributing graphs are shown in Figs.\ \ref{fig:qggdiff} and \ref{fig:qgAg} and their 
computation gives
\begin{align}
K_1&=\frac{x_1 \theta(x_1)}{y_1 (y_1+y_3)^2}-\frac{1}{y_1
   y_3^2 (y_2+y_3)^3}\Big\{y_1 x_2^2 (y_2+3 y_3)-x_2 (y_2+y_3) \left((y_2+y_3)^2-y_1 (y_2+3 y_3)\right)
\nonumber
\\
&\ \ \ \ +(y_2+y_3)^2 \left(y_1 (y_3-y_2)+(y_2+y_3)^2\right)\Big\}\theta(x_1-y_1)
+\frac{x_2^3 (3 y_2+y_3) \theta(-x_2)}{y_2^2 (x_1-y_1)(y_2+y_3)^3}
\nonumber\\
&\ \ \ \ 
+\frac{1}{y_2^2 y_3^2 (x_1-y_1) (y_1+y_3)^2}
\Big\{y_2^2 \big(y_1^2  (y_3-y_2)+y_1 \left(y_2^2+3 y_3^2\right)+2 y_3 \left(y_2^2+y_2 y_3+y_3^2\right)\big)
\nonumber\\
&\ \ \ \ 
-y_2^2 x_2 \big(2 y_1 (y_2-y_3)-2 y_1^2
+y_3(4 y_2+y_3)\big)+y_2^2 x_2^2 (y_1+2 y_3)-x_2^3 (y_1+y_3)^2\Big\}\theta(y_2-x_2) \, ,
\\
K_2&= \frac{x_1 (y_1+2 y_3)\theta(x_1) }{y_3^2 (y_1+y_3)^2}-\frac{x_1 \theta(x_1-y_3)}{y_1 y_3^2}+\frac{x_1 \theta(y_2-x_2)}{y_1 (y_1+y_3)^2}\, ,
\\
K_3&=-\frac{x_1 \theta (x_1)}{y_1 (y_1+y_2)^2}+\frac{1}{y_1 y_2^2 (y_2+y_3)^3}\Big\{y_1 x_2^2 (3
   y_2+y_3)-x_2 (y_2+y_3) \left((y_2+y_3)^2-y_1 (3 y_2+y_3)\right)
\nonumber
\\
& \ \ \ \ +(y_2+y_3)^2 \left(y_1 (y_2-y_3)+(y_2+y_3)^2\right)\Big\}\theta (x_1-y_1) +\frac{1}{y_2^2 y_3^2 (x_1-y_1) (y_1+y_2)^2}\Big\{(y_1+y_2)^2
\nonumber
\\
& \ \ \ \ \times  (x_2-y_3)^3-x_1 y_3^2 (x_1-y_1) (y_1+2 y_2)+3 y_3 (y_1+y_2)^2 (x_2-y_3)^2\Big\}\theta (y_3-x_2)
\nonumber
\\
&-\frac{x_2^3 (y_2+3 y_3) \theta (-x_2)}{y_3^2 (x_1-y_1)(y_2+y_3)^3}\, ,
\\
K_4&=\frac{x_1 (x_2-y_1) (y_2+3 y_3)\theta (x_1)}{ y_3^2 (y_2+y_3)^3}+\frac{x_1 (y_1-x_2) \theta (x_1-y_3)}{ y_2^2 y_3^2}
+
\frac{x_1 (y_1-x_2) (3 y_2+y_3) \theta (y_1-x_2)}{ y_2^2 (y_2+y_3)^3}\, ,
\\
K_5&=\frac{x_1}{ y_2^2
   (y_1+y_2)^2 (y_2+y_3)^3}\Big\{ x_1 (y_1+y_2)^2 (3 y_2+y_3)-(y_2+y_3) \Big[y_3 \left(y_1^2-3 y_2^2\right)+y_2 \big(3 y_1^2\
\nonumber
\\
&+5 y_1 y_2+y_2^2\big)-y_3^2 (y_1+2 y_2)\Big]\Big\}\theta (x_1)+\frac{x_1}{ y_2^2} \bigg\{\frac{x_2-y_1}{y_3^2}-\frac{1}{y_1}\bigg\} \theta (x_1-y_2)
\nonumber
\\
&+\frac{x_1 (y_1-x_2) (y_2+3 y_3) \theta (y_1-x_2)}{ y_3^2 (y_2+y_3)^3}+\frac{x_1\theta (y_3-x_2)}{y_1 (y_1+y_2)^2}\, ,
\\
K_6&=  -\frac{x_1 (y_1+2 y_2) \theta (x_1)}{y_2^2 (y_1+y_2)^2}+\frac{x_1 \theta (x_1-y_2)}{y_1 y_2^2}-\frac{x_1 \theta (y_3-x_2)}{y_1 (y_1+y_2)^2}
\, .
\end{align}
One of the consistency checks on the above kernels is their conformal invariance: they are annihilated by the differential operators in Eqs.\ \re{S0MomDiff} 
and \re{S+MomDiff}. Yet, they when Fourier transformed to the coordinate space look superficially different from the corresponding results obtained 
using the conformal technique \cite{Braun:2009vc}. This is  obvious from the fact that the $C_6$ channel is absent in the latter analysis and 
what is more disconcerting is that the Fourier transform of the light-ray kernels in Ref.\ \cite{Braun:2009vc} for the transitions given in Eq.\ \re{432transition} 
develops logarithmic dependence on the momentum fraction variables. From the momentum-space technique that is employed is this work, it is obvious that 
logarithms simply cannot emerge at one loop merely because one does not have enough integrations to generate them in the first place. This disparity between 
the two results is actually an agreement in disguise. To observe it, one has to use the symmetry of the operators involved in the transition. Namely, the
three-particle operator $\mathcal{O}^{iad}$ defined in Eq.\ \re{432incoming} is symmetric under the interchange of the gluon fields, i.e., simultaneous
exchange $a\leftrightarrow d$ and $y_2\leftrightarrow y_3$. This procedure eliminates the logarithms from the coordinate-space analysis. To get a complete 
agreement, we can redistribute the color structure $[C_6]^{ia}_{ia'd}$ and its corresponding kernel $K_6$ into other channels. The final expressions do
coincide. To be more explicit, we inverse-Fourier transform our results to the coordinate space and list results in Appendix \ref{appc1} since it gives a simplified 
form of the corresponding light-ray transition kernels.

%%%%%%%%%%%%%%%%%%%%%%%%%%%%%%%%%%%%%%%%%%%%%%%%%%%%%%%%%%%%%%%%%%%%%%%%%%
%                                                                                                                   FIGURE  10                                                                                                                                   %
%%%%%%%%%%%%%%%%%%%%%%%%%%%%%%%%%%%%%%%%%%%%%%%%%%%%%%%%%%%%%%%%%%%%%%%%%%
\begin{figure}[H]
    \centering
   \psfrag{i}[bc][bc]{\scriptsize$i$}
	\psfrag{i'}[tc][tc]{\scriptsize$i'$}
	\psfrag{a}[bc][bc]{\scriptsize$a$}
	\psfrag{d}[tc][tc]{\scriptsize$d$}
	\psfrag{a'}[tc][tc]{\scriptsize$a'$}
	\psfrag{k1}[cl][cl]{\scriptsize$k_1$}
	\psfrag{k2}[cl][cl]{\scriptsize$k_2$}
	\psfrag{p1}[cl][cl]{\scriptsize$p_1$}
	\psfrag{p2}[cl][cl]{\scriptsize$p_2$}
	\psfrag{p3}[c][]{\scriptsize$p_3$}
	\psfrag{k2mp2}[cl][cl]{\scriptsize$k_2-p_2$}
	\psfrag{k2mp3}[cl][cl]{\scriptsize$k_2-p_3$}
	\psfrag{g1}[cc][cl]{\scriptsize$C_1$}
	\psfrag{g2}[cc][cl]{\scriptsize$C_3-C_1$}
	\psfrag{g3}[cc][cl]{\scriptsize$-(C_1+C_2)$}
	\psfrag{g4}[cc][cl]{\scriptsize$(C_1+C_2);C_1;C_2$}
	\psfrag{g5}[cc][cl]{\scriptsize$C_2$}
	\psfrag{g6}[cc][cl]{\scriptsize$C_6-C_2-C_5$}
	\psfrag{g7}[cc][cl]{\scriptsize$C_3$}
	\psfrag{g8}[cc][cl]{\scriptsize$C_3+C_4$}
	\psfrag{g9}[cc][cl]{\scriptsize$C_4$}
	\psfrag{g10}[cc][cl]{\scriptsize$C_4-C_5$}
	\psfrag{g11}[cc][cl]{\scriptsize$C_5$}
	\psfrag{g12}[cc][cl]{\scriptsize$C_6-C_5$}
	\psfrag{g13}[cc][cl]{\scriptsize$C_6$}
	\psfrag{g14}[cc][cl]{\scriptsize$C_6+C_3-C_1$}
	\psfrag{g15}[cc][cl]{\scriptsize$C_3+C_4+C_6-C_2-C_5$}
	\psfrag{g16}[cc][cl]{\scriptsize$C_2+C_5-C_1-C_4$}
	   \makebox[\textwidth]{\includegraphics[width=0.85\textwidth]{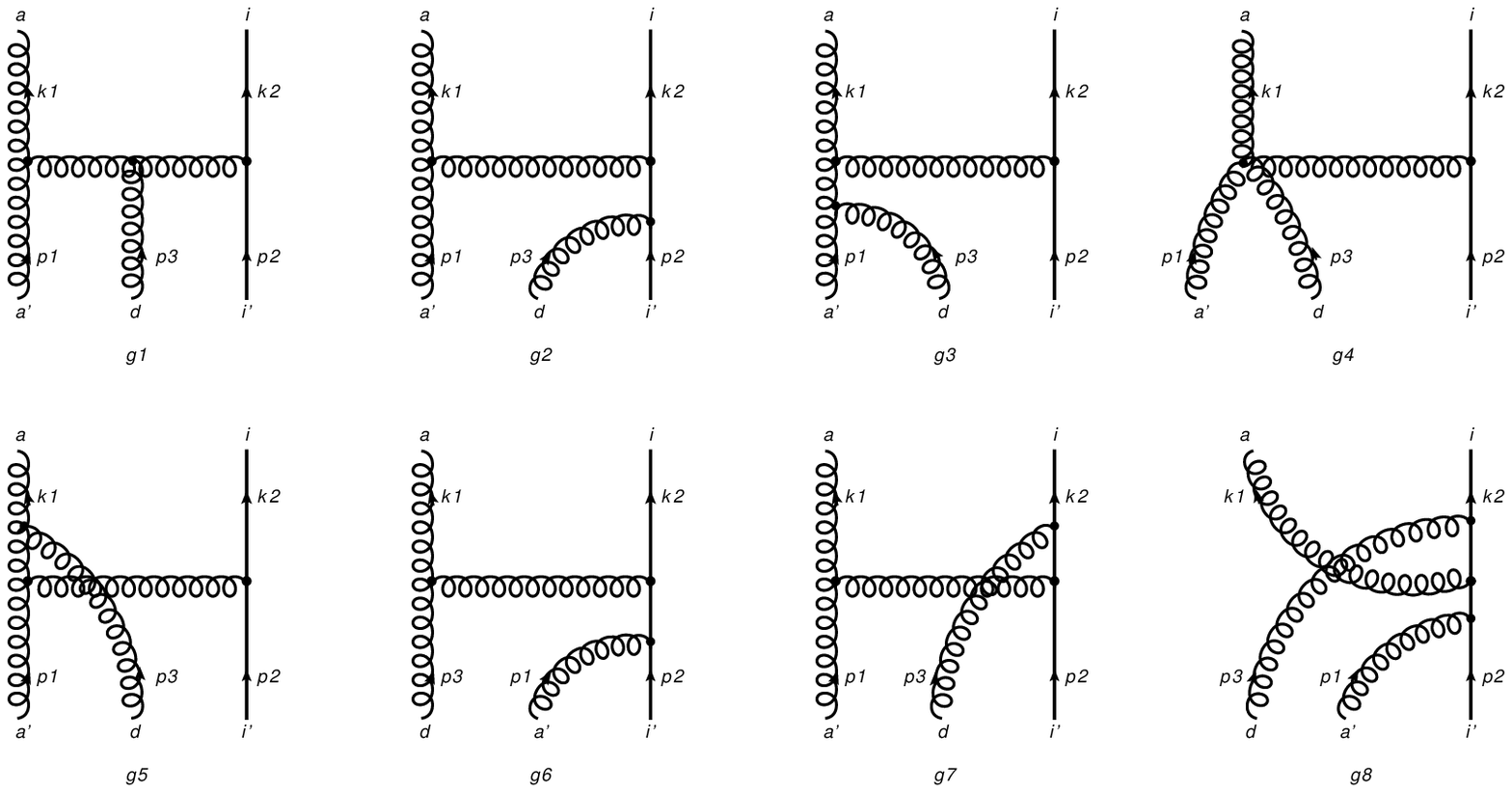}}
      \makebox[\textwidth]{\includegraphics[width=0.85\textwidth]{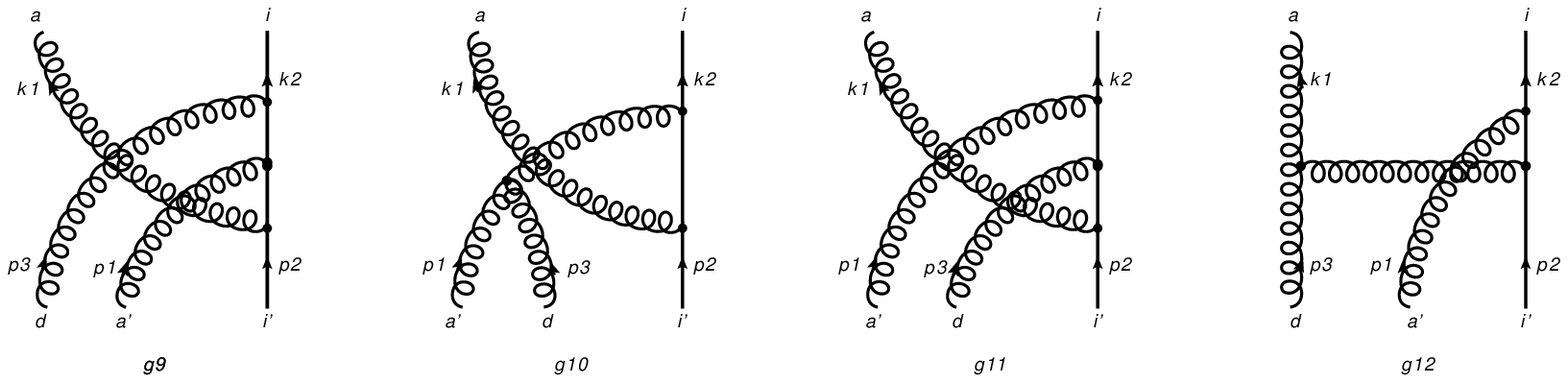}}
	\end{figure}
\begin{figure}[H]
    \centering
         \psfrag{i}[bc][bc]{\scriptsize$i$}
	\psfrag{i'}[tc][tc]{\scriptsize$i'$}
	\psfrag{a}[bc][bc]{\scriptsize$a$}
	\psfrag{d}[tc][tc]{\scriptsize$d$}
	\psfrag{a'}[tc][tc]{\scriptsize$a'$}
	\psfrag{k1}[cl][cl]{\scriptsize$k_1$}
	\psfrag{k2}[cl][cl]{\scriptsize$k_2$}
	\psfrag{p1}[cl][cl]{\scriptsize$p_1$}
	\psfrag{p2}[cl][cl]{\scriptsize$p_2$}
	\psfrag{p3}[c][]{\scriptsize$p_3$}
	\psfrag{k2mp2}[cl][cl]{\scriptsize$k_2-p_2$}
	\psfrag{k2mp3}[cl][cl]{\scriptsize$k_2-p_3$}
	\psfrag{g1}[cc][cl]{\scriptsize$C_1$}
	\psfrag{g2}[cc][cl]{\scriptsize$C_3-C_1$}
	\psfrag{g3}[cc][cl]{\scriptsize$-(C_1+C_2)$}
	\psfrag{g4}[cc][cl]{\scriptsize$(C_1+C_2);C_1;C_2$}
	\psfrag{g5}[cc][cl]{\scriptsize$C_2$}
	\psfrag{g6}[cc][cl]{\scriptsize$C_6-C_2-C_5$}
	\psfrag{g7}[cc][cl]{\scriptsize$C_3$}
	\psfrag{g8}[cc][cl]{\scriptsize$C_3+C_4$}
	\psfrag{g9}[cc][cl]{\scriptsize$C_4$}
	\psfrag{g10}[cc][cl]{\scriptsize$C_4-C_5$}
	\psfrag{g11}[cc][cl]{\scriptsize$C_5$}
	\psfrag{g12}[cc][cl]{\scriptsize$C_6-C_5$}
	\psfrag{g13}[cc][cl]{\scriptsize$C_6$}
	\psfrag{g14}[cc][cl]{\scriptsize$C_6+C_3-C_1$}
	\psfrag{g15}[cc][cl]{\scriptsize$C_3+C_4+C_6-C_2-C_5$}
	\psfrag{g16}[cc][cl]{\scriptsize$C_2+C_5-C_1-C_4$}
      \makebox[\textwidth]{\includegraphics[width=0.85\textwidth]{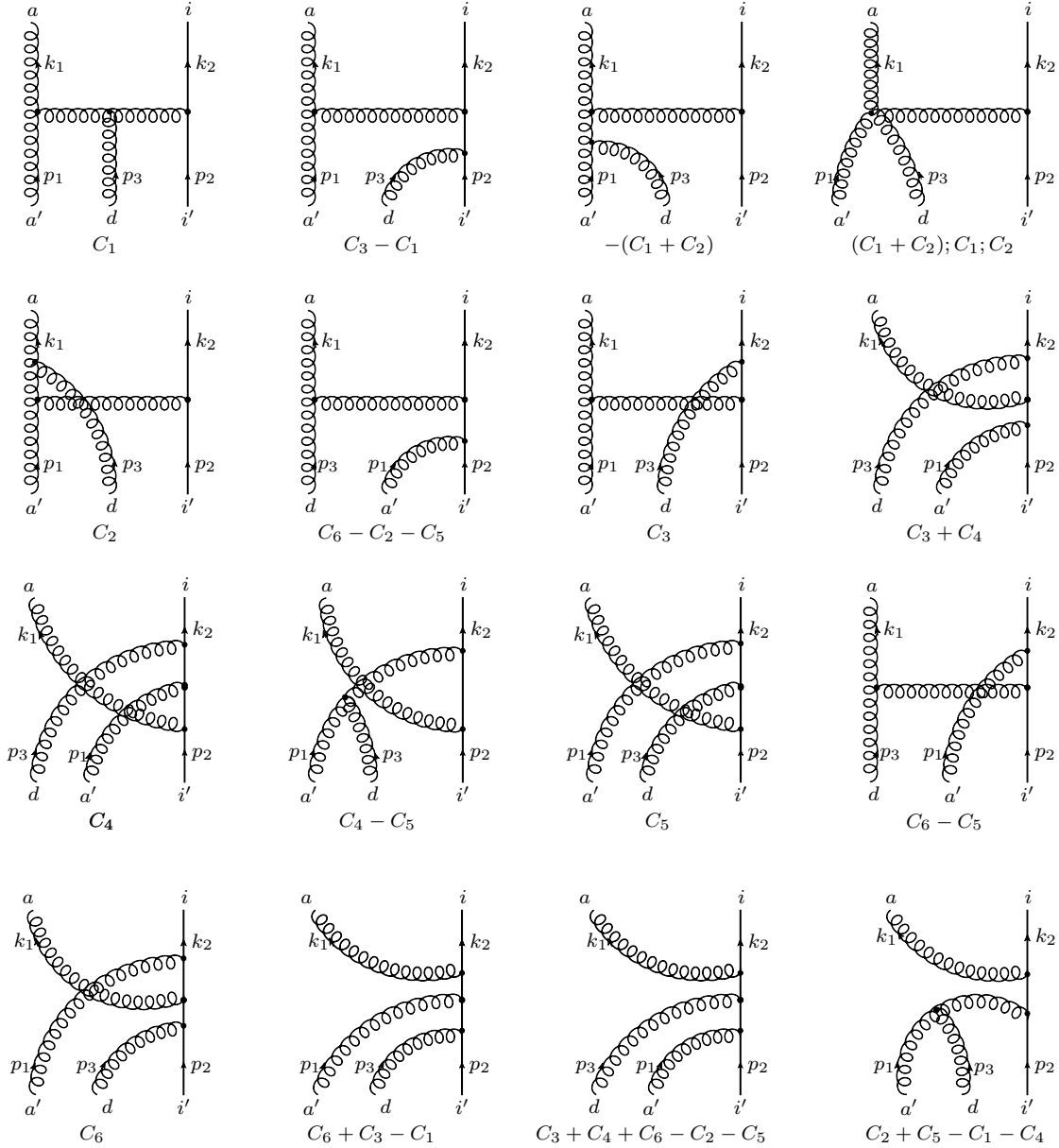}}
	   \caption{Feynman diagrams defining the evolution kernels in Sects.\ \ref{sec63} and \ref{sec434}. }
	   \label{fig:2to33}
\end{figure}

\subsubsection{$\,\bar{f}_{++}\psi_{-}$ and $\tfrac{1}{2}D_{-+}\bar{f}_{++}\psi_{+}$}
\label{sec63}

For the transitions of 
\begin{align}
\mathcal{O}^{ai}(x_1,x_2)=\bar{f}_{++}^a(x_1)\psi_{-}^i(x_2)
\, , \qquad
\mathcal{O}^{ai}(x_1,x_2)=\tfrac{1}{2}D_{-+}\bar{f}_{++}^a(x_1)\psi^i_{+}(x_2)
\end{align}
to 
\begin{align}
\mathcal{O}^{aid}=g\sqrt{2}\bar{f}_{++}^a(y_1)\psi_{+}^i(y_2)\bar{f}^d_{++}(y_3)\, .
\end{align}
we find
\begin{align}
[ \mathcal{K} \, \mathcal{O}]^{ai}(x_1,x_2)
=
\int [\mathcal{D}^3 y]_2 \sum_{c=1}^{6} [C_c]^{ai}_{a'i'd} K_c (x_1, x_2 | y_1, y_2, y_3) \mathcal{O}^{a'i'd}(y_1,y_2,y_3)\, ,
\end{align}
where the color structures are determined by the following tensors
\begin{align}
\label{color2}
[C_1]^{ai}_{a'i'd}&=-i(t^c)_{ii'}f^{cde}f^{aa'e}\, ,&[C_2]^{ia}_{ia'd'}&=-i(t^c)_{ii'}f^{ade}f^{ca'e}\, ,&[C_3]^{ia}_{i'a'd}&=-(t^dt^e)_{ii'}f^{aa'e}\, ,
\nonumber
\\
[C_4]^{ia}_{ia'd}&=i(t^dt^{a'}t^a)_{ii'}\, ,& [C_5]^{ia}_{i'a'd}&=i(t^{a'}t^dt^a)_{ii'}\, ,& [C_6]^{ia}_{i'a'd}&=i(t^{a'}t^at^d)_{ii'}\, .
\end{align}
For the $\mathcal{O}^{ai}(x_1,x_2)=\bar{f}_{++}^a(x_1)\psi_{-}^i(x_2)$ case, the Feynman diagrams responsible for the one-loop evolution are 
presented in Fig.\ \ref{fig:2to33} and produce the following contributions
\begin{align}
\label{628}
K_1&=-\frac{x_1^2 \left(8 y_1^3+9 y_1^2 (y_2+2 y_3)+3 y_1 (y_2+2 y_3)^2+y_3 (y_2+y_3) (y_2+2 y_3)\right)\theta (x_1) }{y_1^2 (y_1+y_3)^3(x_1+x_2)^3}
\nonumber
\\
&+\frac{x_1^2 (y_2+2 y_3) \theta (x_1-y_1)}{y_1^2 y_3^2 (y_2+y_3)^2}-\frac{x_1^2 (y_1+3 y_3) \theta (y_2-x_2)}{y_2 y_3^2 (y_1+y_3)^3}+\frac{x_1^2(y_1+3 (y_2+y_3))\theta (-x_2) }{y_2 (y_2+y_3)^2 (x_1+x_2)^3}\, ,
\\
K_2&=\frac{x_1^2 \left(9 y_3^2 (2 y_1+y_2)+3 y_3 (2
   y_1+y_2)^2+y_1 (y_1+y_2) (2 y_1+y_2)+8 y_3^3\right)\theta (x_1)}{y_3^2 (y_1+y_3)^3 (x_1+x_2)^3}
\nonumber
\\
&-\frac{x_1^2 (2 y_1+y_2) \theta (x_1-y_3)}{y_1^2 y_3^2 (y_1+y_2)^2}+\frac{x_1^2 (3 y_1+y_3) \theta (y_2-x_2)}{y_1^2 y_2 (y_1+y_3)^3}-\frac{x_1^2(3 (y_1+y_2)+y_3) \theta (-x_2)}{y_2 (y_1+y_2)^2 (x_1+x_2)^3}\, ,
\\
K_3&= \frac{x_1^2\left(4 y_1^2+y_1 (5 y_2+2 y_3)+y_2 (y_2+y_3)\right) \theta (x_1) }{y_1^2 (y_1+y_2)^2 (x_1+x_2)^3}-\frac{x_1^2 \theta (x_1-y_1)}{y_1^2 y_2 (y_2+y_3)^2}
\nonumber
\\
&+\frac{(y_3-x_2)\left(x_1^2+x_1 y_2+y_2 (y_1+y_2)\right) \theta (y_3-x_2)}{y_2 y_3^2 (x_1-y_1) (y_1+y_2)^2}
\nonumber
\\
&+\frac{1}{y_3^2 (x_1-y_1) (y_2+y_3)^2 (y_1+y_2+y_3)^3}\Big\{x_1^2 y_1 (y_1 (y_2+2 y_3)+(y_2+y_3)^2 (x_1+x_2)^3
\nonumber
\\
&\ \ \ \ -x_1^3 (y_1 (y_2+2 y_3)+(y_2+y_3) (y_2+4 y_3))+(y_2+y_3)(y_2+4 y_3))\Big\}\theta (-x_2)\, ,
\\
K_4&=\frac{x_1^2  (3 (y_1+y_2)+y_3)\theta (x_1)}{y_2 (y_1+y_2)^2 (y_1+y_2+y_3)^3}-\frac{(x_1-y_2)^2 (3 y_1+y_3) \theta (x_1-y_2)}{y_1^2 y_2 (y_1+y_3)^3}
\nonumber
\\
&+\frac{(x_2-y_3) (y_2 (y_1+y_2)-x_1 (2 y_1+y_2)) \theta (y_3-x_2)}{y_1^2 y_3^2(y_1+y_2)^2}
\nonumber
\\
& -\frac{1}{y_3^2
   (y_1+y_3)^3 (y_1+y_2+y_3)^3}\Big\{x_1^2
\big(9 y_3^2 (2 y_1+y_2)+3 y_3 (2 y_1+y_2)^2+8 y_3^3
\nonumber
\\
& \ \ \ \ +y_1 (y_1+y_2) (2 y_1+y_2)\big)-2 x_1 (y_1+3 y_3) (x_1+x_2)^3
+y_2 (y_1+3 y_3) (x_1+x_2)^3\Big\}\theta (-x_2)\, ,
\\
K_5&=\frac{x_1^2\left(y_1+y_2+3y_3\right)\theta(x_1) }{y_2 y_3^2 (x_1+x_2)^3}-\frac{(x_1-y_2)^2
   (y_1+3 y_3) \theta (x_1-y_2)}{y_2 y_3^2 (y_1+y_3)^3}-\frac{x_1^2 \theta (x_1-y_3)}{ y_2 y_3^2 (y_1+y_2)^2}
\nonumber
\\
&+\frac{(x_2-y_1) \left(x_1^2-x_1y_2+y_2y_3\right)\theta (y_1-x_2)}{y_1^2 y_2 y_3^2 (y_3-x_1)}+ \frac{1}{y_1^2 y_3 (x_1+x_2)^3}\bigg\{\frac{x_2 (x_1-y_2) (x_1+x_2)^2}{y_1+y_3}
\nonumber
\\
& \ \ \ \ +\frac{y_1 (x_1+x_2)}{(x_1-y_3) (y_1+y_2)^2} \Big[x_1^2 (y_3 (y_1+y_2+x_2)+2 x_2 (y_1+y_2))+x_1 y_3 (y_1+y_2) (2 x_2-y_1-y_2)
\nonumber
\\
& \ \ \ \ \ \ +y_3 x_2^2 (y_1+y_2)\Big]-\frac{(x_1+x_2)^2
   \big(x_1^3-x_1^2 (y_2+y_3)+y_3(x_1 y_1-x_2y_2)\big)}{(x_1-y_3) (y_1+y_2)}+\frac{x_1 y_1 x_2 (y_3-y_1)}{y_1+y_3}
\nonumber
\\
&\ \ \ \ -\frac{y_1 (x_1-y_2) (x_1+x_2) (y_1-y_3) \left((x_1+x_2)^2-x_1 (2 y_1+y_2+2
   y_3)\right)}{(y_1+y_3)^3}-x_1 y_1 x_2\bigg\}\theta (-x_2)\, ,
\\
\label{182}
K_6&=-\frac{x_1^2(y_1 (y_2+2 y_3)+(y_2+y_3) (y_2+4 y_3)) \theta (x_1) }{ y_3^2 (y_2+y_3)^2(x_1+x_2)^3}+\frac{x_1^2 \theta (x_1-y_3)}{y_2 y_3^2 (y_1+y_2)^2}
\nonumber
\\
&+\frac{(x_2-y_1) \left(x_1^2+x_1 y_2+y_2 (y_2+y_3)\right) \theta (y_1-x_2)}{y_1^2 y_2 (x_1-y_3) (y_2+y_3)^2}+\frac{1}{y_1^2 (x_1-y_3) (y_1+y_2)^2 (x_1+x_2)^3}\Big\{x_1^3
\nonumber
\\
&\ \ \ \ \ \ \ \ \times\ \left(4 y_1^2+y_1 (5 y_2+2 y_3)+y_2 (y_2+y_3)\right)-x_1^2 y_3
   \left(4 y_1^2+y_1 (5 y_2+2 y_3)+y_2 (y_2+y_3)\right)
\nonumber
\\
&\ \ \ \ \ \ \ \  -(y_1+y_2)^2 (x_1+x_2)^3\Big\}\theta (-x_2)\, .
\end{align}

%%%%%%%%%%%%%%%%%%%%%%%%%%%%%%%%%%%%%%%%%%%%%%%%%%%%%%%%%%%%%%%%%%%%%%%%%%
%                                                                                                                   FIGURE 10                                                                                                                                   %
%%%%%%%%%%%%%%%%%%%%%%%%%%%%%%%%%%%%%%%%%%%%%%%%%%%%%%%%%%%%%%%%%%%%%%%%%%
\begin{figure}[t]
         \psfrag{i}[bc][bc]{\scriptsize$i$}
	\psfrag{i'}[tc][tc]{\scriptsize$i'$}
	\psfrag{a}[bc][bc]{\scriptsize$a$}
	\psfrag{d}[tc][tc]{\scriptsize$d$}
	\psfrag{a'}[tc][tc]{\scriptsize$a'$}
	\psfrag{k1}[cl][cl]{\scriptsize$k_1$}
	\psfrag{k2}[cl][cl]{\scriptsize$k_2$}
	\psfrag{p1}[cl][cl]{\scriptsize$p_1$}
	\psfrag{p2}[cl][cl]{\scriptsize$p_2$}
	\psfrag{p3}[c][]{\scriptsize$p_3$}
	\psfrag{k1mp1}[cl][cl]{\scriptsize$k_1-p_1$}
	\psfrag{k1mp3}[cl][cl]{\scriptsize$k_1-p_3$}
	\psfrag{g1}[cc][cl]{\scriptsize$C_1$}
	\psfrag{g2}[cc][cl]{\scriptsize$C_3-C_1$}
	\psfrag{g3}[cc][cl]{\scriptsize$-(C_1+C_2)$}
	\psfrag{g4}[cc][cl]{\scriptsize$(C_1+C_2);C_1;C_2$}
	\psfrag{g5}[cc][cl]{\scriptsize$C_2$}
	\psfrag{g6}[cc][cl]{\scriptsize$C_6-C_2-C_5$}
	\psfrag{g7}[cc][cl]{\scriptsize$C_3$}
	\psfrag{g8}[cc][cl]{\scriptsize$C_3+C_4$}
	\psfrag{g9}[cc][cl]{\scriptsize$C_4$}
	\psfrag{g10}[cc][cl]{\scriptsize$C_4-C_5$}
	\psfrag{g11}[cc][cl]{\scriptsize$C_5$}
	\psfrag{g12}[cc][cl]{\scriptsize$C_6-C_5$}
	\psfrag{g13}[cc][cl]{\scriptsize$C_6$}
	\psfrag{g14}[cc][cl]{\scriptsize$C_3+C_6-C_1$}
	\psfrag{g15}[cc][cl]{\scriptsize$C_3+C_4+C_6-C_2-C_5$}
	\psfrag{g16}[cc][cl]{\scriptsize$C_2+C_5+C_1-C_4$}
         \makebox[\textwidth]{\includegraphics[width=0.85\textwidth]{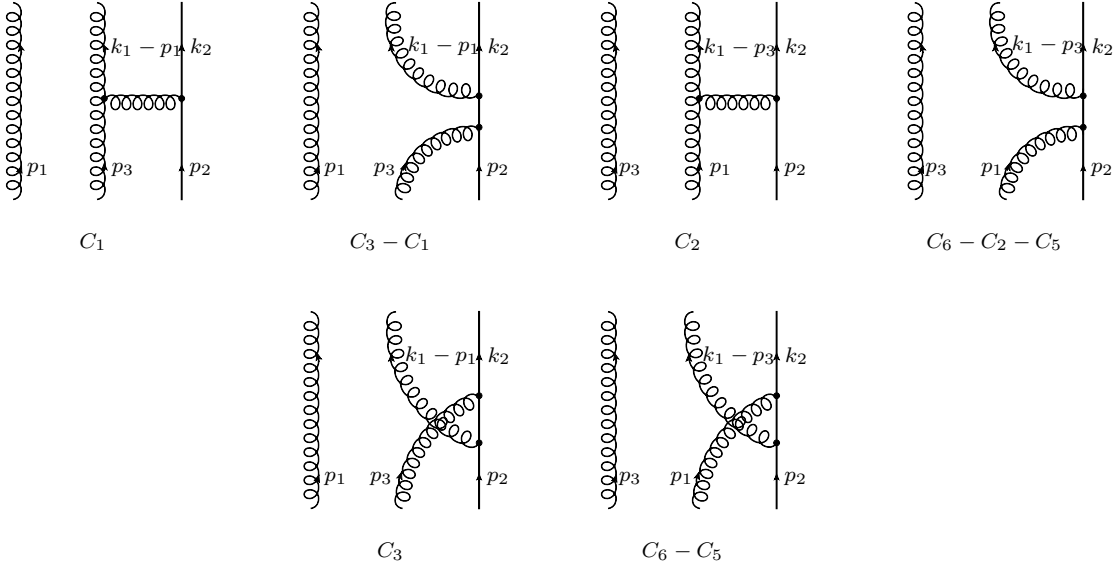}}
         \caption{Graphs producing the contribution of gluon fields in the covariant derivative $D_{-+}$ as for 
         the case of $\tfrac{1}{2}D_{-+}\bar{f}_{++}\psi_-$ in Sect.\ \ref{sec63}, or in the form of $\bar{A}_{\perp}{A}_{\perp}$ for the case of $f_{+-}\psi$ in 
         Sect.\ \ref{sec434}.}
      \label{fig:2to34}
\end{figure}

\noindent For the $\mathcal{O}^{ai}(x_1,x_2)=\tfrac{1}{2}D_{-+}\bar{f}_{++}^a(x_1)\psi^i_{+}(x_2)$ operator, the Feynman graphs describing its transition into 
$\bar{f}^{a'}_{++}(z_1)\psi^{i'}_{+}(z_2)\bar{f}_{++}^d(z_3)$ are shown in Figs.\ \ref{fig:2to33} and \ref{fig:2to34}, such that
\begin{align}
K_1&=\frac{x_1^3}{y_1^2 (y_1+y_3)^3 (x_2-y_2) (x_1+x_2)^3}\Big\{x_2 \Big[8 y_1^3+9 y_1^2
   (y_2+2 y_3)+3 y_1 (y_2+2 y_3)^2+y_3 (y_2+y_3) 
\nonumber
\\
&\ \ \ \ \ \ \ \ \ \ \times (y_2+2 y_3)\Big]+(y_1+y_3)^3 (3y_1+y_2+y_3)\Big\}\theta (x_1)+\frac{1}{y_1^2 y_3^2 (y_2+y_3)^2 (x_2-y_2)}\Big\{x_1^4 (y_2+2 y_3)
\nonumber
\\
&\ \ \ \ -x_1^3 \left(y_1 (y_2+2 y_3)+y_2^2+3 y_2 y_3+3 y_3^2\right)-3 x_1^2 y_1^2 (y_2+2 y_3)+x_1 y_1^2 \big(5 y_1 (y_2+2 y_3)
\nonumber
\\
&\ \ \ \ +2 y_2^2+7 y_2 y_3+8 y_3^2\big)-y_1^2
   \left(2 y_1^2 (y_2+2 y_3)+y_1 \left(y_2^2+4 y_2 y_3+5 y_3^2\right)-y_3 (y_2+y_3)^2\right)\Big\}\theta(x_1-y_1)
\nonumber
\\
&-\frac{1}{y_2 y_3^2 (y_1+y_3)^3}\Big\{x_1^2 y_2 (y_1+3 y_3)-x_1^3(y_1+3
   y_3)+x_1 (y_1+y_3) \left(3 y_1^2+y_1 (y_2+6 y_3)+3 y_3 (y_2+y_3)\right)
\nonumber
\\
&\ \ \ \  -(y_1+y_3)^2 \left(2 y_1^2+y_1 (y_2+2 y_3)-y_2 y_3\right)\Big\}\theta (y_2-x_2)
\nonumber
\\
&+\frac{x_2 \left(x_1^2 (y_1+3 (y_2+y_3))+x_1 (x_1+x_2) (y_1+3 (y_2+y_3))-2 y_1 (x_1+x_2)^2\right)\theta (-x_2)}{y_2 (y_2+y_3)^2 (x_1+x_2)^3}\, ,
\\
K_2&= \frac{x_1^3}{y_3^2 (y_1+y_3)^3 (y_2-x_2) (x_1+x_2)^3}\Big\{x_2 \big(9 y_3^2 (2 y_1+y_2)+3 y_3 (2 y_1+y_2)^2+y_1 (y_1+y_2)
\nonumber
\\
&\ \ \ \ \times  (2 y_1+y_2)+8 y_3^3\big)+(y_1+y_3)^3
   (y_1+y_2+3 y_3)\Big\}\theta (x_1)+\frac{1}{y_1^2 y_3^2 (y_1+y_2)^2 (x_2-y_2)}
\nonumber
\\
&\ \ \ \ \times\Big\{x_1^3 \big(y_1^2+x_2 (2 y_1+y_2)\big)+y_3^2 \big(3 x_1^2 (2 y_1+y_2)-x_1 \left(8 y_1^2+7 y_1 y_2+2 y_2^2\right)
\nonumber
\end{align}
\begin{align}
&\ \ \ \ -y_1 (y_1+y_2)^2\big)-y_3^3 \big(5 x_1 (2 y_1+y_2)-5 y_1^2-4 y_1 y_2-y_2^2\big)+2 y_3^4 (2y_1+y_2)\Big\}\theta (x_1-y_3) 
\nonumber
\\
&\ \ \ \ +\frac{1}{y_1^2 y_2 (y_1+y_3)^3}\Big\{x_1^2 y_2 (3 y_1+y_3)-x_1^3 (3 y_1+y_3)+x_1 (y_1+y_3)\big(y_3 (6 y_1+y_2)
\nonumber
\\
&+3 y_1 (y_1+y_2)+3 y_3^2\big)-(y_1+y_3)^2(y_3 (y_2+2 y_3)-y_1 (y_2-2 y_3))\Big\}\theta (y_2-x_2)
\nonumber
\\
&\ \ \ \ -\frac{x_2\big(x_1^2 (3 (y_1+y_2)+y_3)+x_1 (x_1+x_2) (3 (y_1+y_2)+y_3)-2 y_3 (x_1+x_2)^2\big)}{y_2 (y_1+y_2)^2 (x_1+x_2)^3}\theta (-x_2)\, ,
\\
K_3&=\frac{x_1^3\left(4 y_1^2+y_1 (5 y_2+2 y_3)+y_2 (y_2+y_3)\right)\theta (x_1) }{ y_1^2 (y_1+y_2)^2 (x_1+x_2)^3}-\frac{(x_1-y_1)^2 (x_1+2 y_1) \theta (x_1-y_1)}{y_1^2 y_2(y_2+y_3)^2}
\nonumber
\\
&+\frac{(x_2-y_3)^2 (x_1+2 (y_1+y_2)) \theta (y_3-x_2)}{y_2 y_3^2 (y_1+y_2)^2}-\frac{1}{y_3^2 (y_2+y_3)^2 (x_1+x_2)^3}\Big\{x_1^3 (y_1 (y_2+2 y_3)
\nonumber
\\
&\ \ \ \ +(y_2+y_3) (y_2+4 y_3))-3 x_1 (y_2+2y_3) (x_1+x_2)^3+2 (x_1+x_2)^3 \big(y_1 (y_2+2 y_3)
\nonumber
\\
&\ \ \ \ +(y_2+y_3)^2\big)\Big\}\theta (-x_2)\, ,
 \\
K_4&= -\frac{x_1^3  (3 (y_1+y_2)+y_3)\theta (x_1)}{y_2 (y_1+y_2)^2 (x_1+x_2)^3}+\frac{(x_1-y_2)^3 (3
   y_1+y_3) \theta (x_1-y_2)}{y_1^2 y_2 (y_1+y_3)^3}
\nonumber
\\
&-\frac{(x_2-y_3)^2 \left(2 x_1 y_1+x_1 y_2+y_1^2-y_2^2\right) \theta (y_3-x_2)}{y_1^2 y_3^2 (y_1+y_2)^2}+\frac{x_2}{y_3^2 (y_1+y_3)^3 (y_1+y_2+y_3)^3} 
\nonumber
\\
&\times \Big\{3 x_2 (y_1+y_3)\left(3 y_3 (2 y_1+y_2)+y_1 (y_1+y_2)+5 y_3^2\right) (x_1+x_2)-x_2^2 \big(9 y_3^2 (2 y_1+y_2)
\nonumber
\\
&\ \ \ \ +3 y_3(2 y_1+y_2)^2+y_1 (y_1+y_2) (2 y_1+y_2)+8 y_3^3\big)-6 y_3 (y_1+y_3)^2 (x_1+x_2)^2\Big\}\theta (-x_2)\, ,
\\
K_5&=-\frac{x_1^3\left(y_1+y_2+3y_3\right)\theta (x_1) }{y_2y_3^2 (x_1+x_2)^3}+\frac{(x_1-y_2)^3 (y_1+3 y_3)\theta (x_1-y_2)}{y_2 y_3^2 (y_1+y_3)^3}
\nonumber
\\
&+\frac{(x_1-y_3)^2 (x_1+2 y_3) \theta (x_1-y_3)}{y_2 y_3^2 (y_1+y_2)^2}-\frac{(x_1-y_2+2y_3) \left(x_2-y_1\right)^2 \theta (y_1-x_2)}{y_1^2 y_2 y_3^2}
\nonumber
\\
&+\frac{1}{y_1^2 y_3} \bigg\{\frac{2 x_1^3 y_1^2}{y_2 (x_1+x_2)^3}-\frac{x_1^3 y_1^2}{y_2 (x_1+x_2) (y_1+y_2)^2}-\frac{x_1^3 y_1^2}{y_2 (x_1+x_2)^2 (y_1+y_2)}-\frac{2 y_1^2 (x_1-y_2)^3}{y_2
   (y_1+y_3)^3}
\nonumber
\\
&\ \ \ \ +\frac{3 y_3 \left((y_1+y_2)^2-x_1 (2 y_1+y_2)\right)}{(y_1+y_2)^2}+\frac{(x_1-y_2)^3}{y_2 (y_1+y_3)}+\frac{y_1 (x_1-y_2)^3}{y_2 (y_1+y_3)^2}+\frac{2 y_3^2 (2
   y_1+y_2)}{(y_1+y_2)^2}\bigg\}\theta (-x_2)\, ,
\\
K_6&=\frac{x_1^3  (y_1 (y_2+2 y_3)+(y_2+y_3) (y_2+4 y_3))\theta (x_1)}{y_3^2 (y_2+y_3)^2 (x_1+x_2)^3}-\frac{(x_1-y_3)^2 (x_1+2
   y_3) \theta (x_1-y_3)}{y_2 y_3^2 (y_1+y_2)^2}
\nonumber
\\
&+\frac{(x_2-y_1)^2 (x_1+2 (y_2+y_3)) \theta (y_1-x_2)}{y_1^2 y_2 (y_2+y_3)^2}-\frac{x_2}{y_1^2 (y_1+y_2)^2 (x_1+x_2)^3} \Big\{x_2^2 \big(4 y_1^2+y_1 (5 y_2+2 y_3)
\nonumber
\\
&\qquad +y_2 (y_2+y_3)\big)-3 x_2 \left(4 y_1^2+y_1 (5 y_2+2 y_3)+y_2 (y_2+y_3)\right) (x_1+x_2)
\nonumber
\\
&\qquad+6 y_1 (y_1+y_2)(x_1+x_2)^2\Big\}\theta (-x_2)\, .
\end{align}

 In this transition channel, we encounter the same logarithmic conundrum as in the channel discussed in the previous Sect.\ \ref{sec62} after Fourier transforming 
 corresponding kernels derived in Ref.\ \cite{Braun:2009vc}. Yet again making use of the symmetry under $a\leftrightarrow d$, $y_1\leftrightarrow y_3$, we manage 
 to get rid of all  logarithmic terms and the resulting expression completely agree with our Feynman diagrammatic results. All our results here are conformally invariant 
 as expected. The inverse-Fourier transformed kernels are provided in Appendix \ref{appc2} for comparison with Ref.\ \cite{Braun:2009vc}.
 
 \subsubsection{$f_{+-}\psi_{+}$ and $f_{++}\psi_{-}$}
\label{sec434}

The two-particle blocks
\begin{align}
\mathcal{O}^{ai}(x_1,x_2)=f_{+-}^a(x_1)\psi_{+}^i(x_2)
\, , \qquad
\mathcal{O}^{ai}(x_1,x_2)=f_{++}^a(x_1)\psi^i_{-}(x_2)
\end{align}
undergo a transition to the following three-field operator
\begin{align}
\mathcal{O}^{aid}=g \sqrt{2} f_{++}^a(y_1)\psi_{+}^i(y_2)\bar{f}^d_{++}(y_3)
\, ,
\end{align}
according to
\begin{align}
[ \mathcal{K} \, \mathcal{O} ]^{ai}(x_1,x_2)
=
\int [\mathcal{D}^3 y]_2\sum_{c=1}^{6} [C_c]^{ai}_{a'i'd} K_c (x_1, x_2 | y_1, y_2, y_3) \mathcal{O}^{a'i'd} (y_1,y_2,y_3)\, ,
\end{align}
where the decomposition runs over the color structures  given in Eq.\ \re{612}.

The evolution kernels for the two cases $\mathcal{O}^{ai}(x_1,x_2)= f_{+-}^a(x_1)\psi_{+}^i(x_2)$ and $\mathcal{O}^{ai}(x_1,x_2)= f_{++}^a(x_1)
\psi^i_{-}(x_2)$ read,
\begin{align}
K_1&=\frac{x_1(3y_1+y_3)\theta(x_1)}{y_1^2(y_1+y_3)^3}-\frac{x_1\theta(x_1-y_1)}{y_1^2y_3^2}+\frac{x_1(y_1+3y_3)\theta(x_1-y_1-y_3)}{y_3^2(y_1+y_3)^3}\, ,
\\
K_2&=\frac{x_1(x_1(y_1+3y_3)-2y_3(y_1+y_3))\theta(x_1)}{(x_1-y_1-y_3)y_3^2(y_1+y_3)^3}-\frac{(x_1-y_3)^2\theta(x_1-y_3)}{y_1^2y_3^2(x_1-y_1-y_3)}
\nonumber
\\
&-\frac{((y_1+y_3)^2-x_1(3y_1+y_3))\theta(x_1-y_1-y_3)}{y_1^2(y_1+y_3)^3}\, ,
\\
K_3&=0\, ,
\\
K_4&=0\, ,
\\
K_5&=0\, ,
\\
K_6&=0\, 
\end{align}
and 
\begin{align}
K_1&=\frac{(y_2+2y_3)\theta(x_1-y_1)}{y_3^2(y_2+y_3)^2}-\frac{\theta(x_1-y_1-y_3)}{y_2y_3^2}+\frac{\theta(-x_2)}{y_2(y_2+y_3)^2}\, ,
\\
K_2&=0\, ,
\\
K_3&=-\frac{\theta(x_1-y_1)}{y_2(y_2+y_3)^2}+\frac{\Big(1-\displaystyle\frac{y_2}{x_1-y_1}\Big)\theta(x_1-y_1-y_2)}{y_2y_3^2}-\frac{\Big(y_2+2y_3-\displaystyle\frac{(y_2+y_3)^2}{x_1-y_1}\Big)\theta(-x_2)}{y_3^2(y_2+y_3)^2}\, ,
\\
K_4&=\frac{(x_1(3y_1+y_3)-y_1(y_1+3y_2)-y_3(y_1+y_2))\theta(x_1)}{y_1^2(y_1+y_3)^3}+\frac{(x_2-y_3)\theta(y_3-x_2)}{y_1^2y_3^2}
\nonumber
\\
&+\frac{((y_3-x_2)(y_1+3y_3)-2y_3^2)\theta(-x_2)}{y_3^2(y_1+y_3)^3}
\, , \\
K_5&=\frac{(x_1(y_1+3y_3)-y_1(y_2+y_3)-y_3(3y_2+y_3))\theta(x_1-y_2)}{y_3^2(y_1+y_3)^3}+\frac{(x_2-y_1)\theta(y_1-x_2)}{y_1^2y_3^2}
\nonumber
\\
&-\frac{(2y_1^2-x_1(3y_1+y_3)+3y_1(y_2+y_3)+y_3(y_2+y_3))\theta(-x_2)}{y_1^2(y_1+y_3)^3}\, ,
\\
K_6&=0\, ,
\end{align}
respectively. The first set comes from Fig.\ \ref{fig:2to33}, while the second one from both Figs.\ \ref{fig:2to33} and \ref{fig:2to34}.
This time we have found an exact agreement with the findings of Ref.\ \cite{Braun:2009vc} without further implementation of symmetry properties. 
This can be easily explained by the fact that the operators in this case lack the ``field exchanging'' symmetries. As a result, no redundancies in the evolution 
kernels are allowed to be left over. By taking the heavy quark limit, we also reproduced the results reported in Ref.\ \cite{Offen}.
 
 \subsubsection{$\bar{\psi}_{+}f_{+-}$ and $\tfrac{1}{2}D_{-+}\bar{\psi}_{+}f_{++}$}
 \label{sec435}
 
Next, we calculate
\begin{align}
\mathcal{O}^{ia}(x_1,x_2)=\bar{\psi}_{+}^i(x_1)f_{+-}^a(x_2)
\, , \qquad
\mathcal{O}^{ia}(x_1,x_2)=\tfrac{1}{2}D_{-+}\bar{\psi}^i_{+}(x_1)f_{++}^a(x_2)
\, , 
\end{align}
transition to 
\begin{align}
\mathcal{O}^{iad}=g\sqrt{2}\bar{\psi}_{+}^i(y_1)f_{++}^a(y_2)\bar{f}_{++}^d(y_3)
\, .
\end{align}
It is described by
\begin{align}
[ \mathcal{K} \, \mathcal{O}]^{ia}(x_1,x_2)
=
\int [\mathcal{D}^3y]_2 \sum_{c=1}^{6} [C_n]^{ia}_{a'i'd} K_c (x_1, x_2 | y_1, y_2, y_3) \mathcal{O}^{a'i'd}(y_1,y_2,y_3)\, ,
\end{align}
with the color structures  given in Eq.\ \re{612}.

For $\mathcal{O}^{ia}(x_1,x_2)=\bar{\psi}_{+}^i(x_1)f_{+-}^a(x_2)$ the kernels are
\begin{align}
K_1&=-\frac{x_1x_2(3y_1+y_2+3y_3)\theta(x_1)}{y_1(y_1+y_3)^2(x_1+x_2)^3}+\frac{x_1x_2(y_2+3y_3)\theta(x_1-y_1)}{y_1y_3^2(y_2+y_3)^3}-\frac{x_1x_2(y_1+2y_3)\theta(y_2-x_2)}{y_2^2y_3^2(y_1+y_3)^2}
\nonumber
\\
&+\frac{x_1x_2\big(y_1^2(3y_2+y_3)+3y_1(y_2+y_3)(3y_2+y_3)+2(y_2+y_3)^2(4y_2+y_3)\big)\theta(-x_2)}{y_2^2(x_1+x_2)^3(y_2+y_3)^3}\, ,
\\
K_2&=-\frac{x_1x_2\big(y_1^2+2y_3(y_2+2y_3)+y_1(y_2+5y_3)\big)\theta(x_1)}{y_3^2(y_1+y_3)^2(x_1+x_2)^3}+\frac{x_1x_2\theta(x_1-y_3)}{y_1y_3^2(y_1+y_2)^2}-\frac{x_1x_2\theta(y_2-x_2)}{y_1^2y_2^2(y_1+y_3)^2}
\nonumber
\\
&+\frac{x_1x_2\big(y_1^2+2y_2(2y_2+y_3)+y_1(5y_2+y_3)\big)\theta(-x_2)}{y_2^2(y_1+y_2)^2(x_1+x_2)^3}\, ,
\\
K_3&=\frac{x_1\big(2(x_1+x_2)(y_1+y_2)-x_1(3(y_1+y_2)+y_3)\big)\theta(x_1)}{y_1(y_1+y_2)^2(x_1+x_2)^3}
\nonumber
\\
&-\frac{x_1\big(2y_2(y_2+y_3)
-x_1(3y_2+y_3)+y_1(3y_2+y_3)\big)\theta(x_1-y_1)}{y_1y_2^2(y_2+y_3)^3}
\nonumber
\\
&-\frac{(x_2-y_3)^2\big(y_2^2+x_1(y_1+2y_2)\big)\theta(y_3-x_2)}{y_2^2y_3^2(x_1-y_1)(y_1+y_2)^2}+\frac{1}{y_3^2(y_2+y_3)^3}\Big\{\frac{(y_2+y_3)^3}{x_1-y_1}
\nonumber
\\
&+\frac{x_1^2\big(y_1^2(y_2+3y_3)+3y_1(y_2+y_3)(y_2+3y_3)+2(y_2+y_3)^2(y_2+4y_3)\big)}{(x_1+x_2)^3}
\nonumber
\\
&-\frac{x_1\big(y_1^2(y_2+3y_3)+3(y_2+y_3)^2(y_2+3y_3)+2y_1(y_2+y_3)(2y_2+5y_3)\big)}{(x_1+x_2)^2}\Big\}\theta(-x_2)\, ,
\\
K_4&=\frac{x_1}{y_3^2(x_1+x_2)^3(y_2+y_3)^3}\Big\{x_1\big(y_1^2(y_2+3y_3)+3y_1(y_2+y_3)(y_2+3y_3)
\nonumber
\\
&+2(y_2+y_3)^2(y_2+4y_3)\big)-(x_1+x_2)(y_2+y_3)\big(y_1(y_2+3y_3)+(y_2+y_3)(y_2+5y_3)\big)\Big\}\theta(x_1)
\nonumber
\\
&-\frac{x_1\big(x_1(y_1+2y_2)-y_1(y_2+y_3)-y_2(y_2+2y_3)\big)\theta(x_1-y_3)}{y_2^2y_3^2(y_1+y_2)^2}
\nonumber
\\
&-\frac{x_1(x_2-y_1)(3y_2+y_3)\theta(y_1-x_2)}{y_1y_2^2(y_2+y_3)^3}+\frac{x_1x_2(3(y_1+y_2)+y_3)\theta(-x_2)}{y_1(y_1+y_2)^2(x_1+x_2)^3}\, ,
\\
K_5&=\frac{x_1}{y_3^2}\Big\{\frac{2x_1y_2}{y_1(y_2+y_3)^3}-\frac{2y_2(y_1+y_2)+x_1(y_1+2y_2)}{y_2^2(y_1+y_2)^2}-\frac{3x_1+2y_2}{y_1(y_2+y_3)^2}+\frac{4}{y_1(y_2+y_3)}
\nonumber
\\
&-\frac{2x_1(y_1+y_2)}{y_1(x_1+x_2)^3}+\frac{3x_1+2(y_1+y_2)}{y_1(x_1+x_2)^2}-\frac{4}{y_1(x_1+x_2)}\Big\}\theta(x_1)-\frac{(x_1-y_2)^2\theta(x_1-y_2)}{y_1y_2^2y_3^2}
\nonumber
\\
&+\frac{(x_2-y_3)^2\theta(y_3-x_2)}{y_1(y_1+y_2)^2y_3^2}+\frac{(y_1-x_2)\big(x_1(y_2+3y_3)-(y_2+y_3)^2\big)\theta(y_1-x_2)}{y_1y_3^2(y_2+y_3)^3}
\nonumber
\\
&-\frac{\big((x_1+x_2)^3-2x_1(x_1+x_2)(y_1+y_2+2y_3)+x_1^2(y_1+y_2+3y_3)\big)\theta(-x_2)}{y_1y_3^2(x_1+x_2)^3}\, ,
\\
K_6&=\frac{x_1\big(2y_2(x_1+x_2)(y_1+y_2)-x_1(y_1^2+2y_2(2y_2+y_3)+y_1(5y_2+y_3))\big)\theta(x_1)}{y_2^2(y_1+y_2)^2(x_1+x_2)^3}
\nonumber
\\
&+\frac{(x_1-y_2)^2\theta(x_1-y_2)}{y_1y_2^2(y_1+y_3)^2}-\frac{(x_2-y_3)^2\theta(y_3-x_2)}{y_1y_3^2(y_1+y_2)^2}
\nonumber
\\
&+\frac{x_2\big((x_1+x_2)^2(y_1+2y_3)-x_1(y_1^2+2y_3(y_2+2y_3)+y_1(y_2+5y_3))\big)\theta(-x_2)}\, ,
\end{align}
while for $\mathcal{O}^{ia}(z_1,z_2)=\tfrac{1}{2}D_{-+}\bar{\psi}^i_{+}(z_1)f_{++}^a(z_2)$ they are found to be
\begin{align}
K_1&=\frac{x_1^2}{y_1(y_1+y_3)^2}\Big\{\frac{1}{x_2-y_2}+\frac{2x_1y_2-3(x_1+x_2)(x_1+y_2)+5(x_1+x_2)^2}{(x_1+x_2)^3}\Big\}\theta(x_1)
\nonumber
\\
&+\frac{1}{y_3^2}\Big\{1-\frac{(x_1-y_1)^2}{y_1(x_2-y_2)(y_2+y_3)^3}\big(y_2(x_2-y_3)^2+y_3(x_2-y_3)(3y_1+5y_2-3x_1)
\nonumber
\\
&+2y_3^2(4y_1+5y_2-4x_1)+6y_3^3\big)\Big\}\theta(x_1-y_1)
\nonumber
\\
&-\frac{1}{y_2^2y_3^2(y_1+y_3)^2}\Big\{x_1^3(y_1+2y_3)-(y_1+y_3)\big(y_1^3-y_2^2y_3+2y_1^2(y_2+y_3)+y_1y_3(2y_2+y_3)\big)
\nonumber
\\
&+x_1(x_1+x_2)\big(3y_1^2+2y_3(y_2+y_3)+y_1(y_2+5y_3)\big)
\nonumber
\\
&-x_1^2\big(3y_1^2+4y_3(y_2+y_3)+y_1(2y_2+7y_3)\big)\Big\}\theta(y_2-x_2)
\nonumber
\\
&+\frac{x_2^2}{y_2^2(y_2+y_3)^3(x_1+x_2)^3}\Big\{x_1\big(y_1^2(3y_2+y_3)+3y_1(y_2+y_3)(3y_2+y_3)+2(4y_2+y_3)
\nonumber
\\
&\ \ \ \ \times(y_2+y_3)^2\big)-y_1(x_1+x_2)\big(y_1(3y_2+y_3)+(y_2+y_3)(4y_2+y_3)\big)\Big\}\theta(-x_2)\, ,
\\
K_2&=\frac{x_1^2}{y_3^2(y_1+y_3)^2}\Big\{9+\frac{2x_1-y_1}{x_2-y_2}-\frac{2x_1y_2(y_1+y_2)}{(x_1+x_2)^3}+\frac{3(y_2(y_1+y_2)+x_1(y_1+2y_2))}{(x_1+x_2)^3}
\nonumber
\\
&-\frac{4x_1+5y_1+10y_2}{x_1+x_2}\Big\}\theta(x_1)-\frac{x_1^2-y_3^2}{y_1y_3^2(y_1+y_2)^2(x_2-y_2)}\Big\{x_1^2+3y_1^2+(y_2+y_3)^2
\nonumber
\\
&+y_1(4y_2+3y_3)-x_1\big(3y_1+2(y_2+y_3)\big)\Big\}\theta(x_1-y_3)-\frac{x_2^2(x_1+y_1+y_3)\theta(y_2-x_2)}{y_1y_2^2(y_1+y_3)^2}
\nonumber
\\
&+\frac{x_2^2}{y_2^2(y_1+y_2)^2(x_1+x_2)^3}\Big\{x_1\big(y_1^2+2y_2(2y_2+y_3)+y_1(5y_2+y_3)\big)
\nonumber
\\
&+(x_1+x_2)\big(y_1^2+y_1(4y_2+y_3)+y_2(3y_2+2y_3)\big)\Big\}\theta(-x_2)\, ,
\\
K_3&=\frac{x_1^2\big(x_1(3(y_1+y_2)+y_3)-3(x_1+x_2)(y_1+y_2)\big)\theta(x_1)}{y_1(y_1+y_2)^2(x_1+x_2)^3}-\frac{(x_1-y_1)^2}{y_1y_2^2(y_2+y_3)^3}\Big\{x_1(3y_2+y_3)
\nonumber
\\
&-3y_2(y_2+y_3)-y_1(3y_2+y_3)\Big\}\theta(x_1-y_1)+\frac{(y_3-x_2)^3(y_1+2y_2)\theta(y_3-x_2)}{y_2^2y_3^2(y_1+y_2)^2}
\nonumber
\\
&+\frac{x_2^2}{y_3^2(y_2+y_3)^3(x_1+x_2)^3}\Big\{(x_1+x_2)\big(2(y_2+y_3)^3+3y_1(y_2+y_3)
(y_2+2y_3)+y_1^2(y_2+3y_3)\big)
\nonumber
\\
&-x_1\big(y_1^2(y_2+3y_3)+3y_1(y_2+y_3)(y_2+3y_3)+2(y_2+y_3)^2(y_2+4y_3)\big)\Big\}\theta(-x_2)\, ,
\\
K_4&=\frac{x_1^2}{y_3^2(y_2+y_3)^3(x_1+x_2)^3}\Big\{(x_1+x_2)(y_2+y_3)\big(2(y_2+y_3)(y_2+4y_3)+y_1(2y_2+5y_3)\big)
\nonumber
\\
&-x_1\big(y_1^2(y_2+3y_3)+3y_1(y_2+y_3)(y_2+3y_3)+2(y_2+y_3)^2(y_2+4y_3)\big)\Big\}\theta(x_1)
\nonumber
\\
&+\frac{x_1^2-y_3^2}{y_2^2(y_1+y_2)^2y_3^2}\Big\{x_1(y_1+2y_2)-2y_2(y_2+y_3)-y_1(2y_2+y_3)\Big\}\theta(x_1-y_3)
\nonumber
\\
&-\frac{(x_2-y_1)^2\big((y_2+y_3)(2y_2+y_3)+x_1(3y_2+y_3)\big)\theta(y_1-x_2)}{y_1y_2^2(y_2+y_3)^3}
\nonumber
\\
&+\frac{x_2^2\big((x_1+x_2)(2(y_1+y_2)+y_3)+x_1(3(y_1+y_2)+y_3)\big)\theta(-x_2)}{y_1(y_1+y_2)^2(x_1+x_2)^3}\, ,
\\
K_5&=\frac{x_1^2}{y_3^2}\Big\{\frac{3y_2(y_1+y_2)-x_1(y_1+2y_2)}{y_2^2(y_1+y_2)^2}-\frac{2x_1y_2}{y_1(y_1+y_3)^3}+\frac{3(x_1+y_2)}{y_1(y_2+y_3)^2}-\frac{6}{y_1(y_2+y_3)}
\nonumber
\\
&+\frac{2x_1(y_1+y_2)}{y_1(x_1+x_2)^3}-\frac{3(x_1+y_1+y_2)}{y_1(x_1+x_2)^2}+\frac{6}{y_1(x_1+x_2)}\Big\}\theta(x_1)+\frac{(x_1-y_2)^3\theta(x_1-y_2)}{y_1y_2^2y_3^2}
\nonumber
\\
&+\frac{(x_2-y_3)^3\theta(y_3-x_2)}{y_1y_3^2(y_1+y_2)^2}+\frac{(x_2-y_1)^2\big(y_2(y_2+y_3)-x_1(y_2+3y_3)\big)\theta(y_1-x_2)}{y_1y_3^2(y_2+y_3)^3}
\nonumber
\\
&-\frac{x_2^2\big((x_1+x_2)(y_1+y_2)-x_1(y_1+y_2+3y_3)\big)\theta(-x_2)}{y_1y_3^2(x_1+x_2)^3}\, ,
\\
K_6&=\frac{x_1^2\big(x_1(y_1^2+2y_2(2y_2+y_3)+y_1(5y_2+y_3)-3y_2(x_1+x_2)(y_1+y_2))\big)\theta(x_1)}{y_2^2(y_1+y_2)^2(x_1+x_2)^3}
\nonumber
\\
&-\frac{(x_1-y_2)^3\theta(x_1-y_2)}{y_1y_2^2(y_1+y_3)^2}-\frac{(x_2-y_3)^3\theta(y_3-x_2)}{y_1y_3^2(y_1+y_2)^2}
\nonumber
\\
&+\frac{x_2^2}{y_3^2(y_1+y_3)^2(x_1+x_2)^3}\Big\{(x_1+x_2)\big((y_1+2y_3)(y_1+y_2)+y_3^2\big)
\nonumber\\
&-x_1\big(y_1^2+2y_3(y_2+2y_3)+y_1(y_2+5y_3)\big)\Big\}\theta(-x_2)\, .
\end{align}
Here we again find complete agreement with Ref. \cite{Braun:2009vc}.
 
 \subsubsection{$\bar{\psi}_{+}\psi_{-}$ and $\tfrac{1}{2}D_{-+}\bar{\psi}_{+}\psi_{+}$}
 \label{sec436}
  
Finally, we address the evolution of
\begin{align}
\mathcal{O}^{ij}(x_1,x_2)=\bar{\psi}_{+}^i(x_1)\psi_{-}^j(x_2)
\, , \qquad
\mathcal{O}^{ij}(x_1,x_2)=\tfrac{1}{2}D_{-+}\bar{\psi}^i_{+}(x_1)\psi_{+}^j(x_2)
\end{align}
into
\begin{align}
\mathcal{O}^{ijd}=g\sqrt{2} \bar{\psi}_{+}^i(y_1)\psi_{+}^j(y_2)\bar{f}_{++}^d(y_3)
\, .
\end{align}
This sector is determined by the transition
\begin{align}
[ \mathcal{K} \, \mathcal{O}]^{ij}(x_1,x_2)
=
\int [\mathcal{D}^3 y]_2 \sum_{c=1}^{3} [C_c]^{ij}_{a'i'd} K_c (x_1, x_2 | y_1, y_2, y_3) \mathcal{O}^{i'j'd}(y_1,y_2,y_3)\, ,
\end{align}
expanded over three color structures given in Eq.\ \re{63}, with explicit results for $\mathcal{O}^{ij}(x_1,x_2)=\bar{\psi}^i_{+}(x_1)\psi_{-}^j(x_2)$ 
and $\mathcal{O}^{ij}(z_1,z_2)=\tfrac{1}{2}D_{-+}\bar{\psi}^i_{+}(z_1)\psi_{+}^j(z_2)$  cases being
\begin{align}
K_1&=\frac{x_1(2y_1+y_2+2y_3)\theta(x_1)}{y_1(x_1+x_2)^2(y_1+y_3)^2}-\frac{x_1(y_2+2y_3)\theta(x_1-y_1)}{y_1y_3^2(y_2+y_3)^2}+\frac{x_1(y_1+2y_3)\theta(y_2-x_2)}{y_2y_3^2(y_1+y_3)^2}
\nonumber
\\
&-\frac{x_1\big(y_1+2(y_2+y_3)\big)\theta(-x_2)}{y_2(x_1+x_2)^2(y_2+y_3)^2}\, ,
\\
K_2&=\frac{x_1\big(y_1(y_1+y_2)+2y_3(2y_1+y_2)+3y_3^2\big)\theta(x_1)}{y_3^2(x_1+x_2)^2(y_1+y_3)^2}-\frac{x_1\theta(x_1-y_3)}{y_1y_3^2(y_1+y_2)}+\frac{x_1\theta(y_2-x_2)}{y_1y_2(y_1+y_3)^2}
\nonumber
\\
&-\frac{x_1\theta(-x_2)}{y_2(y_1+y_2)(x_1+x_2)^2}\, ,
\\
K_3&=\frac{x_1\theta(x_1)}{y_1(y_1+y_2)(x_1+x_2)^2}-\frac{x_1\theta(x_1-y_1)}{y_1y_2(y_2+y_3)^2}+\frac{(y_3-x_2)(x_1+y_2)\theta(y_3-x_2)}{y_2y_3^2(x_1-y_1)(y_1+y_2)}
\nonumber
\\
&+\frac{1}{y_3^2(x_1-y_1)(x_1+x_2)^2(y_2+y_3)^2}\Big\{(x_1+x_2)^2(y_2+y_3)^2-x_1^2\big(y_1(y_2+2y_3)
\nonumber
\\
&+(y_2+y_3)(y_2+3y_3)\big)+x_1y_1\big(y_1(y_2+2y_3)+(y_2+y_3)(y_2+3y_3)\big)\Big\}\theta(-x_2)\, ,
\end{align}
and
\begin{align}
K_1&=\frac{x_1^2\big(3y_1(y_1+y_2)+y_2^2+3y_3(2y_1+y_2+y_3)-x_1(2y_1+y_2+2y_3)\big)\theta(x_1)}{y_1(x_2-y_2)(y_1+y_3)^2(x_1+x_2)^2}
\nonumber
\\
&+\frac{1}{y_3^2}\Big\{1-\frac{(x_1-y_1)^2\big(y_2(y_1+y_2)+y_3(2y_1+3y_2)+3y_3^2-x_1(y_2+2y_3)\big)}{y_1(x_2-y_2)(y_2+y_3)^2}\Big\}\theta(x_1-y_1)
\nonumber
\\
&-\frac{1}{y_3^2}\Big\{1-\frac{x_2\big(y_1(y_1+y_3)-x_1(y_1+2y_3)\big)}{y_2(y_1+y_3)^2}\Big\}\theta(y_2-x_2)
\nonumber
\\
&-\frac{x_2\big(y_1(x_1+x_2)-x_1(y_1+2(y_2+y_3))\big)\theta(-x_2)}{y_2(x_1+x_2)^2(y_2+y_3)^2}\, ,
\\
K_2&=\frac{x_1^2}{y_3^2(y_1+y_3)^2}\Big\{\frac{y_1+2y_3}{x_2-y_2}+\frac{y_1(y_1+y_2)+2y_3(2y_1+y_2)+3y_3^2}{(x_1+x_2)^2}\Big\}\theta(x_1)
\nonumber
\\
&-\frac{(x_2+y_1)(x_1^2-y_3^2)\theta(x_1-y_3)}{y_1y_3^2(x_2-y_2)(y_1+y_2)}-\frac{x_2(x_1+y_1+y_3)\theta(y_2-x_2)}{y_1y_2(y_1+y_3)^2}+\frac{x_2(2x_1+x_2)\theta(-x_2)}{y_2(y_1+y_2)(x_1+x_2)^2}\, ,
\\
K_3&=\frac{x_1^2\theta(x_1)}{y_1(y_1+y_2)(x_1+x_2)^2}-\frac{(x_1-y_1)^2\theta(x_1-y_1)}{y_1y_2(y_2+y_3)^2}+\frac{(x_2-y_3)^2\theta(y_3-x_2)}{y_2y_3^2(y_1+y_2)}
\nonumber
\\
&-\frac{x_2}{y_3^2(x_1+x_2)^2(y_2+y_3)^2}\Big\{(x_1+x_2)\big((y_2+y_3)^2+y_1(y_2+2y_3)\big)
\nonumber
\\
&-x_1\big(y_1(y_2+2y_3)+(y_2+y_3)(y_2+3y_3)\big)\Big\}\theta(-x_2)\, ,
\end{align}
respectively. Again, when Fourier transformed to the coordinate space, we found complete agreement with the conformally approach of Ref.\ 
\cite{Braun:2009vc}. This completes our study of flavor-nonsinglet transitions.
 
\section{Outlook and Conclusion}
 
In this paper, we generalized the formalism suggested by Bukhvostov-Frolov-Lipatov-Kuraev for renormalization of quasipartonic operators to include
nonquasipartonic operators as well. The advantage of the method is that at one-loop order, the procedure is purely algebraic requiring straightforward
though quite tedious manipulations with Dirac and Lorentz structure of Feynman graphs. The focus of the present study was the evolution equations 
for non-singlet twist-four operators. Their basis consists of four-particle quasipartonic  and three-particle good-good-bad light-cone operators. While the 
former were studied at length in existing literature, the latter were addressed here starting from Feynman graphs, providing an explicit brute force calculation
of these evolution kernels. The main ingredients for these transitions are good-bad two-to-two and two-to-three components. A crucial role in both cases 
is placed by proper use of QCD equations of motion which induce extra contribution that are required for proper closure of evolutions equations. Since 
the basis of twist-four operators is built from conformal primary fields, the resulting evolution kernels should obey a very stringent consistency constraint
of being conformally invariant. This was explicitly confirmed by our analysis. 

We provided a Fourier transform from the momentum to coordinate space and back and checked our findings against the only available earlier results
for nonquasipartonic operators that were derived for light-ray operators making use of the conformal symmetry and dynamical part of the Poincar\'e algebra. 
We found agreement in all cases and also provided a simplified form of light-ray kernels in certain channels that made use of the exchange symmetry of the
operators involved.

\section*{Acknowledgments}

We would like to thank A.~Manashov for very instructive discussions and clarification with regards to results of Ref.\ \cite{Braun:2009vc} and N.~Offen for 
helpful correspondence about heavy light-ray operators and, last but not least, M. Ramsey-Musolf and M. Glatzmaier for their interest in the project at its 
early stages. This work was supported by the U.S. National Science Foundation under the grant PHY-1068286.

\appendix

\section{Sample calculations in light cone gauge}
\label{samplecal}

%%%%%%%%%%%%%%%%%%%%%%%%%%%%%%%%%%%%%%%%%%%%%%%%%%%%%%%%%%%%%%%%%%%%%%%%%%
%                                                                                                                   FIGURE                                                                                                                                      %
%%%%%%%%%%%%%%%%%%%%%%%%%%%%%%%%%%%%%%%%%%%%%%%%%%%%%%%%%%%%%%%%%%%%%%%%%%
\begin{figure}[t]
\label{diag1}
    \centering
	\psfrag{k1}[cl][cl]{\small$k_1$}
	\psfrag{k2}[cl][cl]{\small$k_2$}
	\psfrag{p1}[cl][cl]{\small$p_1$}
	\psfrag{p2}[cl][cl]{\small$p_2$}
	\psfrag{g}[cc][cl]{\small$\frac{\sqrt{2}}{4}\gamma^+\gamma^-(1-\gamma^5)$}
	\psfrag{a mu}[cr][cr]{\small$a\,\,\mu$}
	\psfrag{b nu}[cl][cc]{\small$b\,\,\nu$}
  \makebox[\textwidth]{\includegraphics[width=0.2\paperwidth]{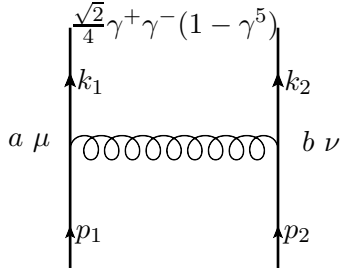}}
      \caption{Feynman Diagram responsible for $\chi_+\otimes\psi_-\rightarrow\chi_+\otimes\psi_-+\chi_-\otimes\psi_+$ in Eq.\ \re{524}.}
      \label{fig:sample}
\end{figure}

In this appendix we provide an explicit calculation of the transitions kernel for good-bad two-to-two quark transitions
$\chi_+\otimes\psi_-\rightarrow\chi_+\otimes\psi_-+\chi_-\otimes\psi_+$, shown in Fig.\ \ref{fig:sample}. The operator in question can be written at one loop
in the form
\begin{align}
\label{Gintegrand}
\mathcal{O}(&x_1,x_2) 
=
\int\frac{d^4p_1}{(2\pi)^4}\frac{d^4p_2}{(2\pi)^4}\frac{d^4k_1}{(2\pi)^4} \frac{d^4k_2}{(2\pi)^4}\delta(k_1^{+}-x_1)\delta(k_2^{+}-x_1)\delta(p_1^+-y_1)
\delta(p_2^+-y_2)\\
&\times 
\bar{\psi}(p_2) 
\big \{ \mathcal{V}_{\nu}^b(k_3,p_2,-k_2)  i \mathcal{P}(-k_2)\frac{\sqrt{2}}{4} \gamma^{+}\gamma^{-}(1+\gamma^5) i\mathcal{P}(k_1) \, \mathcal{V}_{\mu}^a(-k_3,-k_1,p_1) 
(-i) \Delta_{\mu\nu}^{ab}(k_3) \big \} \psi(p_1)
\, , \nonumber
\end{align}
where in the gluon propagator in the light-cone gauge was introduced in Eq.\ \re{GluonPropagator}, while for reader's convenience we provide below 
expressions for the quark propagator and and the vertex function, respectively,
\begin{align*}
\mathcal{P}(k)=\frac{\slashed{k}}{k^2 + i 0} \, , \qquad
\mathcal{V}_\mu^a(k_1,k_2,k_3) =i g t^a \gamma^\mu (2\pi)^4 \delta ^4(k_1+k_2+k_3) \, .
\end{align*}
Denoting the string introduced in curly brackets as $\mathcal{N}/\mathcal{D}$, we can work out the denominator $\mathcal{D}$ stemming from the propagators as 
$\mathcal{D}=(p_1+p_2-k_1)^2(p_1-k_1)^2 k_1^2$. Choosing the loop momentum as $k = k_1$, we expand $\mathcal{D}$ in inverse powers of the transverse 
momentum $k_\perp$ and find immediately for the leading and first subleasing contributions
\begin{align}
\frac{1}{\mathcal{D}} 
&=  
\frac{1}{k_{\perp}^6}
\frac{1}{[k^{+}\beta-1][(k^+ -p_1^+ - p_2^+)\beta - 1][(k^+ -p_1^+)\beta-1]}
\\
&\times
\left[ 
1 - \frac{2 (p_1^{\perp}+p_2^{\perp})\cdot k^{\perp}}{k_\perp^2[(k^+ - p^+_1 - p^+_2) \beta - 1]}-\frac{2 p_1^{\perp}\cdot k^{\perp}}{k_{\perp}^2 [(k^+ - p_1^+)\beta-1]}
\right]
+
O (1/k^8_{\perp}) 
\, , \nonumber
\end{align}
We will parametrize the contributions of the first, second and third terms in the square brackets as $\mathcal{A}$, $\mathcal{B}$ and $\mathcal{C}$ contributions, 
respectively, i.e., $\mathcal{A} - \mathcal{B} - \mathcal{C}$. 

To clarity the manipulations involved in the analysis, the numerator
\begin{align}
\label{Numerator}
\mathcal{N} 
&
= 
- \frac{i\sqrt{2}}{4} g^2 t^a \otimes t^a
\\
&
\times
\bar{\psi}(p_2) [ \gamma^{\nu} ( \slashed{k} - \slashed{p}{}_1 - \slashed{p}{}_2)\gamma^+ \gamma^- \slashed{k}\gamma^{\mu} ] (1+\gamma^5) \psi(p_1)
\left( 
g_{\mu\nu} + \frac{(k-p_1)_{\mu}n_{\nu} + (k-p_1)_{\nu}n_{\mu}}{(p_1-k)^{+}} 
\right)
\, , \nonumber
\end{align}
will be calculated term by term. To start with, notice that  $p_1^{-}$ and $p_2^{-}$ can be automatically neglected in the calculation 
as they vanish for Fourier transform of light-ray operators that we consider in this work. Let us start with the $g_{\mu\nu}$ piece and denote
its contraction with the strong of Dirac matrices in Eq.\ \re{Numerator} as $\mathcal{I}_1$. Then after Sudakov decomposition of all momenta 
and little Dirac algebra, we find after rescaling the $k^-$ momentum component according to Eq.\ \re{beta}
\begin{align}
\mathcal{I}_1
= 
g_{\mu\nu}\gamma^{\nu}(\slashed{k}-\slashed{p}{}_1-\slashed{p}{}_2)\gamma^{+}\gamma^{-}\slashed{k}\gamma^{\mu} 
\simeq 
4k_\perp^2 [\beta (k^{+}-p_1^{+}-p_2^{+})-1] 
\end{align}
where we have neglected all terms that do not produce any divergences, i.e., terms scaling as $k_{\perp}^n$ with $n<2$. Next, we turn to the second 
$(k-p_1)_{\mu}n_{\nu}$ and third $(k-p_1)_{\nu}n_{\mu}$ terms. For their contraction with the scare bracket, we find in a completely analogous manner
\begin{align}
\mathcal{I}_2
&= \gamma_{\nu}(\slashed{k}-\slashed{p}{}_1-\slashed{p}{}_2)\gamma^{+}\gamma^{-}\slashed{k}\gamma_{\mu}(k-p_1)^{\mu}n^{\nu} 
\nonumber\\
&\simeq 2(k^+-p_1^+-p_2^+)[2k^-(k^+-p_1^+)-k^2_{\perp}]\gamma^+\gamma^--2\beta k_\perp^2 (k^+-p_1^+-p_2^+)\gamma^+\slashed{p}{}_1^{\perp}
\, , \\
\mathcal{I}_3
& = \gamma_{\nu}(\slashed{k}-\slashed{p}{}_1-\slashed{p}{}_2)\gamma^{+}\gamma^{-}\slashed{k}\gamma_{\mu}(k-p_1)^{\nu}n^{\mu}
\nonumber\\
&\simeq 
4 k_{\perp} \cdot( p_{1\perp} + p_{2\perp})\slashed{k}_{\perp}\gamma^+
-
2 k^2_{\perp }\slashed{p}{}_2 \gamma^+
-
2 k^2_{\perp}[\beta(k^+-p_1^+-p_2^+)-1]\gamma^+\slashed{k}_{\perp}
\, .
\end{align}
Now, combining the above in the integrand, we trace only terms with $1/k_\perp^2$ behavior since these are the only contributions yielding logarithmic
divergence. Integrating over the longitudinal $k^+$ component with the help of the Dirac delta function in Eq.\ \re{Gintegrand}
\begin{align}
\mathcal{I}_1 \mathcal{A}
& = \frac{1}{8 \pi^2}\int_{0}^{2\pi} \frac{d \varphi}{2 \pi} \int \frac{d\beta}{2\pi}\int^{\mu^2} d k^2_\perp \, k^2_{\perp}  
\frac{4k^2_{\perp}[\beta(x_1-p_1^+-p_2^+)-1]}{k^6_{\perp}[\beta x_1 -1][\beta (x_1-p_1^+)-1][\beta(x_1-p_1^+-p_2^+)-1]}
\nonumber\\
&= \frac{1}{\pi^2} \ln \mu \int \frac{d\beta}{2\pi}\frac {1}{[\beta x_1-1][\beta(x_1-p_1^+)-1]}
=
\frac{i}{\pi^2} \ln \mu \, \vartheta^0_{11}(x_1,x_1-p_1^+)
\, ,
\end{align}
where in the last step we restored the omitted causal $i 0$ prescription in the longitudinal denominators use the defining integral Eq.\ \re{Generalizedstep} for the 
generalized  step functions. Similarly, we find
\begin{align}
\mathcal{I}_2 \mathcal{A}
&= 
\frac{i}{2 \pi^2} \ln \mu \frac{x_1-p_1^+-p_2^+}{p_1^+-x_1}\gamma^+\gamma^-\vartheta^0_{11}(x_1,x_1-p_1^+-p_2^+)
\\
&-
\frac{i}{2\pi^2} \ln \mu
\frac{\gamma^+\slashed{p}{}_{2 \perp}}{p_1^+-x_1}
[\vartheta^0_{11}(x_1,x_1-p_1^+)
+
\vartheta^0_{111}(x_1,x_1-p_1^+,x_1-p_1^+-p_2^+)
]
\, . \nonumber
\end{align}
To proceed further with other contributions, we compute the following integral first
\begin{align}
\int^{2\pi}_0 \frac{d\varphi}{2\pi} \int^{\mu^2} \frac{dk_{\perp}^2}{k_{\perp}^4}  p_{\perp}\cdot k_{\perp} k_{\perp}^\alpha
=
\ln \mu \, p_\perp^\alpha
\, .
\end{align}
Here we used the fact that the integrand does not have any vectors but $k_\perp$ so that one can immediately calculate the average in the
two-dimensional transverse plane $k_\perp^\alpha k_\perp^\beta \to k_\perp^2 \delta^{\alpha \beta}/2$.
Thus we obtain
\begin{align}
\mathcal{I}_3\mathcal{A} 
&
= - \frac{i}{2 \pi^2} \ln \mu
\frac{ \gamma^+\slashed{p}{}_{1\perp}+p_2^+\gamma^-\gamma^+ }{p_1^+ - x_1} \vartheta^0_{111}(x_1,x_1-p_1^+,x_1-p_1^+-p_2^+)
\, \\
\mathcal{I}_3 \mathcal{B}
&=  \frac{i}{2 \pi^2} \ln \mu
\frac{\gamma^+(\slashed{p}{}_1^{\perp}+\slashed{p}{}_2^{\perp})}{p_1^+-x_1}\vartheta^0_{111}(x_1,x_1-p_1^+,x_1-p_1^+-p_2^+)
\, \\
\mathcal{I}_3 \mathcal{C}
&= \frac{i}{2 \pi^2} \ln \mu
\frac{\gamma^+\slashed{p}{}_1^{\perp}}{p_1^+-x_1}\vartheta^0_{12}(x_1,x_1-p_1^+)
\, .
\end{align}
Putting all the pieces together, we get
\begin{align}
\mathcal{G}&
= \frac{\alpha_s}{\pi} t^a \otimes t^a \ln \mu  \int dy_1dy_2\int \frac{dp_1^-d^2p_{1\perp}}{(2\pi)^4}\int \frac{dp_1^-d^2 p_{2\perp}}{(2\pi)^4} \delta(x_1+x_2-y_1-y_2)
\nonumber\\
&\times\bar{\psi}(p_1)
\Bigg\{
2 \vartheta^0_{11}(x_1,x_1-y_1) - \frac{2y_2}{y_1-x_1}\vartheta^0_{111}(x_1,x_1-y_1,-x_2)
\nonumber\\
&\qquad\qquad
+
\gamma^+\gamma^-
\bigg[
\frac{y_2}{y_1-x_1}\vartheta^0_{111}(x_1,x_1-y_1,-x_2)
-
\frac{x_2}{y_1-x_1}\vartheta^0_{11}(x_1,x_1-y_1-y_2)
\bigg]
\nonumber\\
&\qquad\qquad
+
\frac{\gamma^+\slashed{p}{}_1^{\perp}}{y_1-x_1}[\vartheta^0_{12}(x_1,x_1-y_1)-\vartheta^0_{11}(x_1,x_1-y_1)-\vartheta^0_{111}(x_1,x_1-y_1,-x_2)]
\nonumber\\
&\qquad\qquad
+
\frac{\gamma^+\slashed{p}{}_2^{\perp}}{y_1-x_1}\vartheta^0_{111}(x_1,x_1-y_1,-x_2)
\Bigg\} \frac{\sqrt{2}(1 + \gamma_5)}{4} \psi(p_2)
\, .
\end{align}
Finally, using equations of motion for the (anti)quark fields, with neglected gluon field since we after the two-to-two transitions only,
$(p_2^+\gamma^-+\slashed{p}{}_{2 \perp}) \psi(p_2)=0$ and $\bar{\psi}(p_1)(p_1^+\gamma^-+\slashed{p}{}_{1 \perp})=0$, we can 
trade transverse momenta accompanying the good component of the quark to the bad quark fields. This way we arrive at Eqs.\ \re{528}-\re{531}.

\section{Flavor singlet $2\rightarrow 2$ transitions}
\label{appA}

%%%%%%%%%%%%%%%%%%%%%%%%%%%%%%%%%%%%%%%%%%%%%%%%%%%%%%%%%%%%%%%%%%%%%%%%%%
%                                                                                                                   FIGURE                                                                                                                                      %
%%%%%%%%%%%%%%%%%%%%%%%%%%%%%%%%%%%%%%%%%%%%%%%%%%%%%%%%%%%%%%%%%%%%%%%%%%
 \begin{figure}[t]
    \centering
         \psfrag{i}[bc][bc]{\scriptsize$i_1$}
	\psfrag{j}[bc][bc]{\scriptsize$i_2$}	
	\psfrag{i'}[tc][tc]{\scriptsize$i'_1$}	
	\psfrag{j'}[tc][tc]{\scriptsize$i'_2$}
	\psfrag{a}[tc][tc]{\scriptsize$a$}
	\psfrag{b}[tc][tc]{\scriptsize$b$}
	\psfrag{k1}[cl][cl]{\scriptsize$k_1$}
	\psfrag{k2}[cl][cl]{\scriptsize$k_2$}
	\psfrag{p1}[cl][cl]{\scriptsize$p_1$}
	\psfrag{p2}[cl][cl]{\scriptsize$p_2$}
	\psfrag{g1}[cc][cl]{\scriptsize$C_4;\tilde{C_4}$}
	\psfrag{g2}[cc[cc]{\scriptsize$C_5-C_6$}
	\psfrag{g3}[cc][cl]{\scriptsize$C_5$}
	\psfrag{g4}[cc][cl]{\scriptsize$C_6$}
  	 \makebox[\textwidth]{\includegraphics[scale=0.74]{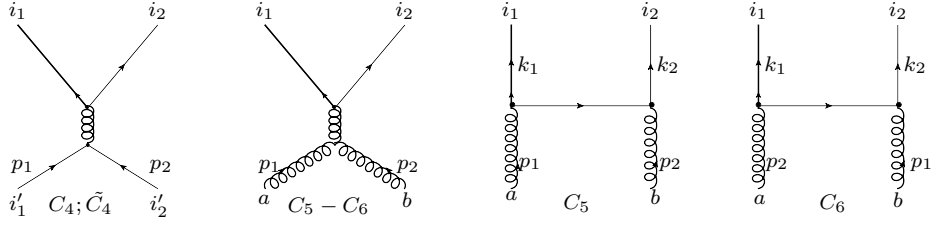}}
	  \caption{Two-to-two quark-gluon transitions in Eq.\ \re{57}. The  color structures $C_c$ are defined in Eq. \re{2to2color}.}
	  \label{fig:2to2quarkgluontrans}
\end{figure} 

We complement the non-singlet analysis performed in the body of the paper with particle results involving the singlet sector. In all cases we found
agreement with corresponding expressions reported in Ref.\ \cite{Braun:2009vc}. 

\subsection{Quasi-partonic operators}

To start with, we present the quasipartonic quark-antiquark to gluon-gluon kernels and gluon-gluon transitions as well.

\subsubsection{
$\mathcal{O}^{i_1 i_2}(x_1, x_2)
=
\{
\psi_{+}^{i_1} \bar{\chi}_{+}^{i_2 }
,\,
\bar{\psi}_+^{i_1} \chi_+^{i_2}
,\,
\psi_+^{i_1} \bar{\psi}_+^{i_2}
,\,
\bar{\chi}_+^{i_1}\chi_+^{i_2}
\} (x_1, x_2)
$
}

In the singlet sector, the quark-antiquark evolution will produce extra annihilation-type contributions shown by the first two graph in Fig.\ \ref{fig:2to2quarkgluontrans}
\begin{align}
\label{57}
[\mathcal{K\,O}]^{i_1i_2}(x_1,x_2)&=...-\int[\mathcal{D}^2y]_2K_1(x_1,x_2|y_1,y_2)\sum_{f}\Big\{[C_4]^{i_1i_2}_{i'_1i'_2}\bar{\psi}_+^{i'_1f}(y_1)\psi_+^{i'_2f}(y_2)
\nonumber
\\
&\qquad\qquad+[\tilde{C}_4]^{i_1i_2}_{i'_1i'_2}\chi^{i'_1f}_+(y_1)\bar{\chi}^{i'_2f}_+(y_2)\Big\}
\nonumber
\\
&-i\int [\mathcal{D}y^2]_2\Big\{ [C_5]^{i_1i_2}_{ab}K_2+[C_6]^{i_1i_2}_{ab}K_3\Big\}(x_1,x_2|y_1,y_2)f^a_{++}(y_1)\bar{f}^b_{++}(y_2)
\, ,
\end{align}
in addition to already computed transitions, denoted above by ellipses, and given in Eqs.\ \re{55} and\ \re{56}. The index $f$ runs over all quark flavors. The last line
represents transitions into gluons, exhibited by the last two graphs in Fig.\ \re{fig:2to2quarkgluontrans}. The color structures are displayed  in Eq.\ \re{2to2color}. The 
transition kernels then read
\begin{align}
K_1(x_1,x_2;y_1,y_2)&=\frac{4x_1x_2\vartheta^0_{11}(x_1,-x_2)}{(x_1+x_2)^2}\, ,
\\
K_2(x_1,x_2;y_1,y_2)&=\frac{2x_2}{y_1y_2}\bigg\{\frac{x_1(y_2-y_1)}{(x_1+x_2)^2}\vartheta^0_{11}(x_1,-x_2)
- \vartheta^0_{111}(x_1,x_1-y_1,-x_2)-\vartheta^0_{11}(x_1-y_1,-x_2)\bigg\}\, ,
\\
K_3(x_1,x_2;y_1,y_2)&=2\frac{x_1-y_2}{y_1y_2}\vartheta^0_{111}(x_1,x_1-y_2,-x_2)+\frac{2x_1x_2(y_1-y_2)}{y_1y_2(x_1+x_2)^2}\vartheta^0_{11}(x_1,-x_2)\, .
\end{align}

%%%%%%%%%%%%%%%%%%%%%%%%%%%%%%%%%%%%%%%%%%%%%%%%%%%%%%%%%%%%%%%%%%%%%%%%%%
%                                                                                                                   FIGURE                                                                                                                                      %
%%%%%%%%%%%%%%%%%%%%%%%%%%%%%%%%%%%%%%%%%%%%%%%%%%%%%%%%%%%%%%%%%%%%%%%%%%
 \begin{figure}[t]
    \centering
         \psfrag{i}[bc][bc]{\scriptsize$i$}
	\psfrag{i'}[tc][tc]{\scriptsize$i'$}
	\psfrag{a}[bc][bc]{\scriptsize$a$}
	\psfrag{b}[bc][bc]{\scriptsize$b$}
	\psfrag{a'}[tc][tc]{\scriptsize$a'$}
	\psfrag{b'}[tc][tc]{\scriptsize$b'$}
	\psfrag{k1}[cl][cl]{\scriptsize$k_1$}
	\psfrag{k2}[cl][cl]{\scriptsize$k_2$}
	\psfrag{p1}[cl][cl]{\scriptsize$p_1$}
	\psfrag{p2}[cl][cl]{\scriptsize$p_2$}
	\psfrag{g1}[cc][cl]{\scriptsize$C_7$}
	\psfrag{g2}[cc][cl]{\scriptsize$C_8$}
	\psfrag{g3}[cc][cl]{\scriptsize$C_7;\,C_8;\, C_7+C_8$}
  	 \makebox[\textwidth]{\includegraphics[scale=0.54]{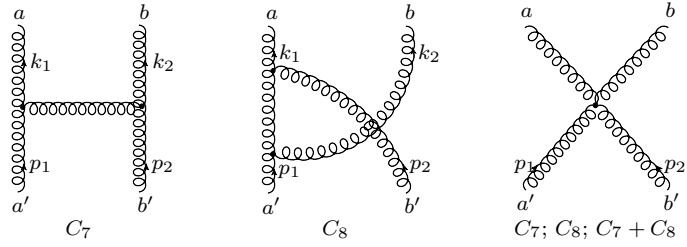}}
	  \caption{Two-to-two quasipartonic gluon-gluon transition of Sect.\ \ref{quasigluon1}. The induced color-flow structure is defined in Eq.\ \re{2to2color}.}
	  \label{fig:2to2gluongluona}
\end{figure} 

In the following two subsection, we will list our results of the evolution kernels for the pure gluonic transitions. 

\subsubsection{$\mathcal{O}^{ab}(x_1,x_2)=\{ f^a_{++}f^b_{++},\, \bar{f}^a_{++}\bar{f}^b_{++}\}(x_1,x_2) $}
\label{quasigluon1}

For gluon blocks of the same chirality, the nonvanshing Feynman graphs that induce the transition
\begin{align}
[\mathcal{K\,O}]^{ab}(x_1,x_2)&=\int (\mathcal{D}^2y)_2\{[C_7]^{ab}_{a'b'}K_1
+
[C_8]^{ab}_{a'b'}K_2\}(x_1,x_2|y_1,y_2)\mathcal{O}^{a'b'}(y_2,y_2)\, ,
\end{align}
are given in Fig.\ \ref{fig:2to2gluongluona} and produce
\begin{align}
K_1(x_1,x_2;y_1,y_2)&=\frac{x_1^3+x_1^2(2x_2-y_1+y_2)-x_2y_1(x_1+2x_2)}{(x_1-y_1)y_1y_2}\vartheta^0_{111}(x_1,x_1-y_1,-x_2) 
\nonumber
\\
&+\frac{x_1x_2(x_1+y_1)(x_2+y_2)}{(x_1-y_1)y_1y_2}\vartheta^1_{111}(x_1,x_1-y_1,-x_2)+\frac{x_1x_2}{y_1y_2}\vartheta^0_{11}(x_1,-x_2) \, ,
\\
K_2(x_1,x_2;y_1,y_2)&=\frac{x_1^3+x_1^2(2x_2-y_2+y_1)-x_2y_2(x_1+2x_2)}{(x_1-y_2)y_1y_2}\vartheta^0_{111}(x_1,x_1-y_2,-x_2) 
\nonumber
\\
&+\frac{x_1x_2(x_1+y_2)(x_2+y_1)}{(x_1-y_2)y_1y_2}\vartheta^1_{111}(x_1,x_1-y_2,-x_2)+\frac{x_1x_2}{y_1y_2}\vartheta^0_{11}(x_1,-x_2) \, .
\end{align}
Here we again observe the ``exchange symmetry" elaborated in details in Sects.\ \ref{sec62} and \ref{sec63}. In the present case, it implies the 
simultaneous interchange of $a\leftrightarrow b$ and $z_1\leftrightarrow z_2$.

\subsubsection{$\mathcal{O}^{ab}(x_1,x_1)= \{f^a_{++}\bar{f}^b_{++}\}(x_1,x_2)$}
\label{quasigluon2}

\begin{figure}[t]
    \centering
         \psfrag{i}[bc][bc]{\scriptsize$i$} 
	\psfrag{i'}[tc][tc]{\scriptsize$i'$}
	\psfrag{a}[bc][bc]{\scriptsize$a$}
	\psfrag{b}[bc][bc]{\scriptsize$b$}
	\psfrag{a'}[tc][tc]{\scriptsize$a'$}
	\psfrag{b'}[tc][tc]{\scriptsize$b'$}
	\psfrag{k1}[cl][cl]{\scriptsize$k_1$}
	\psfrag{k2}[cl][cl]{\scriptsize$k_2$}
	\psfrag{p1}[cl][cl]{\scriptsize$p_1$}
	\psfrag{p2}[cl][cl]{\scriptsize$p_2$}
	\psfrag{g1}[cc][cl]{\scriptsize$C_7$}
	\psfrag{g2}[cc][cl]{\scriptsize$C_8$}
	\psfrag{g3}[cc][cl]{\scriptsize$-(C_7+C_8)$}
	\psfrag{g4}[cc][cl]{\scriptsize$C_7;\, C_8;\, C_7+C_8$}
  	 \makebox[\textwidth]{\includegraphics[scale=0.74]{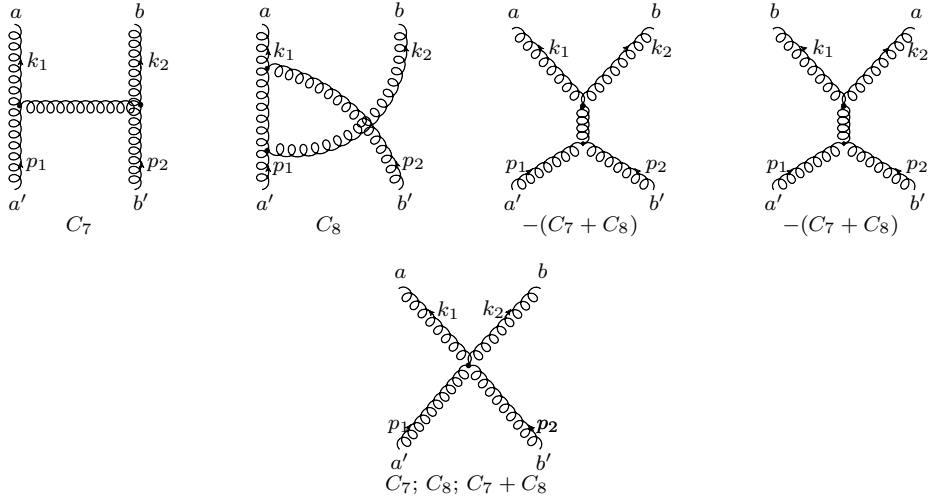}}
	  \caption{Two-to-two transition of quasipartonic gluon-gluon fields in sect.\ \ref{quasigluon2} where the color structures are define in Eq.\ \re{2to2color}.}
	  \label{fig:2to2gluongluonb}
\end{figure}

Finally, the opposite-chirality gluon sector evolves as
\begin{align}
[\mathcal{K\,O}]^{ab}(x_1,x_2)&=\int [\mathcal{D}^2y]_2
\{
[C_7]^{ab}_{a'b'} K_1
+
[C_8]^{ab}_{a'b'}K_2\}(x_1,x_2|y_1,y_2)\mathcal{O}^{a'b'}(y_2,y_2)\, ,
\end{align}
according to nontrivial Feynman diagrams in Fig.\ \ref{fig:2to2gluongluonb} with
\begin{align}
K_1(x_1,x_2;y_1,y_2)&=\frac{x_1^2x_2+x_1(x_2-2y_1)y_1+2(x_2-y_1)^2y_1}{(y_1-x_1)y_1y_2}\vartheta^0_{111}(x_1,x_1-y_1,-x_2) 
\nonumber
\\
&+\frac{x_1x_2(x_1+2x_2-y_1)(x_1+y_1)}{(x_1-y_1)y_1y_2}\vartheta^1_{111}(x_1,x_1-y_1,-x_2) 
\nonumber
\\
&-\frac{x_1x_2(x_1^2+x_2(3x_2-2y_1)+2x_1(2x_2+y_1))}{(x_1+x_2)^2y_1y_2}\vartheta^0_{11}(x_1,-x_2)\, ,
\\
K_2(x_1,y_1;y_1,y_2)&=\frac{2(x_1-y_2)^2}{y_1y_2}\vartheta^0_{111}(x_1,x_1-y_2,-x_2)
\nonumber
\\
&+\frac{2x_1x_2(x_2(x_2-y_1)+x_1(x_2+y_1))}{(x_1+x_2)^2y_1y_2}\vartheta^0_{11}(x_1,-x_2)\, .
\end{align}

%%%%%%%%%%%%%%%%%%%%%%%%%%%%%%%%%%%%%%%%%%%%%%%%%%%%%%%%%%%%%%%%%%%%%%%%%%
%                                                                                                                   FIGURE                                                                                                                                      %
%%%%%%%%%%%%%%%%%%%%%%%%%%%%%%%%%%%%%%%%%%%%%%%%%%%%%%%%%%%%%%%%%%%%%%%%%%
All other quasipartonic singlet transitions can be found in the literature \cite{BFLK,Bukhvostov:1983te,BelExact98,Braun:2009vc}

\subsection{Non-Quasipartonic operators}
\label{appnonquasi}

In this Appendix, we complement  non-quasipartonic operators with purely gluonic transitions, thus extending the consideration of Sect.\ \ref{2to2nonquasi}.

\subsubsection{Gluon-gluon transitions of same chiralities}
\label{nonquasigg1}

Extending the class of non-singlet operators Eq.\ \re{524} to gluons, we introduce two doublets of gluonic blocks,
\begin{align}
\label{Oabp}
\bit{\mathcal{O}}_+^{ab}
=\Bigg\{
\begin{pmatrix}
f_{+-}^a\otimes f_{++}^b
\\
f_{++}^a\otimes f_{+-}^b
\end{pmatrix}
\Bigg\},
\qquad
\bit{\mathcal{O}}_-^{ab}
=\Bigg\{
\begin{pmatrix}
f_{++}^a\otimes \bar{D}_{-+}f_{++}^b
\\
\bar{D}_{-+}f_{++}^a\otimes f_{++}^b
\end{pmatrix}
\Bigg\}\, .
 \end{align}
Then the transition equation can be written as in the quasipartonic case
 \begin{align}
[\mathcal{K}\,\bit{\mathcal{O}}_+]^{ab}(x_1,x_2)=-[C_7]^{ab}_{a'b'}\int(\mathcal{D}^2y)_2\bit{K}(x_1,x_2|y_1,y_2)\bit{\mathcal{O}}_+^{a'b'}(y_1,y_2),
\end{align}
though the kernels are now matrix valued and obviously have  different components
 \begin{align}
 K_{11}&=\frac{x_1x_2 \left(y_1^2-2 y_1 y_2-2 y_2^2\right) \vartheta^0_{11} (x_1,-x_2)}{ y_1 y_2 (y_1+y_2) (x_2-y_2)}
 \nonumber
 \\
 &+\frac{x_1 \left(2 x_1 y_2 (y_1+y_2)+y_1^2 x_2\right) \vartheta^0_{11} (x_1,x_1-y_1)}{
   y_2 (y_1+y_2)^2 (y_2-x_2)}
\nonumber
\\
 &-\frac{x_2 \left(2 y_1 (y_1+y_2)^3-x_1 \left(2 y_1^3+5 y_1^2 y_2-2 y_2^3\right)\right) \vartheta^0_{11} (x_1-y_1,-x_2)}{ y_1 y_2 (y_1+y_2)^2 (y_2-x_2)}\, ,
 \\
 K_{12}&= \frac{x_1 \left(x_2y_1^2+2x_1y_2(y_1+y_2)\right) \vartheta^0_{11} (x_1,x_1-y_1)}{y_1 (x_1-y_1) (y_1+y_2)^2}
 \nonumber
 \\
 &-\frac{x_1 x_2 (3 y_1+2 y_2) \vartheta^0_{11} (x_1,-x_2)}{ y_1 (x_1-y_1)
   (y_1+y_2)}+\frac{x_1 y_2 x_2 (3 y_1+2 y_2) \vartheta^0_{11} (x_1-y_1,-x_2)}{y_1 (x_1-y_1) (y_1+y_2)^2}\, ,
   \\
   K_{21}&= \frac{x_1 x_2 y_1(2 y_1+3 y_2) \vartheta^0_{11} (x_1,x_1-y_1)}{ y_2 (y_1+y_2)^2 (x_2-y_2)}-\frac{x_1x_2 (2 y_1+3 y_2) \vartheta^0_{11} (x_1,-x_2)}{ y_2 (y_1+y_2) (x_2-y_2)}
   \nonumber
   \\
   &+\frac{x_2
   \left(2 y_1^2 x_2+y_2^2 (y_1-x_2)+2 y_1 y_2 x_2+y_2^3\right) \vartheta^0_{11} (x_1-y_1,-x_2)}{ y_2 (y_1+y_2)^2 (x_2-y_2)}\, ,
   \\
  K_{22}&=   \frac{x_1 \left(x_2 \left(2 y_1^3-5 y_1 y_2^2-2 y_2^3\right)+2 y_2 (y_1+y_2)^3\right) \vartheta^0_{11} (x_1,x_1-y_1)}{ y_1 y_2 (y_1+y_2)^2 (y_2-x_2)}
  \nonumber
  \\
  &+\frac{x_1 x_2 \left(2 y_1(y_1+y_2)-y_2^2\right) \vartheta^0_{11} (x_1,-x_2)}{ y_1 y_2 (y_1+y_2) (x_2-y_2)}
   \nonumber
   \\
   &+\frac{x_2 \left(2 y_1^2 x_2+y_2^2 (y_1-x_2)+2 y_1 y_2 x_2+y_2^3\right) \vartheta^0_{11} (x_1-y_1,-x_2)}{y_1 (y_1+y_2)^2 (x_2-y_2)}\, .
 \end{align}
For $\bit{\mathcal{O}}_-^{ab}$ operator set, we find  
\begin{align}
[\mathcal{K}\,\bit{\mathcal{O}}_-]^{ab}(x_1,x_2)=-[C_7]^{ab}_{a'b'}\int(\mathcal{D}^2y)_2\bit{K}(x_1,x_2|y_1,y_2)\bit{\mathcal{O}}_-^{a'b'}(y_1,y_2),
\end{align}
where
 \begin{align}
 K_{11}&= \frac{x_2^2 \left(2 y_1^3 x_2+2 y_1^2 y_2 x_2+y_2^3 (y_1-2 x_2)+y_2^2 x_2 (x_2-y_1)+y_2^4\right) \vartheta^0_{11} (x_1,-x_2)}{y_1 y_2^3 (y_1+y_2)
   (x_2-y_2)}
   \nonumber
   \\
   &+\frac{x_2^2 \left(-x_2 \left(2 y_1^2+3 y_1 y_2+2 y_2^2\right)+y_2^2 (y_1+y_2)+y_2 x_2^2\right) \vartheta^0_{11} (x_1-y_1,-x_2)}{ y_1 y_2 (y_1+y_2)^2 (y_2-x_2)}
   \nonumber
   \\
   &+\frac{1}{y_1 y_2^3 (y_1+y_2)^2 (y_2-x_2)}\Big\{y_1 x_2^3 \left(2 y_1^3+2 y_1^2 y_2-y_1 y_2^2-2 y_2^3\right)+y_2^3 x_2^2 (3 y_1+2 y_2) (y_1+y_2)
   \nonumber
   \\
   &-4 y_2^3 x_2 (y_1+y_2)^3+2 y_2^3 (y_1+y_2)^4+y_1 y_2^2 x_2^4\Big\} \vartheta^0_{11}(x_1,x_1-y_1)\, ,
   \\
   K_{12}&= \frac{\left(x_1^2 y_2+x_1 y_1 (2 y_1+y_2)-2 y_1 (y_1+y_2)^2\right) \vartheta^0_{211} (x_1,x_1-y_1,-x_2)}{ y_1 y_2 (y_2-x_2)}-\frac{x_1x_2^2 \vartheta^0_{11} (x_1,-x_2)}{ y_1 y_2
   (y_1+y_2)}
   \nonumber
   \\
   &+\frac{x_1 (x_1+y_1) (y_2+x_2) \vartheta^0_{21} (x_1,x_1-y_1)}{ y_1 y_2 (x_2-y_2)}+\frac{2x_2^2 \vartheta^0_{111} (x_1,x_1-y_1,-x_2)}{y_2(x_2-y_2)}\, ,
 \\
 %%%%%%%%%%%
 K_{21}&=\frac{2x_1^2 \vartheta^0_{111} (x_1,x_1-y_1,-x_2)}{y_1 (y_2-x_2)}+\frac{x_2
   (x_1+y_1) (y_2+x_2) \vartheta^0_{12} (x_1-y_1,-x_2)}{ y_1 y_2 (y_2-x_2)}
 \nonumber
 \\
 &-\frac{\left(y_2 x_2 (y_1+2 y_2)-2 y_2 (y_1+y_2)^2+y_1 x_2^2\right) \vartheta^0_{112} (x_1,x_1-y_1,-x_2)}{y_1 y_2 (y_2-x_2)}-\frac{x_1^2 x_2 \vartheta (x_1,-x_2)}{ y_1 y_2 (y_1+y_2)}\, ,
 \\
 %%%%%%%%%%%%%%%%%%
 K_{22}&=\frac{x_1^2 \left(2 y_2 \left(y_2 (x_1+y_1)+y_1^2\right)-y_1 y_2 x_2-y_1 x_2^2\right) \vartheta^0_{11} (x_1,y_2-x_2)}{ y_1 y_2 (y_1+y_2)^2 (y_2-x_2)}
 \nonumber
 \\
 &+\frac{x_1^2 \left(2 x_1
   y_2^2 (y_1+y_2)-y_1^2 y_2 x_2+y_1^2 x_2^2\right) \vartheta^0_{11}(x_1,-x_2)}{y_1^3 y_2 (y_1+y_2) (y_2-x_2)}
   \nonumber
   \\
   &-\frac{1}{y_1^3 y_2 (y_1+y_2)^2 (y_2-x_2)}\Big\{x_1^4 y_1^2 y_2-x_1^3 y_2 \left(2 y_1^3+y_1^2 y_2-2 y_1 y_2^2-2
   y_2^3\right)
   \nonumber
   \\
   &\quad+x_1^2 y_1^3 (y_1+y_2) (2 y_1+3 y_2)-2 y_1^3 (y_1+y_2)^3 (x_1- x_2)\Big\}\vartheta^0_{11} (x_1-y_1,-x_2) \, .
 \end{align}
 Notice that the graphs defining these transitions are the same as the quasipartonic case.
 
\subsubsection{Gluon-gluon transitions of opposite chiralities}
\label{nonquasigg2}

For opposite-helicity gluon operators, we introduce
\begin{align}
\label{FGab}
\bit{\mathcal{O}}^{ab}
=
\begin{pmatrix}
f_{+-}^a\otimes \bar{f}_{++}^b
\\
f_{++}^a\otimes \tfrac{1}{2}D_{-+}\bar{f}_{+-}^b
\, ,
\end{pmatrix}
 \end{align}
and focus on the mixing within the group $\mathcal{O}^{ab}$ and disregard transitions into singlet quark opertaors. The transition then reads
\begin{align}
[ \mathcal{K} \, \bit{\mathcal{O}} ]^{ab}(x_1,x_2)
=
-
\int [\mathcal{D}^2y]_2
\Bigg\{ [C_7]^{ab}_{a'b'} \bit{K} (x_1, x_2 | y_1, y_2) 
+
[C_8]^{ab}_{a'b'} \widetilde{\bit{K}} (x_1, x_2 | y_1, y_2) 
\Bigg\}
\bit{\mathcal{O}}^{a'b'}(y_1,y_2)
\, .
\end{align}  
with the matrix elements being
\begin{align}
 K_{11}&=\frac{x_1}{y_1 y_2
   (y_1+y_2)^3 (y_2-x_2)} \Big\{2 x_2^3 \left(y_1^2-y_1 y_2+y_2^2\right)-3 y_1 x_2^2 \left(y_1^2+2 y_1 y_2-y_2^2\right)
   \nonumber
   \\
   &\quad+4 y_1 y_2 x_2 (y_1+y_2)^2+y_1 y_2 (y_1+y_2)^3\Big\} \vartheta^0_{11} (x_1,-x_2)
   \nonumber
   \\
   &-\frac{1}{ y_1 y_2 (y_1+y_2)^2 (y_2-x_2)}\Big\{x_2^3 \left(y_1
   y_2-2 y_1^2+2 y_2^2\right)+y_1 x_2^2 \left(2 y_1^2+3 y_1 y_2-y_2^2\right)
   \nonumber
   \\
   &\quad+3 y_1 y_2^2 x_2 (y_1+y_2)+y_1 y_2^2 (y_1+y_2)^2-2 y_2 x_2^4\Big\} \vartheta^0_{11} (x_1-y_1,-x_2)
   \nonumber
   \\
   &+\frac{x_1 \left(x_2^2 (3 y_1+2 y_2)-2 y_1 y_2 x_2+y_1 y_2 (y_1+y_2)-2 x_2^3\right) \vartheta^0_{11} (x_1,x_1-y_1)}{ y_2 (y_1+y_2)^2 (y_2-x_2)}\, ,
 \\
   K_{12}&=\frac{\left((y_1+y_2)^2-2 y_2 x_2+x_2^2\right) \vartheta^0_{11} (x_1,-x_2)}{ (y_1+y_2) (y_2-x_2)}
   \nonumber
   \\
   &-\frac{x_1 \left(x_2^2 (3 y_1+2 y_2)-2 y_1 y_2 x_2+y_1 y_2 (y_1+y_2)-2 x_2^3\right) \vartheta^0_{11}
   (x_1,x_1-y_1)}{ y_2 (y_1+y_2)^2 (y_2-x_2)}
   \nonumber
   \\
   &-\frac{x_1 \left(y_2^3 (y_1+y_2)+2 x_2^3 (y_1+2 y_2)-y_2 x_2^2 (2 y_1+y_2)-2 y_2^3 x_2\right) \vartheta^0_{11} (x_1-y_1,-x_2)}{ y_2^2 (y_1+y_2)^2
   (y_2-x_2)}\, ,
  \\
   K_{21}&= \frac{x_1^2 \left(-2 y_1^2 y_2-x_2^2 (3 y_1+2 y_2)+y_1 y_2 x_2+2 x_2^3\right) \vartheta^0_{11} (x_1,x_1-y_1)}{y_2 (y_1+y_2)^2 (y_2-x_2)}
   \nonumber
   \\
   &+\frac{x_1^2}{y_1 y_2 (y_1+y_2)^3
   (y_2-x_2)} \Big\{x_2^2 \left(3 y_1^3+12 y_1^2 y_2+5 y_1 y_2^2+2 y_2^3\right)-2 x_2^3 \left(y_1^2-y_1
   y_2+y_2^2\right)
   \nonumber
   \\
   &\quad+y_1 y_2 x_2 (y_1-3 y_2) (y_1+y_2)+2 y_1 y_2 (y_1-y_2) (y_1+y_2)^2\Big\} \vartheta^0_{11} (x_1,-x_2)
   \nonumber
   \\
   &+\frac{\left(2 x_1^3 x_2^2+x_1^2 y_1 \left(2 y_1 y_2-y_2 x_2+x_2^2\right)+2 y_1 y_2 (y_1+y_2)^2 (x_2-y_2)\right)\vartheta^0_{11} (x_1-y_1,-x_2) }{ y_1 (y_1+y_2)^2 (x_2-y_2)}\, ,
  \\
   K_{22}&=\frac{x_1^2 \left(2 x_2^3-2 y_1^2 y_2-x_2^2 (3 y_1+2 y_2)+y_1 y_2 x_2\right) \vartheta^0_{11} (x_1,y_2-x_2)}{ y_1 y_2 (y_1+y_2)^2 (y_2-x_2)}
   \nonumber
   \\
   &+\frac{x_1^2}{ y_1^2 y_2 (y_1+y_2)^3 (y_2-x_2)} \Big\{4 y_1^2 y_2 (y_1+y_2)^2+x_2^2 \left(3
   y_1^3+14 y_1^2 y_2+7 y_1 y_2^2+2 y_2^3\right)
   \nonumber
   \\
   &\quad-2 y_1 x_2^3 (y_1-2 y_2)+y_1 y_2 x_2 (3 y_1-y_2) (y_1+y_2)\Big\} \vartheta^0_{11} (x_1,-x_2)
   \nonumber
   \\
   &-\frac{1}{ y_1^2 y_2^2 (y_1+y_2)^2 (y_2-x_2)} \Big\{2 y_2^5 \left(x_1^2+y_1^2\right)+3 x_1^2 y_1 y_2^2 (5 x_1-7 y_1) (x_1-y_1)
   \nonumber
   \\
   &\quad-2 x_1^2 y_1^2 (x_1-y_1)^3-2 x_1^2 y_1 y_2 (2 x_1-5 y_1) (x_2-y_2)^2+2 y_2^4 \left(-2 x_1^3+5
   x_1^2 y_1+2 y_1^3\right)
   \nonumber
   \\
   &\quad+y_2^3 \left(2 x_1^4-21 x_1^3 y_1+23 x_1^2 y_1^2+2 y_1^4\right)\Big\}\vartheta^0_{11}
   (x_1-y_1,-x_2)\, ,
  \\
   \widetilde{K}_{11}&=\frac{2x_1 y_1 (x_2-y_1)^2 \vartheta^0_{11}
   (x_1-y_2,-x_2)}{y_2^2 (y_1+y_2)^2}+\frac{2x_1 (x_1-y_2)^2 \vartheta^0_{11} (x_1,x_1-y_2)}{y_2 (y_1+y_2)^2}
   \nonumber
   \\
   &-\frac{2x_1 \left(x_2^2 \left(y_1^2+2 y_1 y_2+4 y_2^2\right)+y_1^2 (y_1+y_2)^2-2 y_1 x_2 (y_1+y_2)^2\right) \vartheta^0_{11} (x_1,-x_2)}{y_2^2 (y_1+y_2)^3}
   \\
   \widetilde{K}_{12}&= \frac{2(x_2-y_1)^2 \left(y_1^2 (y_1+y_2)-x_2 \left(y_1^2+y_1 y_2+y_2^2\right)\right) \vartheta^0_{11} (x_1-y_2,-x_2)}{y_1 y_2^3 (y_1+y_2)^2}
   \nonumber
   \\
   &-\frac{2x_1}{y_1 y_2^3 (y_1+y_2)^3} \Big\{y_1^3 (y_1+y_2)^2+x_2^2 \left(y_1^3+3 y_1^2 y_2+4 y_1 y_2^2-y_2^3\right)
   \nonumber
   \\
   &\quad-y_1 x_2 (y_1+y_2)^2 (2 y_1+y_2)\Big\} \vartheta^0_{11} (x_1,-x_2)-\frac{2x_1 (x_1-y_2)^2
   \vartheta^0_{11} (x_1,x_1-y_2)}{y_1 y_2 (y_1+y_2)^2}\, ,
\\
   \widetilde{K}_{21}&=\frac{2x_1x_2^2\left((y_2^2-y_1^2)y_2-x_1(y_1^2-y_1y_2+y_2^2)\right)\vartheta^0_{11}(x_1,-x_2)}{y_1y_2(y_1+y_2)^3}
   \nonumber
   \\
   &-\frac{2(x_1-y_2)^2\left((x_1+y_1-y_2)\vartheta^0_{111}(x_1,x_1-y_2,-x_2)+y_1\vartheta^0_{112}(x_1,x_1-y_2,-x_2)\right)}{y_1y_2}\, ,
\\
   \widetilde{K}_{22}&=\frac{-2x_1^2}{y_1 y_2^3
   (y_1+y_2)^3} \Big\{y_1^2 (y_1-y_2) (y_1+y_2)^2-2 x_2 \left(y_1^4+2 y_1^3 y_2-y_1 y_2^3\right)
   \nonumber
   \\
   &\quad+x_2^2 \left(y_1^3+3 y_1^2 y_2+4 y_1 y_2^2-y_2^3\right)\Big\} \vartheta^0_{11} (x_1,-x_2)+\frac{2(x_2-y_1)^2}{y_1 y_2^3 (y_1+y_2)^2} \Big\{x_2^2 \left(y_1^2+y_1 y_2+y_2^2\right)
   \nonumber
   \\
   &\quad-2 y_1^2 x_2 (y_1+y_2)+y_1 (y_1-y_2)
   (y_1+y_2)^2\Big\} \vartheta^0_{11} (x_1-y_2,-x_2)
   \nonumber
   \\
   &-\frac{2x_1^2 (x_1-y_2)^2 \vartheta^0_{11} (x_1,x_1-y_2)}{y_1 y_2 (y_1+y_2)^2}\, .
 \end{align}
This concludes our discussion for the two-to-two gluonic transitions in the singlet sector. All the results presented here coincide with the ones 
given in Ref. \cite{Braun:2009vc}. 

\section{Fourier transform}
\label{FourierAppendix}

As an example, we start with a coordinates-space kernel
\begin{align}
\label{421}
[\mathcal{H}_{12}](z_1,z_2)=z_{12}^2\int^1_0d\alpha\int^1_{\bar{\alpha}}d\beta\,\frac{\bar{\alpha}\bar{\beta}}{\alpha}\,\varphi(z_{12}^{\alpha},z_2,z^{\beta}_{21})\, ,
\end{align}
where $z_{ij}^{\alpha}=\bar{\alpha}z_i+\alpha z_j=(1-\alpha)z_i+\alpha z_j$. Then the Fourier transform takes form
\begin{align}
\label{422}
K (x_1,x_2 | y_1,y_2,y_3)
&=
\partial_{x_1}^2\int^1_0d\alpha\int^1_{\bar{\alpha}}d\beta\,\frac{\bar{\alpha}\bar{\beta}}{\alpha}\,\delta[x_1-\bar\alpha y_1-\beta y_3]
\nonumber\\
&=\partial^2_{x_1}\int^1_0d\alpha\frac{1}{y_3}\bar{\alpha}\big(\bar{\alpha}y_1-x_1+y_3\big)\vartheta^0_{11}(x_1-\bar{\alpha}(y_1+y_3),x_1-y_3-\bar{\alpha}y_1)
\nonumber
\\
&=\frac{1}{y_3}\partial^2_{x_1}\bigg\{\bigg[(x_1-2y_1-y_3)\frac{y_1}{y_3}\bigg(\frac{x_1-y_1-y_3}{y_1+y_3}-\frac{x_1-y_1-y_3}{y_1}\bigg)
\nonumber
\\
&-\frac{y_1^2}{2y_3}\bigg(\bigg(\frac{x_1-y_1-y_3}{y_1+y_3}\bigg)^2-\bigg(\frac{x_1-y_1-y_3}{y_1}\bigg)^2\bigg)\bigg]\vartheta^0_{11}(x_1-y_3,x_1-y_1-y_3)
\nonumber
\\
&\qquad\,\,\,+\bigg[\frac{x_1(x_1-2y_1-y_3)}{y_1+y_3}+\frac{y_1}{2}\bigg(1-\bigg(\frac{x_1-y_1-y_3}{y_1+y_3}\bigg)^2\bigg)\bigg]\vartheta^0_{11}(x_1,x_1-y_3)
\nonumber
\\
&+\frac{x_1-y_1-y_3}{y_3^2}\bigg[\ln\bigg(\frac{y_1+y_3-x_1}{y_1}\bigg)\big(\theta(x_1-y_1-y_3)-\theta(x_1-y_3)\big)
\nonumber
\\
&\qquad\,\,\,-\ln\bigg(\frac{y_1+y_3-x_1}{y_1+y_3}\bigg)\big(\theta(x_1-y_1-y_3)-\theta(x_1)\big)\bigg]\bigg\}
\nonumber
\\
&=\frac{x_1(y_1+2y_3)-y_3(y_1+y_3)}{y_3^2(y_1+y_3)^2(x_1-y_1-y_3)}\theta(x_1)-\frac{(x_1-y_3)\theta(x_1-y_3)}{y_1y_3^2(x_1-y_1-y_3)}+\frac{\theta(x_1-y_1-y_3)}{y_1(y_1+y_3)^2}\, ,
\end{align}
where we have employed the results in Eq.\ \re{fouriertransform}. In the last step of differentiation, we have dropped all the terms proportional to the Dirac delta 
function and its derivatives since we focus on expressions away from the kinematical boundaries which are sufficient to confront against the light-ray results of
Ref.\ \cite{Braun:2009vc}. In principle, it is very straightforward to recover the contact terms as well. In this case, all the calculations follow through as in Eq.\ \re{422} 
until the last step of differentiation. Then one gets
\begin{align}
K&=\frac{x_1(y_1+2y_3)-y_3(y_1+y_3)}{y_3^2(y_1+y_3)^2(x_1-y_1-y_3)}\theta(x_1)-\frac{(x_1-y_3)\theta(x_1-y_3)}{y_1y_3^2(x_1-y_1-y_3)}+\frac{\theta(x_1-y_1-y_3)}{y_1(y_1+y_3)^2}
\nonumber
\\
&+\frac{2\Big[x_1(y_1+2y_3)+(y_1+y_3)^2\ln\big(\frac{x_2-y_2}{y_1+y_3}\big)\Big]\delta(x_1)}{y_3^2(y_1+y_3)^2}-\frac{2\Big[x_1-y_3+y_1\ln\big(\frac{x_2-y_2}{y_1}\big)\Big]\delta(x_1-y_3)}{y_1y_3^2}
\nonumber
\\
&-\frac{2\Big[y_3\big(y_1^2+(x_2+y_1-y_2)y_3\big)+y_1(y_1+y_3)^2\ln\big(\frac{y_1}{y_1y_3}\big)\Big]\delta(y_2-x_2)}{y_1y_3^2(y_1+y_3)^2}
\nonumber
\end{align}
\begin{align}
&+\frac{\Big[x_1(x_1(y_1+2y_3)-2(y_1+y_3)^2)-2(x_2-y_2)(y_1+y_3)^2\ln\big(\frac{x_2-y_2}{y_1+y_3}\big)\Big]\delta'(x_1)}{2y_3^2(y_1+y_3)^2}
\nonumber
\\
&-\frac{\Big[x_2^2-y_1^2-2x_2y_2+y_2^2-2y_1(x_2-y_2)\ln\big(\frac{x_2-y_2}{y_1}\big)\Big]\delta'(x_1-y_3)}{2y_1y_3^2}
\nonumber
\\
&+\frac{(x_2-y_2)\Big[y_3(2y_1^2+(x_2+2y_1-y_2)y_3)+2y_1(y_1+y_3)^2\ln\big(\frac{y_1}{y_1+y_3}\big)\Big]\delta'(y_2-x_2)}{2y_1y_3^2(y_1+y_3)^2}\, ,
\end{align}
where we  invoked the momentum conservation condition $x_1+x_2=y_1+y_2+y_3$ to simplify the expression. The extra delta-function terms can be recovered
within the momentum formalism as well by properly taking into account field renormalization as discussed in Sect.\ \ref{1LoopSection}.

\section{Equation-of-motion graphs}
\label{EOMAppendix}

In the preceding Appendix \ref{samplecal} illustrating the good-bad two-to-two transitions, we already had to rely on the use of equations of motion to produce 
correct evolution kernels. We ignored however the effects of the gluon field. In the analysis of the two-to-three transitions, we have to restore this neglected
contributions. Here we demonstrate how this can be achieved. We will choose a nontrivial diagram from Sect.\ \ref{sec63} to illustrate our point. For 
simplicity we only consider quark fields. For gluons, similar logic is applicable in a straightforward fashion though the algebra gets a bit more involved. 
We start with the equation for a massless quark field,
\begin{align}
\slashed{D}\psi(z)=\left(\slashed{\partial}-ig \slashed{A}(z)\right)\psi(z)=0
\end{align}
that translates in momentum space to
\begin{align}
\label{C3}
\slashed{p}\psi(p) = - g 
\int d^4 p' \slashed{A}(p')\psi(p - p')
\, .
\end{align} 
As before, the fields' momenta do not possess the ``$-$'' components. 

A way to extract the gluon splitting off the quark line due to the equation of motion is to collect terms proportional to $\slashed{p}$. In practice, however, this 
proves to be extremely difficult. The trick instead is to explicitly spill a gluon off the quark line with ordinary Feynman rules, $\psi (p) \to \psi (p_1) A (p_2)$ as 
shown in the graph
\begin{figure}[H]
    \centering
    \label{fig2}
	\psfrag{p1}[cl][cl]{\small$p_1$}
	\psfrag{p2}[cl][cl]{\small$p_2$}
	\psfrag{p1pp2}[cr][cr]{\small$p_1+p_2$}
	\psfrag{a mu}[cr][cr]{\small$a\,\,\mu$}
	\psfrag{b nu}[cl][cc]{\small$b\,\,\nu$}
  \makebox[\textwidth]{\includegraphics[width=0.15\paperwidth]{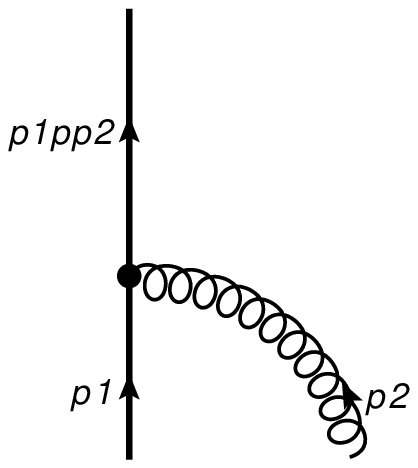}}
\end{figure} 
\noindent The emerging quark propagator $\mathcal{P} (p_1 + p_2)$ has a pole as it goes on-shell due to the collinearity of emitted gluon off a collinear 
quark. However, we can introduce nonvanishing transverse momenta for outgoing quarks and gluons and then collect the terms of the form $(p_1+p_2)^2=
(p_1+p_2)_\perp^2$ to cancel the singular denominator in $\mathcal{P} (p_1 + p_2)$.

As an explicit example, let us consider the diagram
 \begin{figure}[H]
    \centering
\label{fig2}
     \psfrag{i}[bc][bc]{\small$i$}
	\psfrag{i'}[tc][tc]{\small$i'$}
	\psfrag{a}[bc][bc]{\small$a$}
	\psfrag{d}[tc][tc]{\small$d$}
	\psfrag{a'}[tc][tc]{\small$a'$}
	\psfrag{k1}[cl][cl]{\small$k_1$}
	\psfrag{k2}[cl][cl]{\small$k_2$}
	\psfrag{p1}[cl][cl]{\small$p_1$}
	\psfrag{p2}[cl][cl]{\small$p_2$}
	\psfrag{p3}[c][]{\small$p_3$}
	\psfrag{k1mp1}[cl][cl]{\small$k_1-p_1$}
	\psfrag{k1mp3}[cl][cl]{\small$k_1-p_3$}
	\psfrag{g1}[cc][cl]{\small$C_1$}
	\psfrag{g2}[cc][cl]{\small$C_1+C_2$}
	\psfrag{g3}[cc][cl]{\small$-(C_1+C_3)$}
	\psfrag{g4}[cc][cl]{\small$(C_1+C_3);C_1;(-C_3)$}
	\psfrag{g5}[cc][cl]{\small$C_2$}
	\psfrag{g6}[cc][cl]{\small$C_3$}
	\psfrag{g7}[cc][cl]{\small$C_3+C_6-C_5$}
	\psfrag{g8}[cc][cl]{\small$C_4$}
	\psfrag{g9}[cc][cl]{\small$C_4-C_5$}
	\psfrag{g10}[cc][cl]{\small$C_5$}
	\psfrag{g11}[cc][cl]{\small$C_6-C_5$}
	\psfrag{g12}[cc][cl]{\small$C_6$}
	\psfrag{g13}[cc][cl]{\small$C_6$}
    \makebox[\textwidth]{\includegraphics[width=0.15\paperwidth]{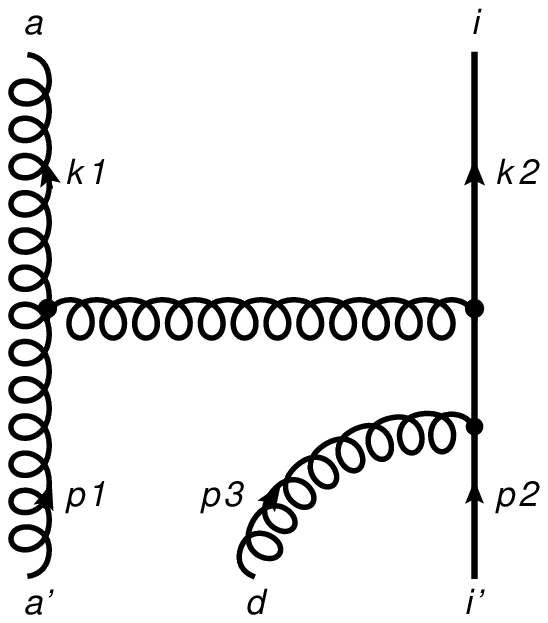}}
\end{figure}
\noindent that corresponds to the transition of $\bar{f}_{++}^a(y_1)\psi_-(y_2)$ into $\bar{f}_{++}^a(y_1)\psi_+(y_2)\bar{f}^d_{++}(y_3)$. In 
the light cone gauge, this can be translated into the expression
\begin{align}
\mathcal{I}&
=
\frac{\sqrt[4]{8}
\gamma_{\perp\sigma}\gamma_{\perp}^{\alpha}\bar{\gamma}_{\perp}\gamma^+\gamma^-\slashed{k}{}_2
\gamma^{\nu}\slashed{q}{}_{\perp}\slashed{A}{}_{\perp}
}{4
k^2k_2^2(p_1-k)^2q^2_{\perp}
}
\big((2p_1-k)^{\lambda}g^{\mu\sigma}+(2k-p_1)^{\sigma}g^{\mu\lambda}-(k+p_1)^{\mu}g^{\lambda\sigma}\big)
\nonumber\\
&\times
\left(
g_{\alpha\lambda}-\frac{k_{\alpha}
n_{\lambda}+k_{\lambda}n_{\alpha}}{k^+}
\right)
\left(
g_{\mu\nu}-\frac{(p_1-k)_{\mu}n_{\nu}+(p_1-k)_{\nu}n_{\mu}}{(p_1-k)^+}
\right)
\, ,
\end{align}
wherer $k\equiv k_1, q\equiv p_2+p_3$. Performing the Sudakov decomposition for all momenta and performing some algebra, we find
\begin{align}
\mathcal{I}
&= -\frac{g^3\sqrt[4]{2}\sqrt{2}\bar{A}_{\perp}\Gamma\ln\mu}{8\pi^2(x_1-y_1)y_1y_3}
\big[
x_1\vartheta^0_{111}(x_1,x_1-y_1,-x_2)+\vartheta^0_{112}(x_1,x_1-y_1,-x_2)
\nonumber\\
&\qquad\qquad\qquad\quad
+(x_1^2+(x_2-y_1)y_1+x_1(x_2+y_1)\vartheta^1_{112}(x_1,x_1-y_1,-x_2))
\nonumber\\
&\qquad\qquad\qquad\quad
+(x_1+y_1)\vartheta^1_{111}(x_1,x_1-y_1,-x_2)
\big] (t^et^d)_{ii'}f^{aa'e}
\, ,
\end{align}
where we have restored the color structures and $\Gamma =\gamma^+(1-i\gamma^1\gamma^2)$ is the matrix structure corresponding to  the projection onto 
operator $\bar{f}_{++}^a(y_1)\psi_+(y_2)\bar{f}^d_{++}(y_3)$. Here only terms proportional to $q_{\perp}^2$ are kept since they are the ones corresponding to 
the equation of motion operators that we are dealing with. We should point out  that our management of the equation-of-motion diagrams coincide with the 
procedures described in Ref.\ \cite{Rohrwildthesis}. 

\section{Light-ray kernels in section \ref{sec62} and \ref{sec63}}
\label{appC}

Here we convert our diagrammatic results given in Sects.\ \ref{sec62} and \ref{sec63} into the coordinate space. This is done by inverse-Fourier transform 
the corresponding kernels in momentum space making use of formulas in Sect.\ \ref{Fourier} and applied in Appendix \ref{FourierAppendix}.

\subsection{Coordinate kernes for Sect.\ \ref{sec62}}
\label{appc1}

We start with the operators involving transverse derivatives\footnote{Here, $\mathbb{O}^{ia}$ and $\mathbb{O}^{i'a'd}$ corresponds to operator 
$X$ and $Y$ in Ref.\ \cite{Braun:2009vc}, respectively.} $\mathbb{O}^{ia}(z_1,z_2)=\tfrac{1}{2}D_{-+}\bar{\psi}^i_{+}(z_1)\bar{f}_{++}^a(z_2)$ and
 $\mathbb{O}^{ia}(z_1,z_2)=\bar{\psi}^i_{+}(z_1)\tfrac{1}{2}D_{-+}\bar{f}_{++}^a(z_2)$ evolving into 
$\mathbb{O}^{iad} (z_1, z_2, z_3)=g \sqrt{2}\bar{\psi}_{+}^i(z_1)\bar{f}_{++}^a(z_2)\bar{f}^d_{++}(z_3)$. For both of them, the transition
gets decomposed into the same color-flow structures and  reads 
\begin{align}
[\mathbb{H}^{(2 \to 3)} \mathbb{O}]^{ia}(z_1,z_2)&=\sum\limits_{c=1}^{6} [C_c]^{ia}_{a'i'd} [ \mathbb{H}_c \mathbb{O}^{i'a'd}] (z_1,z_2,z_3)\, ,
\end{align}
where $C_c$ are defined in Eq.\ \re{612}. The kernels for the first operator, i.e., $\tfrac{1}{2}D_{-+}\bar{\psi}^i_{+}(z_1)\bar{f}_{++}^a(z_2)$
read
\begin{align}
[\mathbb{H}_1 \mathbb{O}]^{ia}(z_1,z_2)&=z_{12}\bigg\{\int^1_0d\beta\bar{\beta}\, \mathbb{O}^{i'a'd}(z_1,z_2,z^{\beta}_{12})+\int^1_0d\alpha\int^1_{\bar{\alpha}}d\beta\,\frac{\bar{\beta}^2}{\beta}\, \mathbb{O}^{i'a'd}(z^{\alpha}_{21},z_2,z^{\beta}_{12})
\nonumber
\\
&\ \ \ \ \ \ \ \ \ +2\int^1_0d\alpha\int^1_{\bar{\alpha}}d\beta\,\alpha\bar{\beta}\, \mathbb{O}^{i'a'd}(z_1,z^{\alpha}_{12},z^{\beta}_{21})\bigg\}\, ,
\\
[\mathbb{H}_2 \mathbb{O}]^{ia}(z_1,z_2)&=z_{12}\int^1_{0}d\alpha\int^1_{\bar{\alpha}}d\beta\,\frac{\bar{\alpha}\beta}{\alpha}\bigg(2-\frac{\bar{\alpha}\bar{\beta}}{\alpha\beta}\bigg)
 \mathbb{O}^{i'a'd}(z^{\alpha}_{12},z_2,z^{\beta}_{21})\, ,
\\
[\mathbb{H}_3 \mathbb{O}]^{ia}(z_1,z_2)&=-z_{12}\bigg\{\int^1_0d\beta\bar{\beta}\, \mathbb{O}^{i'a'd}(z_1,z^{\beta}_{12},z_2)+\int^1_0d\alpha\int^1_{\bar{\alpha}}d\beta\,\frac{\bar{\beta}^2}{\beta}\, \mathbb{O}^{i'a'd}(z^{\alpha}_{21},z^{\beta}_{12},z_2)
\nonumber
\\
&\ \ \ \ \ \ \ \ \ +2\int^1_0d\alpha\int^1_{\bar{\alpha}}d\beta\,\alpha\bar{\beta}\, \mathbb{O}^{i'a'd}(z_1,z^{\beta}_{21},z^{\alpha}_{12})\bigg\}\, ,
\\
[\mathbb{H}_4 \mathbb{O}]^{ia}(z_1,z_2)&=-2z_{12}\int^1_{0}d\alpha\int^1_{\bar{\alpha}}d\beta\,\bar{\alpha}\beta\,
 \mathbb{O}^{i'a'd}(z_2,z^{\alpha}_{12},z^{\beta}_{21})\, ,
\\
[\mathbb{H}_5 \mathbb{O}]^{ia}(z_1,z_2)&=z_{12}\bigg\{\int^1_{0}d\alpha\int^1_{\bar{\alpha}}d\beta\frac{\bar{\alpha}\beta}{\alpha}\bigg(2-\frac{\bar{\alpha}\bar{\beta}}{\alpha\beta}\bigg) \mathbb{O}^{i'a'd}(z_{12}^{\alpha},z_{21}^{\beta},z_2)
\nonumber
\\
&-2\int^1_{0}d\alpha\int^{\bar{\alpha}}_0d\beta\,\bar{\alpha}\beta\, \mathbb{O}^{i'a'd}(z_2,z^{\alpha}_{12},z^{\beta}_{21})\bigg\}\, ,
\\
[\mathbb{H}_6 \mathbb{O}]^{ia}(z_1,z_2)&=-z_{12}\int^1_{0}d\alpha\int^1_{\bar{\alpha}}d\beta\frac{\bar{\alpha}\beta}{\alpha}\bigg(2-\frac{\bar{\alpha}\bar{\beta}}{\alpha\beta}\bigg)\mathbb{O}^{i'a'd}(z_{12}^{\alpha},z_{21}^{\beta},z_2)\, .
\end{align}
Here the symmetry of $a\leftrightarrow d, w_2\leftrightarrow w_3$ described in the main text of Sect.\ \ref{sec62} becomes manifest. Notice that the 
kernels $\mathbb{H}_1$ and $\mathbb{H}_3$ can be mapped into each other by a simple exchange of the gluon fields 
$\mathbb{O}(w_1,w_2,w_3)\leftrightarrow \mathbb{O}(w_1,w_3,w_2)$. This serves as another check for our kernels.

For the $\mathbb{O}^{ia}(z_1,z_2)=\bar{\psi}^i_{+}(z_1)\tfrac{1}{2}D_{-+}\bar{f}_{++}^a(z_2)$ case, we find
\begin{align}
[\mathbb{H}_1 \mathbb{O}]^{ia}(z_1,z_2)&=z_{12}\bigg\{\int^1_0d\alpha\int^1_{\bar{\alpha}}d\beta\,\frac{\bar{\alpha}^2\beta}{\alpha} \mathbb{O}^{i'a'd}(z_1,z^{\beta}_{12},z^{\alpha}_{21})-\int^1_0d\beta\,\bar{\beta}\, \mathbb{O}^{i'a'd}(z_1,z_2,z^{\beta}_{12})
\nonumber
\\
&\quad\quad\quad
+\int^1_0d\alpha\int^1_{\bar{\alpha}}d\beta\frac{\alpha\bar{\beta}^2}{\beta}\bigg(2-\frac{\bar{\alpha}\bar{\beta}}{\alpha\beta}\bigg) \mathbb{O}^{i'a'd}(z_1,z^{\alpha}_{12},z^{\beta}_{21})
\nonumber
\\
&\quad\quad\quad+\int^1_{0}d\alpha\int^{\bar{\alpha}}_0d\beta\,\beta\, \mathbb{O}^{i'a'd}(z^{\alpha}_{12},z_2,z^{\beta}_{21})\bigg\}
\, ,
\\
[\mathbb{H}_2 \mathbb{O}]^{ia}(z_1,z_2)&=z_{12}\int^1_0d\alpha\int^1_{\bar{\alpha}}d\beta\,\beta\, \mathbb{O}^{i'a'd}(z^{\alpha}_{12},z_2,z^{\beta}_{21})\, ,
\\
[\mathbb{H}_3 \mathbb{O}]^{ia}(z_1,z_2)&=-z_{12}\bigg\{\int^1_0d\alpha\int^1_{\bar{\alpha}}d\beta\,\frac{\bar{\alpha}^2\beta}{\alpha}\mathbb{O}^{i'a'd}(z_1,z^{\alpha}_{21},z^{\beta}_{12})-\int^1_0d\beta\,\bar{\beta}\,\mathbb{O}^{i'a'd}(z_1,z^{\beta}_{12},z_2)
\nonumber
\\
&\quad\qquad\quad
+\int^1_0d\alpha\int^1_{\bar{\alpha}}d\beta\frac{\alpha\bar{\beta}^2}{\beta}\bigg(2-\frac{\bar{\alpha}\bar{\beta}}{\alpha\beta}\bigg)\mathbb{O}^{i'a'd}(z_1,z^{\beta}_{21},z^{\alpha}_{12})
\nonumber
\\
&\quad\qquad\quad+\int^1_{0}d\alpha\int^{\bar{\alpha}}_0d\beta\,\beta\,\mathbb{O}^{i'a'd}(z^{\alpha}_{12},z^{\beta}_{21},z_2)\bigg\}\, ,
\\
[\mathbb{H}_4 \mathbb{O}]^{ia}(z_1,z_2)&=-z_{12}\bigg\{\int^1_0d\alpha\int^1_{\bar{\alpha}}d\beta\,\bar{\alpha}\bar{\beta}\,\mathbb{O}^{i'a'd}(z_2,z_{12}^{\alpha},z_{21}^{\beta})+\int^1_0\int^{\bar{\alpha}}_0d\beta\,\bar{\alpha}\bar{\beta}\,\mathbb{O}^{i'a'd}(z_2,z^{\beta}_{21},z^{\alpha}_{12})\bigg\}\, ,
\\
[\mathbb{H}_5 \mathbb{O}]^{ia}(z_1,z_2)&=-z_{12}\bigg\{\int^1_0d\alpha\int^1_{\bar{\alpha}}d\beta\,\bar{\alpha}\bar{\beta}\,\mathbb{O}^{i'a'd}(z_2,z_{21}^{\beta},z_{12}^{\alpha})+\int^1_0\int^{\bar{\alpha}}_0d\beta\,\bar{\alpha}\bar{\beta}\,\mathbb{O}^{i'a'd}(z_2,z^{\alpha}_{12},z^{\beta}_{21})
\nonumber
\\
&\quad\qquad\quad+\int^1_{0}d\alpha\int^1_{\bar{\alpha}}d\beta\,\beta\,\mathbb{O}^{i'a'd}(z^{\alpha}_{12},z^{\beta}_{21},z_2)\bigg\}\, ,
\\
[\mathbb{H}_6 \mathbb{O}]^{ia}(z_1,z_2)&=-z_{12}\int^1_{0}d\alpha\int^1_{\bar{\alpha}}d\beta\,\beta\,\mathbb{O}^{i'a'd}(z^{\alpha}_{12},z^{\beta}_{21},z_2)\, .
\end{align}

\subsection{Coordinate kernels for Sect.\ \ref{sec63}}
\label{appc2}

As in Sect.\ \ref{appc1}, we present here coordinate-space transition of $\mathbb{O}^{ai}(z_1,z_2)=\bar{f}_{++}^a(z_1)\psi_{-}^i(z_2)$
and $\mathbb{O}^{ai}(z_1,z_2)=\tfrac{1}{2}D_{-+}\bar{f}_{++}^a(z_1)\psi^i_{+}(z_2)$ into three-particle operator
$\mathbb{O}^{aid}=g\sqrt{2}\bar{f}_{++}^a(z_1) \psi_{+}^i(z_2)\bar{f}^d_{++}(z_3)$. The action of the Hamiltonian yields the decomposition
\begin{align}
[\mathbb{H}^{(2 \to 3)}\mathbb{O}]^{ai}(z_1,z_2)&
= \sum\limits_{c=1}^{6}[C_c]^{ia}_{a'i'd} [\mathbb{H}_c \mathbb{O}^{a'i'd}](z_1,z_2,z_3)\, ,
\end{align}
where the color structures are introduced in Eq.\ \re{color2}. 

Then for $\bar{f}_{++}^a(z_1)\psi_{-}^i(z_2)$, we get
\begin{align}
[\mathbb{H}_1 \mathbb{O}]^{ai}(z_1,z_2)&=-z_{12}^2\bigg\{2\int^1_0d\alpha\int^{\bar{\alpha}}_0d\beta\int^{\bar{\alpha}}_{\beta}d\gamma\, \bar{\alpha}\gamma\, \mathbb{O}^{a'i'd}(z_{12}^{\alpha},z_{21}^{\beta},z_{21}^{\gamma})
\nonumber
\\
&\quad\quad\quad+\int^1_0d\alpha\int^1_{\bar{\alpha}}d\beta\,\bar{\beta}\, \mathbb{O}^{a'i'd}(z_1,z_{12}^{\alpha},z_{21}^{\beta})\bigg\}\, ,
\\
[\mathbb{H}_2 \mathbb{O}]^{ai}(z_1,z_2)&=-z_{12}^2\bigg\{2\int^1_0d\alpha\int^{\bar{\alpha}}_0d\beta\int^{1}_{\bar{\alpha}}d\gamma\, \bar{\alpha}\gamma\, \mathbb{O}^{a'i'd}(z_{12}^{\alpha},z_{21}^{\beta},z_{21}^{\gamma})
\nonumber
\\
&\quad\qquad\quad-\int^1_0d\alpha\int^{\alpha}_0d\beta\,\alpha\, \mathbb{O}^{a'i'd}(z_{21}^{\alpha},z_{21}^{\beta},z_1)\bigg\}\, ,
\\
[\mathbb{H}_3 \mathbb{O}]^{ai}(z_1,z_2)&=z_{12}^2\bigg\{2\int^1_0d\alpha\int^{\bar{\alpha}}_0d\beta\int^{\beta}_0d\gamma\, \bar{\alpha}\gamma\, \mathbb{O}^{a'i'd}(z_{12}^{\alpha},z_{21}^{\beta},z_{21}^{\gamma})
\nonumber
\\
&\quad\qquad\quad+\int^1_0d\alpha\int_{\bar{\alpha}}^1d\beta\,\frac{\bar{\alpha}\bar{\beta}}{\alpha}\, \mathbb{O}^{a'i'd}(z_1,z_{21}^{\alpha},z_{12}^{\beta})\bigg\}\, ,
\\
[\mathbb{H}_4 \mathbb{O}]^{ai}(z_1,z_2)&=2z_{12}^2\int^1_0d\alpha\int_{\bar{\alpha}}^1d\beta\int^{\bar{\alpha}}_0d\gamma\, \bar{\alpha}\gamma\, \mathbb{O}^{a'i'd}(z_{12}^{\alpha},z_{21}^{\beta},z_{21}^{\gamma})\, ,
\\
[\mathbb{H}_5 \mathbb{O}]^{ai}(z_1,z_2)&=z_{12}^2\bigg\{2\int^1_0d\alpha\int_{\bar{\alpha}}^1d\beta\int^1_{\bar{\alpha}}d\gamma\, \bar{\alpha}\gamma\, \mathbb{O}^{a'i'd}(z_{12}^{\alpha},z_{21}^{\beta},z_{21}^{\gamma})
\nonumber
\\
&\quad\qquad\quad+\int^1_0d\alpha\int^1_{\bar{\alpha}}d\beta\,
\frac{\bar{\alpha}\bar{\beta}}{\alpha}\,  \mathbb{O}^{a'i'd}(z^{\beta}_{12},z_{21}^{\alpha},z_1)\bigg\}\, ,
\\
[\mathbb{H}_6 \mathbb{O}]^{ai}(z_1,z_2)&=-z_{12}^2\bigg\{2\int^1_0d\alpha\int_{\bar{\alpha}}^1d\beta\int^1_{\beta}d\gamma\, \bar{\alpha}\gamma\, \mathbb{O}^{a'i'd}(z_{12}^{\alpha},z_{21}^{\beta},z_{21}^{\gamma})
\nonumber
\\
&\quad\qquad\quad+\int^1_0d\alpha\int^1_{\bar{\alpha}}d\beta\,
\frac{\bar{\alpha}\bar{\beta}}{\alpha}\, \mathbb{O}^{a'i'd}(z^{\beta}_{12},z_{21}^{\alpha},z_1)\bigg\}\, .
\end{align}
Here the $w_1\leftrightarrow w_3$ symmetry is readily observed. While for $\tfrac{1}{2}D_{-+}\bar{f}_{++}^a(z_1)\psi^i_{+}(z_2)$ case, the transitions are
\begin{align}
[\mathbb{H}_1 \mathbb{O}]^{ai}(z_1,z_2)&=z_{12}\bigg\{\int^1_0d\beta\,\bar{\beta}\,\mathbb{O}^{a'i'd}(z_1,z_2,z_{12}^{\beta})+\int^1_0d\alpha\int^1_{\bar{\alpha}}d\beta\frac{\alpha\bar{\beta}^2}{\beta}\bigg(2-\frac{\bar{\alpha}\bar{\beta}}{\alpha\beta}\bigg)\mathbb{O}^{a'i'd}(z^{\alpha}_{21},z_2,z_{12}^{\beta})
\nonumber
\\
&\quad\qquad+\int^1_0d\alpha\int^{\bar{\alpha}}_0d\beta\int^{\bar{\alpha}}_{\beta}d\gamma\,\bar{\alpha}\gamma
\bigg(4-\frac{\alpha\gamma}{\bar{\alpha}\bar{\gamma}}\bigg)\,\mathbb{O}^{a'i'd}(z^{\alpha}_{12},z^{\beta}_{21},z^{\gamma}_{21})
\nonumber
\\
&\quad\qquad+\int^1_0d\alpha\int^{\bar{\alpha}}_0d\beta\int^1_{\bar{\alpha}}d\gamma\,\bar{\alpha}\gamma
\bigg(2+\frac{\bar{\alpha}\bar{\gamma}}{\alpha\gamma}\bigg)\,\mathbb{O}^{a'i'd}(z^{\gamma}_{21},z^{\beta}_{21},z^{\alpha}_{12})
\nonumber
\\
&\quad\qquad+\int^1_0d\alpha\int^1_{\bar{\alpha}}d\beta\frac{\bar{\alpha}^2\beta}{\alpha}\mathbb{O}^{a'i'd}(z^{\beta}_{21},z_2,z_{12}^{\alpha})\bigg\}\, ,
\\
[\mathbb{H}_2 \mathbb{O}]^{ai}(z_1,z_2)&=-z_{12}\bigg\{\int^1_0d\beta\,\bar{\beta}\,\mathbb{O}^{a'i'd}(z_{12}^{\beta},z_2,z_1)+\int^1_0d\alpha\int^1_{\bar{\alpha}}d\beta\frac{\alpha\bar{\beta}^2}{\beta}\bigg(2-\frac{\bar{\alpha}\bar{\beta}}{\alpha\beta}\bigg)\mathbb{O}^{a'i'd}(z_{12}^{\beta},z_2,z^{\alpha}_{21})
\nonumber
\\
&\quad\qquad+\int^1_0d\alpha\int^{\bar{\alpha}}_0d\beta\int^{\bar{\alpha}}_{\beta}d\gamma\,\bar{\alpha}\gamma
\bigg(4-\frac{\alpha\gamma}{\bar{\alpha}\bar{\gamma}}\bigg)\,\mathbb{O}^{a'i'd}(z^{\gamma}_{21},z^{\beta}_{21},z^{\alpha}_{12})
\nonumber
\\
&\quad\qquad+\int^1_0d\alpha\int^{\bar{\alpha}}_0d\beta\int^1_{\bar{\alpha}}d\gamma\,\bar{\alpha}\gamma
\bigg(2+\frac{\bar{\alpha}\bar{\gamma}}{\alpha\gamma}\bigg)\,\mathbb{O}^{a'i'd}(z^{\alpha}_{12},z^{\beta}_{21},z^{\gamma}_{21})
\nonumber
\\
&\quad\qquad+\int^1_0d\alpha\int^1_{\bar{\alpha}}d\beta\frac{\bar{\alpha}^2\beta}{\alpha}\mathbb{O}^{a'i'd}(z_{12}^{\alpha},z_2,z^{\beta}_{21})\bigg\}\, ,
\\
[\mathbb{H}_3 \mathbb{O}]^{ai}(z_1,z_2)&=z_{12}\bigg\{\int^1_0d\alpha\int^{\bar{\alpha}}_0d\beta\int^{\beta}_0d\gamma\,\bar{\alpha}\gamma
\bigg(4-\frac{\alpha\gamma}{\bar{\alpha}\bar{\gamma}}\bigg)\,\mathbb{O}^{a'i'd}(z^{\alpha}_{12},z^{\beta}_{21},z^{\gamma}_{21})
\nonumber
\\
&\quad\qquad+\int^1_0d\alpha\int^1_{\bar{\alpha}}d\beta\int^1_{\beta}d\gamma\,\bar{\alpha}\gamma
\bigg(2+\frac{\bar{\alpha}\bar{\gamma}}{\alpha\gamma}\bigg)\,\mathbb{O}^{a'i'd}(z^{\gamma}_{21},z^{\beta}_{21},z^{\alpha}_{12})\bigg\}\, ,
\\
[\mathbb{H}_4 \mathbb{O}]^{ai}(z_1,z_2)&=-z_{12}\bigg\{\int^1_0d\alpha\int_{\bar{\alpha}}^1d\beta\int^{\bar{\alpha}}_0d\gamma\,\bar{\alpha}\gamma
\bigg(4-\frac{\alpha\gamma}{\bar{\alpha}\bar{\gamma}}\bigg)\,\mathbb{O}^{a'i'd}(z^{\alpha}_{12},z^{\beta}_{21},z^{\gamma}_{21})
\nonumber
\\
&\quad\qquad+\int^1_0d\alpha\int^1_{\bar{\alpha}}d\beta\int^1_{\bar{\alpha}}d\gamma\,\bar{\alpha}\gamma
\bigg(2+\frac{\bar{\alpha}\bar{\gamma}}{\alpha\gamma}\bigg)\,\mathbb{O}^{a'i'd}(z^{\gamma}_{21},z^{\beta}_{21},z^{\alpha}_{12})
\nonumber
\\
&\quad\qquad-\int^1_0d\alpha\int^1_{\bar{\alpha}}d\beta\int^1_{\beta}d\gamma\,\bar{\alpha}\gamma
\bigg(2+\frac{\bar{\alpha}\bar{\gamma}}{\alpha\gamma}\bigg)\,\mathbb{O}^{a'i'd}(z^{\gamma}_{21},z^{\beta}_{21},z^{\alpha}_{12})\bigg\}\, ,
\\
[\mathbb{H}_5 \mathbb{O}]^{ai}(z_1,z_2)&=-z_{12}\bigg\{\int^1_0d\alpha\int^{\bar{\alpha}}_0d\beta\int^{\beta}_0d\gamma\,\bar{\alpha}\gamma
\bigg(4-\frac{\alpha\gamma}{\bar{\alpha}\bar{\gamma}}\bigg)\,\mathbb{O}^{a'i'd}(z^{\gamma}_{21},z^{\beta}_{21},z^{\alpha}_{12})
\nonumber
\\
&\quad\qquad+\int^1_0d\alpha\int_{\bar{\alpha}}^1d\beta\int^{\bar{\alpha}}_0d\gamma\,\bar{\alpha}\gamma
\bigg(4-\frac{\alpha\gamma}{\bar{\alpha}\bar{\gamma}}\bigg)\,\mathbb{O}^{a'i'd}(z^{\gamma}_{21},z^{\beta}_{21},z^{\alpha}_{12})
\nonumber\\
&\quad\qquad+\int^1_0d\alpha\int^1_{\bar{\alpha}}d\beta\int^1_{\bar{\alpha}}d\gamma\,\bar{\alpha}\gamma
\bigg(2+\frac{\bar{\alpha}\bar{\gamma}}{\alpha\gamma}\bigg)\,\mathbb{O}^{a'i'd}(z^{\alpha}_{12},z^{\beta}_{21},z^{\gamma}_{21})\bigg\}\, ,
\\
[\mathbb{H}_6 \mathbb{O}]^{ai}(z_1,z_2)&=z_{12}\bigg\{\int^1_0d\alpha\int^1_{\bar{\alpha}}d\beta\int^1_{\beta}d\gamma\,\bar{\alpha}\gamma
\bigg(2+\frac{\bar{\alpha}\bar{\gamma}}{\alpha\gamma}\bigg)\,\mathbb{O}^{a'i'd}(z^{\alpha}_{12},z^{\beta}_{21},z^{\gamma}_{21})
\nonumber
\\
&\quad\qquad+\int^1_0d\alpha\int^{\bar{\alpha}}_0d\beta\int^{\beta}_0d\gamma\,\bar{\alpha}\gamma
\bigg(4-\frac{\alpha\gamma}{\bar{\alpha}\bar{\gamma}}\bigg)\,\mathbb{O}^{a'i'd}(z^{\gamma}_{21},z^{\beta}_{21},z^{\alpha}_{12})\bigg\}\, .
\end{align}

 %%%%%%  Bibliography %%%%%%%%%%%%%%%%%%%%%%%%%%%%%%%%%%%%%%%%%%%%

%%%%%%%%%%%%%%%%%%%%%%%%%%%%%%%%%%%%%%%%%%%%%%%%%%%%%%%%%%%%%%%%
\end{document}